\documentclass[10pt,journal,compsoc]{IEEEtran}

\ifCLASSOPTIONcompsoc
  \usepackage[nocompress]{cite}
\else
  \usepackage{cite}
\fi

\usepackage{amsmath}
\usepackage{amsthm}
\usepackage{amssymb}
\usepackage{amsfonts}
\usepackage{textcomp}
\usepackage{xcolor}
\usepackage{array,float,url,graphicx,wrapfig,verbatim,color,comment}
\usepackage[style=base]{caption}
\usepackage[style=base]{subcaption}
\usepackage{amsfonts,bm}
\usepackage{algorithm,algorithmic}
\usepackage{amssymb}
\usepackage{pgfgantt}
\usepackage{multirow}
\usepackage{soul}
\usepackage{bbding}
\usepackage{enumerate}
\usepackage{booktabs}
\usepackage{tabularx}
\usepackage{bbm}
\usepackage{xr}
\makeatletter
\newcommand*{\addFileDependency}[1]{
  \typeout{(#1)}
  \@addtofilelist{#1}
  \IfFileExists{#1}{}{\typeout{No file #1.}}
}
\makeatother

\newtheorem{lemma}{\textbf{Lemma}}
\newtheorem{theorem}{\textbf{Theorem}}

\newtheorem{corollary}{\textbf{Corollary}}
\newtheorem{assumption}{\textbf{Assumption}}
\theoremstyle{definition}
\newtheorem{remark}{\textbf{Remark}}

\newcounter{relctr} 
\everydisplay\expandafter{\the\everydisplay\setcounter{relctr}{0}} 

\newcommand\labelrel[2]{%
  \begingroup
    \refstepcounter{relctr}%
    \stackrel{\textnormal{(\alph{relctr})}}{\mathstrut{#1}}%
    \originallabel{#2}%
  \endgroup
}
\AtBeginDocument{\let\originallabel\label} 

\ifCLASSINFOpdf
\else
\fi

\begin{document}
\title{Scalable and Low-Latency Federated Learning with Cooperative Mobile Edge Networking}
\author{Zhenxiao~Zhang,~\IEEEmembership{Student~Member,~IEEE,} Zhidong~Gao,~\IEEEmembership{Student~Member,~IEEE,} Yuanxiong~Guo,~\IEEEmembership{Senior~Member,~IEEE,} and Yanmin~Gong,~\IEEEmembership{Senior~Member,~IEEE}
\IEEEcompsocitemizethanks{\IEEEcompsocthanksitem Z. Zhang, Z. Gao, and Y. Gong are with the Department of Electrical and Computer Engineering, The University of Texas at San Antonio, San Antonio, TX, 78249. Y. Guo is with the Department of Information Systems and Cyber Security, The University of Texas at San Antonio, San Antonio, TX, 78249.
E-mail: \{zhidong.gao@my., zhenxiao.zhang@my., yuanxiong.guo@, yanmin.gong@\}utsa.edu}
\IEEEcompsocitemizethanks{\IEEEcompsocthanksitem Z. Zhang and Z. Gao both contributed equally to this work.}

}

\markboth{}%
{Shell \MakeLowercase{\textit{et al.}}: Scalable and Low-Latency Federated Learning with Cooperative Mobile Edge Networking}

\IEEEtitleabstractindextext{%
\begin{abstract}

Federated learning (FL) enables collaborative model training without centralizing data. However, the traditional FL framework is cloud-based and suffers from high communication latency. On the other hand, the edge-based FL framework that relies on an edge server co-located with mobile base station for model aggregation has low communication latency but suffers from degraded model accuracy due to the limited coverage of edge server. In light of high-accuracy but high-latency cloud-based FL and low-latency but low-accuracy edge-based FL, this paper proposes a new FL framework based on cooperative mobile edge networking called cooperative federated edge learning (CFEL) to enable both high-accuracy and low-latency distributed intelligence at mobile edge networks. Considering the unique two-tier network architecture of CFEL, a novel federated optimization method dubbed cooperative edge-based federated averaging (CE-FedAvg) is further developed, wherein each edge server both coordinates collaborative model training among the devices within its own coverage and cooperates with other edge servers to learn a shared global model through decentralized consensus. Experimental results based on benchmark datasets show that CFEL can largely reduce the training time to achieve a target model accuracy compared with prior FL frameworks.  
\end{abstract}

}

\maketitle

\IEEEdisplaynontitleabstractindextext

%
\IEEEpeerreviewmaketitle

\IEEEraisesectionheading{\section{Introduction}\label{sec:introduction}}

The proliferation of edge devices such as smartphones and Internet-of-things (IoT) devices, each equipped with rich sensing, computation, and storage resources, leads to tremendous data being generated on a daily basis at the network edge. At the same time, artificial intelligence (AI) and machine learning (ML) are advancing rapidly and enable efficient knowledge extraction from large volumes of data. The convergence of 5G networks and AI/ML leads to many emerging applications with significant economic and societal impacts such as autonomous driving \cite{zhang2021real}, augmented reality \cite{sheth2017augmented}, real-time video analytics \cite{ananthanarayanan2017real}, mobile healthcare \cite{subasi2018iot}, and smart manufacturing \cite{tao2018data}. A salient feature of these emerging domains is the large and continuously streaming data that these applications generate, which must be processed efficiently enough to support real-time learning and decision making based on these data. 

The standard ML paradigm requires centralizing the data at the cloud, which involves large amounts of distributed data transferred from the network edge to the cloud with high communication cost and privacy risk. An alternative paradigm is \emph{Federated Learning (FL)}, which enables edge devices to collaboratively learn a shared prediction model under the orchestration of the cloud while keeping all the personal data that may contain private information on device \cite{mcmahan2017communication}. Compared with the traditional centralized ML, FL is capable of reducing communication cost, improving latency, and enhancing data privacy while obtaining an accurate shared learning model for on-device inference, and therefore has received significant attention recently \cite{kairouz2021advances}. 

Despite of its great potential, FL faces a major bottleneck in communication efficiency. Specifically, in the current cloud-based FL framework, edge devices need to repeatedly download the global model from the remote cloud and upload local model updates of large data size (e.g., million of parameters for modern DNN models) to the cloud for many times in order to learn an accurate shared model. Although communication compression techniques such as quantization and sparsification \cite{konecny2016federated} have been developed to improve the communication efficiency of FL, due to the long-distance and limited-bandwidth transmissions between an edge device and the remote cloud, the model training in cloud-based FL is inevitably slow and fails to meet the latency requirements of delay-sensitive intelligent applications. 

As more computing and storage resources are being deployed at the mobile network edge in 5G-and-beyond networks, the edge-based FL framework, where an edge server co-located with mobile base station serves as the aggregator to coordinate FL among its proximate edge devices, is gaining popularity \cite{tran2019federated,zhu2019broadband,zhu2020one,chen2020joint,yang2019scheduling,amiri2020machine,ren2020accelerating,zhu2020toward,ren2020scheduling,zeng2021energy}. Although this framework can speed up model training by mitigating the cloud bottleneck and saving long-distance data transmission, an edge server can only access a limited number of edge devices and their collected data. As the ML model performance highly depends on the data volume, edge-based FL cannot meet the accuracy requirements of AI-powered applications that could be safety-critical such as autonomous driving and mobile healthcare. To address the limited coverage issue of edge-based FL framework, hierarchical FL framework \cite{liu2020client} that relies on the cloud to coordinate multiple edge servers has been proposed, but it still suffers from high communication latency with the cloud. 

In light of the high-accuracy but high-latency cloud-based FL and low-latency but low-accuracy edge-based FL, this paper proposes a new FL framework called cooperative federated edge learning (CFEL) to achieve both high-accuracy and low-latency training over wireless edge networks. The key idea of CFEL is to leverage a network of cooperative edge servers located at the wireless edge, rather than relying on a central cloud server or multiple independent edge servers, to facilitate FL among large numbers of edge devices distributed over a wide area. By eliminating the costly communication with the cloud, CFEL can achieve lower model training latency than cloud-based FL, and by tapping into more data from a larger set of edge devices, CFEL can obtain higher model accuracy than edge-based FL. Moreover, due to the distributed system nature of CFEL, there does not exist a single bottleneck, making the framework more scalable than previous frameworks. Although promising, CFEL contains multiple cooperative aggregators rather than a single aggregator as assumed in prior FL frameworks, making the classic federated averaging (FedAvg) algorithm \cite{mcmahan2017communication} not directly applicable. To address that, we further design an efficient federated optimization method for CFEL dubbed cooperative edge-based federated averaging (CE-FedAvg), wherein each edge server first obtains an edge model from the set of edge devices associated to it using FedAvg and then cooperates with other edge servers to learn a shared global model through decentralized consensus. %
%
%
%

In summary, the main contributions of this paper are as follows:
\begin{itemize}
    \item We propose CFEL, a novel FL framework at mobile edge networks, to achieve both high-accuracy and low-latency model training based on cooperative mobile edge networking. CFEL is more scalable than prior FL frameworks by exploiting multiple aggregators and eliminating a single point of failure. 
    
    \item Considering the unique network architecture of CFEL, we design a new federated optimization method named CE-FedAvg that can learn a shared global model efficiently over the collective dataset of all edge devices under the orchestration of a distributed network of cooperative edge servers.
 
 
    \item We prove the convergence of CE-FedAvg theoretically and derive its convergence rates under general assumptions about the loss function, data distribution, and network topology. The obtained convergence guarantees are tighter than those in literature and provide new insights about the algorithm design. 
   
    
    \item We conduct extensive experiments based on common FL benchmark datasets and demonstrate that CFEL can learn an accurate model within a shorter time than other FL frameworks at mobile edge networks. 
\end{itemize}

\section{Related works}
\begin{table}[t]
\centering
  \caption{Comparison of algorithms in multi-server FL setting.}
  \label{tab:comparison}
    \setlength{\tabcolsep}{1mm}{
   \begin{tabular}{ccccc}
   \toprule
  Algorithm & non-IID & non-convex & \begin{tabular}{@{}c@{}}fault \\ tolerance\end{tabular} &\begin{tabular}{@{}c@{}}local \\ aggregation benefit\end{tabular}\\
  \midrule
  {Hier-FAvg}\cite{liu2020client} &\Checkmark&\Checkmark&\XSolid&\XSolid\\
    Hier-FAvg\cite{wang2022demystifying} &\Checkmark&\Checkmark&\XSolid&\Checkmark\\
    P-FedAvg \cite{zhong2021p} & \Checkmark & \XSolid & \Checkmark &\XSolid\\
    MLL-SGD\cite{castiglia2020multi} & \XSolid & \Checkmark & \Checkmark &\Checkmark\\
    SE-FEEL\cite{sun2021semi} &\Checkmark&\Checkmark&\Checkmark&\XSolid\\
    Ours&\Checkmark&\Checkmark&\Checkmark&\Checkmark\\
    \bottomrule
\end{tabular}}
\end{table}

FL at mobile edge networks suffers from high training latency due to limited communication bandwidth. To address this issue, various communication-efficient distributed learning algorithms have been proposed to improve the communication efficiency of FL. Specifically, McMahan et al. \cite{mcmahan2017communication} proposed FedAvg to reduce the number of communication rounds by running multiple steps of SGD update on devices before aggregating their updates at the server to compute the new model. Various communication compression techniques such as sparsification \cite{wang2018atomo} and quantization \cite{reisizadeh2020fedpaq} were also designed to reduce the size of messages transmitted between the server and devices in each communication round of FL. Considering the resource constraints of mobile edge networks, learning and resource allocation were jointly optimized in \cite{tran2019federated,chen2020joint,yang2019scheduling,ren2020accelerating,ren2020scheduling,zeng2021energy} to minimize the training latency of FedAvg at mobile edge networks. All of the aforementioned studies assume a single server that aggregates model updates from all devices in each communication round. However, since the coverage of a single edge server is inherently limited, the proposed solutions cannot scale to a large number of devices.  

A few recent studies \cite{castiglia2020multi,sun2021semi,liu2020client,zhong2021p,wang2022demystifying} have considered multiple edge servers for FL at mobile edge networks, each responsible for aggregating model updates from a subset of devices. In particular, hierarchical FL and the associated hierarchical federated averaging  (Hier-FAvg) optimization algorithm were developed in \cite{liu2020client,wang2022demystifying} that relies on a central entity (e.g., the cloud) to coordinate multiple edge servers in a star topology. As the central entity can become the bottleneck and suffer from a single point of failure, the fault-tolerance and scalability of hierarchical FL is still a concern. Alternatively, decentralized coordination among edge servers without relying on a central entity like the setting of CFEL has been considered in \cite{castiglia2020multi,sun2021semi,zhong2021p}. Castiglia et al.\cite{castiglia2020multi} proposed Multi-Level Local SGD (MLL-SGD) in a two-tier communication network with heterogeneous workers, but it only considers the IID data distribution. Zhong et al.\cite{zhong2021p} proposed a similar algorithm called P-FedAvg, but it only considers the convex model, and the global and local model aggregations operate at the same frequency. The concurrent work \cite{sun2021semi} is mostly related to ours, but as elaborated later, our convergence result is much tighter than theirs and gives new insights on why frequent local model aggregation helps and which system design works better. A detailed comparison between our algorithm and prior algorithms under the same system setting is summarized in Table \ref{tab:comparison}.
\section{System Model and Problem Formulation}

\begin{figure}[t]
\centering
\includegraphics[width=0.48\textwidth]{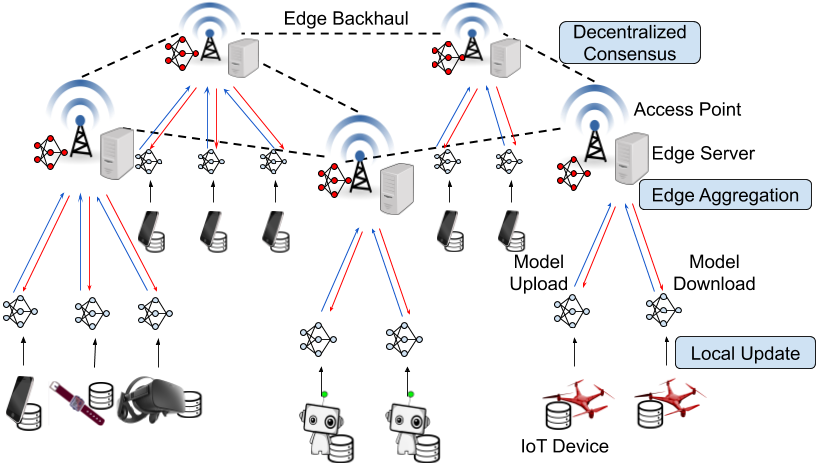}
\caption{CFEL: Cooperative Federated Edge Learning.}\label{fig:system}
\end{figure}

Consider a CFEL system depicted in Fig.~\ref{fig:system}. Assume a set of $m$ clusters in the system. Each cluster $i \in [m]$ contains a single edge server co-located with the base station and a set of devices $\mathcal{S}_i$ with $n_i = \lvert \mathcal{S}_i\rvert$. Devices in $\mathcal{S}_i$ only communicate with the server in the same cluster using the device-edge links. Define the set of all devices in the system as $\mathcal{S} = \cup_{i = 1}^{m} \mathcal{S}_i$, and the total number of devices $n = \lvert \mathcal{S}\rvert$. The edge servers communicate with each other over the edge backhaul. The communication pattern of edge backhual is represented as an undirected and connected graph $\mathcal{G} = \{V, E\}$, where $V$ denotes the set of of all edge servers, and each edge in the graph $(i, j) \in E$ denotes the link between edge servers $i$ and $j$. Let $\mathcal{N}_i = \{j: (i, j) \in E\}$ be the set of neighbors of server $i$ in the graph $\mathcal{G}$. A list of main notations used in the paper is summarized in Table~\ref{tab:notations}. Furthermore, let $\|\cdot\|,\|\cdot\|_{\textup{F}}$ and $\|\cdot\|_{\text{op}}$ denote the $\ell_2$ vector norm, Frobenius norm and matrix operator norm, respectively.

The goal of FL is to find a global model $\mathbf{x} \in \mathbb{R}^d$ that solves the following optimization problem: 
\begin{equation}\label{prob:overall}
\min_{\mathbf{x}} F(\mathbf{x}) = \frac{1}{n}\sum_{k=1}^{n} F_k(\mathbf{x}),
\end{equation}
where $F_k(\mathbf{x}) = \mathbb{E}_{z \sim \mathcal{D}_k}[\ell_k(\mathbf{x}; z)]$ is the local objective function of device $k$, and $\mathcal{D}_k$ is the data distribution of device $k$. Here $\ell_k$ is the loss function defined by the learning task, and $z$ represents a data sample from distribution $\mathcal{D}_k$. 


To solve \eqref{prob:overall} while satisfying the communication constraints of CFEL, we decompose the problem into multiple subproblems, each for a cluster. The local objective function of the $i$-th cluster is defined as
\begin{equation}\label{prob:edge}
\min_{\mathbf{x}} f_i(\mathbf{x}) = \frac{1}{n_i}\sum_{k \in \mathcal{S}_i}  F_k(\mathbf{x}),
\end{equation}
which represents the average loss over all devices in cluster $i$. Then the global objective function~\eqref{prob:overall} can be rewritten as:
\begin{equation}\label{prob:global_edge}
\min_{\mathbf{x}} F(\mathbf{x}) = \sum_{i=1}^m \frac{n_i}{n} f_i(\mathbf{x}).
\end{equation}
In CFEL, devices in the systems collaboratively solve the above optimization problem under the coordination of the edge servers in their clusters without sharing the raw data. 

\section{Learning Algorithm Design for CFEL}

Since the CFEL system in Fig.~\ref{fig:system} contains multiple clusters without a central aggregator, the classic FedAvg algorithm is not directly applicable. In this section, we propose a new federated optimization method called \emph{Cooperative Edge-based Federated Averaging} (CE-FedAvg) to efficiently solve \eqref{prob:global_edge}. 

\begin{table}[t]
  \caption{Summary of main notations.}
  \label{tab:notations}
  \centering
  \begin{tabular}{cc}
    \toprule
    Notation & Definition\\
    \midrule
    $i, j$ &  Index for cluster\\
    $k, k^{\prime}$ & Index for device\\
    $l$ & Index for global round\\
    $r$ & Index for edge round\\
    $s$ & Index for local iteration\\
    $t$ & Index for global iteration\\
    $n$ & Total number of devices\\
    $m$ & Total number of edge servers/clusters\\
    $[m]$ & \{1, 2, \ldots, $m$\}\\
    $\mathcal{S}_i$ & Set of devices in cluster $i$\\
    $\mathcal{S}$ & Set of all devices\\
    $n_i$ & Number of devices in cluster $i$\\
    $\mathcal{G}$ & Communication graph for edge backhaul\\ 
    $\mathbf{y}_{l, r}^{(i)}$ &  Edge model of cluster $i$\\ 
    $\mathcal{D}_k$ & Data distribution of device $k$\\
    $F_k(\cdot)$ & Local objective function of device $k$\\
    $f_i(\cdot)$ & Local objective function of cluster $i$ \\
    $\mathbf{x}_{l, r, s}^{(k)}$ & Local model of device $k$\\
    $\mathbf{g}_k$ & Stochastic gradient of device $k$\\
    $\eta$ & Local learning rate\\
    $\tau$ & Intra-cluster aggregation period\\
    $q \tau$ & Inter-cluster aggregation period\\
    $\mathcal{N}_i$ & Set of neighbors of edge server $i$\\
    $\mathbf{H}$ & Mixing matrix\\
    $\pi$ & Number of gossip steps per round\\
    $\zeta$ & Second largest eigenvalue of $\mathbf{H}$\\
    $\sigma^2$ & Bounded variance\\
    $\epsilon^2$ & Inter-cluster divergence\\
    $\epsilon_i^2$ & Intra-cluster divergence of cluster $i$\\
    $i_{k}$ & Cluster index of device $k$\\
  \bottomrule
\end{tabular}
\end{table}

\subsection{Algorithm Description}

Algorithm~\ref{algorithm-3} describes our proposed CE-FedAvg algorithm for CFEL. The overall training process of CE-FedAvg is divided into multiple global rounds wherein each cluster first performs $q$ edge rounds of intra-cluster collaboration independently and then communicates with other clusters for inter-cluster collaboration. 

At the beginning of the $r$-th edge round in the $l$-th global round (i.e., $(r, l)$-th round), the edge server in each cluster $i$ first broadcasts its current edge model $\mathbf{y}_{l, r}^{(i)}$ to the associated devices $\mathcal{S}_i$ under its coverage in the system. Then, devices in each cluster initialize their local models to be the received edge model and run $\tau$ iterations of SGD to update their local models in parallel. Let $\mathbf{x}_{l, r, s}^{(k)}$ denote the local model of device $k$ at the $s$-th local iteration of $(r, l)$-th round. We have the following update equations for each device $k \in \mathcal{S}_i$:
\begin{gather}\label{eq:localSGD}
\mathbf{x}_{l, r, 0}^{(k)} \gets \mathbf{y}_{l, r}^{(i)},\\
\mathbf{x}_{l,r,s+1}^{(k)} \gets \mathbf{x}_{l,r,s}^{(k)} - \eta \mathbf{g}_k(\mathbf{x}_{l, r, s}^{(k)}), \forall s = 0, \ldots, \tau - 1, 
\end{gather}
where $\eta$ is the local learning rate, and $\mathbf{g}_k(\mathbf{x}_{l, r, s}^{(k)})$ is the stochastic gradient computed over a mini-batch $\theta_k$ sampled from the local data distribution $\mathcal{D}_k$. Next, their final updated local models $\{\mathbf{x}_{l, r, \tau}^{(k)}, \forall k \in S_i\}$ are sent to the edge server $i$ for intra-cluster model aggregation, and each edge server $i \in [m]$ updates its edge model $\mathbf{y}_{l, r+1}^{(i)}$ by averaging the received local models from all associated devices as follows:
\begin{equation}\label{eq:ave_intra}
\mathbf{y}_{l, r + 1}^{(i)} \gets \frac{1}{n_i}\sum_{k \in \mathcal{S}_i}\mathbf{x}_{l, r, \tau}^{(k)}. 
\end{equation}
Then, the same procedure repeats in the next edge round $r + 1$.

After $q$ edge rounds, the edge servers communicate with each other over the edge backhaul for inter-cluster model aggregation by averaging their models with neighboring servers in $\pi$ times using gossip protocol as follows:
\begin{equation}\label{eq:ave_gossip}
\mathbf{y}_{l + 1, 0}^{(i)} \gets\sum_{j \in \{i\}\cup\mathcal{N}_i}\mathbf{H}_{j, i}^{\pi} \mathbf{y}_{l, q}^{(j)}.
\end{equation}
Here $\mathcal{N}_i = \{j: (j, i) \in E\}$ denotes the neighbors of server $i$ in the graph $\mathcal{G}$, and $\mathbf{H} \in [0, 1]^{m \times m}$ denotes the mixing matrix with each element $\mathbf{H}_{j, i}$ being the weight assigned by server $i$ to server $j$. Note that $\mathbf{H}_{j,i} > 0$ only if servers $i$ and $j$ are directly connected in the edge backhaul. Finally, the algorithm goes to the next global round $l + 1$ until $p$ global rounds in total.

\begin{algorithm}[t]
\caption{Proposed CE-FedAvg Algorithm.}\label{algorithm-3}
\begin{algorithmic}[1]
    \STATE Initialization: initial edge models $\mathbf{y}_{0, 0}^{(i)}$, $\forall i \in [m]$, edge backhaul graph $\mathcal{G}$, mixing matrix $\mathbf{H} \in [0, 1]^{m \times m}$, intra-cluster aggregation period $\tau$, inter-cluster aggregation period $q\tau$, and number of gossip steps $\pi$. 
    \FOR{each global round $l = 0, \ldots, p - 1$}
        \FOR{each cluster $i \in [m]$ \textbf{in parallel}}
            \FOR{each edge round $r = 0, \ldots, q - 1$}
                \FOR{each device $k \in \mathcal{S}_i$ \textbf{in parallel}}
                    \STATE $\mathbf{x}_{l, r, 0}^{(k)} \gets \mathbf{y}_{l, r}^{(i)}$ \label{li:broadcast}
                    \FOR{$s = 0, \ldots, \tau - 1$}
                        \STATE Compute a stochastic gradient $\mathbf{g}_k$ over a mini-batch $\theta_k$ sampled from $\mathcal{D}_k$ \label{comp_sgd}
                        \STATE $\mathbf{x}_{l, r, s + 1}^{(k)} \gets \mathbf{x}_{l, r, s}^{(k)} - \eta \mathbf{g}_k(\mathbf{x}_{l, r, s}^{(k)})$ \label{local_upd} 
                    \ENDFOR
                \ENDFOR
             \STATE $\mathbf{y}_{l, r + 1}^{(i)} \gets \frac{1}{n_i}\sum_{k \in \mathcal{S}_i}\mathbf{x}_{l, r, \tau}^{(k)}$ \label{li:ave_intra}
            \ENDFOR
            \STATE $\mathbf{y}_{l + 1, 0}^{(i)} \gets\sum_{j \in \{i\}\cup\mathcal{N}_i}\mathbf{H}_{j, i}^{\pi} \mathbf{y}_{l, q}^{(j)}$ \label{li:ave_gossip}
        \ENDFOR
    \ENDFOR
\end{algorithmic}
\end{algorithm}

Notably, CE-FedAvg inherits the privacy benefits of classic FL schemes by keeping the original data on device and sharing only model parameters. Furthermore, CE-FedAvg is compatible with existing privacy-preserving techniques in FL such as secure aggregation \cite{bonawitz2017practical,guo2018practical}, differential privacy \cite{hu2020personalized,hu2020federated,hu2021concentrated}, and shuffling \cite{girgis2021shuffled} since only the sum rather than individual values is needed for the intra-cluster and inter-cluster model aggregations. 

\subsection{Runtime Analysis of CE-FedAvg} 
We now present a runtime analysis of CE-FedAvg. Here, the communication time of downloading models from the edge server by each device is ignored because the download bandwidth is usually much larger than upload bandwidth for the device-to-edge communication in practice \cite{kairouz2021advances}. Similarly, the computation time for model aggregation at edge servers is ignored because the involved computation workload is rather small compared to the computation capabilities of edge servers. 

In each global round of CE-FedAvg, the total delay consists of the computation time for performing $q \tau$ steps of SGD update, the communication time for performing $q$ rounds of intra-cluster model aggregation, and the communication time for performing one round of inter-cluster model aggregation consisting of $\pi$ steps of gossip averaging. Therefore, the total runtime of CE-FedAvg after $p$ global rounds can be estimated as
\begin{equation}\label{eq:time}
p \times \left[\max_k \frac{q\tau C}{c_k} + \frac{q W}{b_{d2e}} + \frac{\pi W}{b_{e2e}} \right],
\end{equation}
where $C$ is the computation workload of performing one step of SGD update, $c_k$ is the processing capability of device $k$, $W$ is the model size, $b_{\text{d2e}}$ is the uplink bandwidth from device to edge server, and $b_{\text{e2e}}$ is the bandwidth between two connecting edge servers in the backhaul. 

\subsection{Prior Algorithms as Special Cases} 

When the topology of edge backhaul $\mathcal{G}$ is fully connected and the edge models from all servers are averaged in each global aggregation round, CE-FedAvg essentially reduces to Hier-FAvg \cite{wang2022demystifying} with the same model update rule. Also, when there exists only one cluster, and all devices send their local models to a single edge server for model aggregation after $\tau$ local iterations (i.e., $m=1$), CE-FedAvg reduces to FedAvg \cite{mcmahan2017communication}. Moreover, when each cluster only contains one edge device, and each device communicates with its neighboring device after $q \tau$ iterations (i.e., $n=m$), CE-FedAvg reduces to decentralized local SGD \cite{wang2021cooperative}. Therefore, the existing algorithms can be viewed as special cases of CE-FedAvg. %
%
%
However, due to the generality of CE-FedAvg, its convergence analysis presents significant new challenges. As one of the main contributions in this paper, the convergence analysis of CE-FedAvg will be elaborated in the next section. 

\section{Convergence Analysis of CE-FedAvg}

In this section, we first describe the convergence results of CE-FedAvg with respect to the gradient norm of the objective function $F(\cdot)$ and compare CE-FedAvg with prior learning algorithms. Then we analyze the impact of various learning parameters on the convergence rates of CE-FedAvg.

\subsection{Assumptions}
Before stating our results, we make the following assumptions to facilitate our convergence analysis.


\begin{assumption}[Smoothness]\label{ass:smoothness}
Each local objective function $F_k:\mathbb{R}^d\rightarrow \mathbb{R}$ is $L$-smooth for all $k\in \mathcal{S}$ , i.e., 
\[
\|\nabla F_k(\mathbf{x}) - \nabla F_k(\mathbf{x}^\prime)\| \leq L \|\mathbf{x} - \mathbf{x}^{\prime}\|,    \; \forall \mathbf{x},\mathbf{x}^{\prime}\in \mathbb{R}^d. 
\]
\end{assumption}

\begin{assumption}[Unbiased Gradient and Bounded Variance]\label{ass:gradient}
The local mini-batch stochastic gradient is an unbiased estimator of the local gradient: $\mathbb{E}_{\theta_k}[\mathbf{g}_k(\mathbf{x})] = \nabla F_k(\mathbf{x})$ and has bounded variance: $\mathbb{E}_{\theta_k}[\|\mathbf{g}_ k(\mathbf{x}) - \nabla F_k(\mathbf{x})\|^2] \leq \sigma^2, \forall \mathbf{x} \in \mathbb{R}^d, k \in \mathcal{S}$.
\end{assumption}

\begin{assumption}[Lower Bounded]\label{ass:lowerbounded}
 There exists a constant $F_{\inf}$ such that
\[
F(\mathbf{x}) \geq F_{\inf}, \forall \mathbf{x}\in\mathbb{R}^d.
\]
\end{assumption}


\begin{assumption}[Mixing Matrix]\label{ass:mixing}
The graph $\mathcal{G}: = (V, E)$ is strongly connected and the mixing matrix $\mathbf{H} \in [0, 1]^{m \times m}$ defined on it satisfies the following:
\begin{enumerate}
    \item If $(i, j) \in E$, then $\mathbf{H}_{i, j} > 0$; otherwise, $\mathbf{H}_{i, j} = 0$. 
    
    \item $\mathbf{H}$ is doubly stochastic, i.e., $\mathbf{H}^\intercal= \mathbf{H}$.
    
    \item The magnitudes of all eigenvalues except the largest one are strictly less than 1, i.e., $\zeta =  \max \{{|\lambda_2(\mathbf{H})|,|\lambda_n(\mathbf{H})|}\}<\lambda_1(\mathbf{H})=1$
    
\end{enumerate}
\end{assumption}

\begin{assumption}[Bounded Intra-Cluster Divergence]\label{ass:gradient_local}
For each cluster $i \in V$, there exists a constant $\epsilon_{i}\geq 0$ such that $\forall \mathbf{x}\in\mathbb{R}^d$, 
\[
\frac{1}{n_{i}}\sum_{{k}\in \mathcal{S}_{i}}\|\nabla f_{i}(\mathbf{x}) - \nabla F_{k}(\mathbf{x})\|^{2} \leq \epsilon_{i}^{2}.
\]
If the local objective functions of edge devices are identical to each other within a cluster, then we have $\epsilon_i^2 = 0$. 
\end{assumption}

\begin{assumption}[Bounded Inter-Cluster Divergence]\label{ass:gradient_global}
There exists a constant $\epsilon \geq 0$ such that $\forall \mathbf{x}\in\mathbb{R}^d$, 
\[
\sum_{i=1}^m \frac{n_i}{n} \|\nabla f_i(\mathbf{x}) - \nabla F(\mathbf{x})\|^2 \leq \epsilon^2. 
\]
If the local objective functions of clusters are identical to each other, then we have $\epsilon^2=0$.
\end{assumption}


Assumptions \ref{ass:smoothness}, \ref{ass:gradient}, and \ref{ass:lowerbounded} are standard in the analysis of SGD \cite{bottou2018optimization}, \cite{guo2022hybrid}. Assumption~\ref{ass:mixing} follows the decentralized optimization literature \cite{koloskova2020unified} and ensures that the gossip step converges to the average of all the vectors shared between the nodes in the graph $\mathcal{G}$. Here, smaller $\zeta$ indicates better connectivity between edge servers. For example, for complete graphs and bipartite graphs, $\zeta=0$ and $\zeta=1$, respectively. Assumptions~\ref{ass:gradient_local} and \ref{ass:gradient_global} capture the dissimilarities of local objectives within a single and across different clusters due to data heterogeneity, respectively.


Note that most prior work in literature \cite{yu2019linear,wang2021cooperative,sun2021semi} uses the following global divergence assumption to capture the data heterogeneity:
\begin{assumption}[Bounded Global Divergence]\label{ass:gradient_single_glob}
There exists a constant $\hat{\epsilon}\geq 0$ such that $\forall \mathbf{x}\in\mathbb{R}^d$, 
\[
\frac{1}{n}\sum_{k=1}^{n}\|\nabla F_k(\mathbf{x})-\nabla F(\mathbf{x})\|^2\leq \hat{\epsilon}^2.
\] 
\end{assumption}
To see the relationship between Assumptions~\ref{ass:gradient_local}--\ref{ass:gradient_global} and Assumption~\ref{ass:gradient_single_glob}, we can split the global divergence into the intra-cluster and inter-cluster divergences as follows:
\begin{multline}\label{eq_global_inter_intra}
\frac{1}{n}\sum_{k=1}^{n} \|\nabla F_k(\mathbf{x})-\nabla F(\mathbf{x})\|^2 = \sum_{i=1}^m\frac{n_i}{n}\|\nabla f_i(\mathbf{x}) - \nabla F(\mathbf{x})\|^2 \\ +\sum_{i=1}^m\frac{n_i}{n} \times \frac{1}{n_i}\sum_{k\in\mathcal{S}_i}\|\nabla f_i(\mathbf{x})-\nabla F_k(\mathbf{x})\|^2.
\end{multline}
As discussed later in Section~\ref{sec:discussion}, by decomposing the global divergence bound into two components, our assumptions enable a tighter convergence analysis for CE-FedAvg to capture the benefit of local aggregation in accelerating convergence. 

\subsection{Update Rule for CE-FedAvg Algorithm}

Since edge servers are essentially stateless in CE-FedAvg, we focus on how device models evolve in the convergence analysis. We define $t = l q\tau+r\tau+s$, where $ l \in [0, p-1]$, $r \in [0, q-1]$ and $s \in [0, \tau-1]$, as the global iteration index, and $T= pq\tau$ as the total number of global training iterations in Algorithm~\ref{algorithm-3}. Then we can rewritten the local model $\mathbf{x}_{l, r, s}^{(k)}$ as $\mathbf{x}_t^{(k)}$. Without loss of generality, we denote the range of device indices for cluster $i \in [m]$ as $\left[\sum_{j \leq i - 1} n_{j} + 1, \sum_{j \leq i} n_{j}\right]$ with $n_0 = 0$.

The system behavior of CE-FedAvg can be summarized by the following update rule for device models:
\begin{equation}\label{eq_update_rule}
\mathbf{X}_{t+1}=(\mathbf{X}_t-\eta\mathbf{G}_t)\mathbf{W}_t,
\end{equation}
where $\mathbf{X}_t=[\mathbf{x}_t^{(1)},\dots,\mathbf{x}_t^{(n)}]\in \mathbb{R}^{d\times n}$, $\mathbf{G}_t = [\mathbf{g}_1(\mathbf{x}_t^{(1)}),\dots,\mathbf{g}_n(\mathbf{x}_t^{(n)})] \in \mathbb{R}^{d\times n}$, and $\mathbf{W}_t \in \mathbb{R}^{n\times n}$ is a time-varying operator capturing the three stages in CE-FedAvg: SGD update, intra-cluster model aggregation, and inter-cluster model aggregation. Specifically, $\mathbf{W}_t$ is defined as follows:
\begin{align}
\mathbf{W}_t = \begin{cases}
\mathbf{B}^\intercal \text{diag}(\mathbf{c}) \mathbf{H}^{\pi}\mathbf{B}, &  (t+1) \bmod q\tau = 0\\
\mathbf{B}^\intercal \text{diag}(\mathbf{c}) \mathbf{B}, & (t+1) \bmod \tau = 0 \\
& \text{ and } (t+1) \bmod q\tau \neq 0\\
\mathbf{I}_{n \times n}, & \text{otherwise,}
\end{cases}
\end{align}
where $\mathbf{B} \in \{0, 1\}^{m \times n}$ is a binary matrix with each element $\mathbf{B}_{i, k}$ denoting if device $k$ belongs to cluster $i$ (i.e., $\mathbf{B}_{i, k} = 1$) or not (i.e., $\mathbf{B}_{i, k} = 0$), $\mathbf{c} = [1/n_1, \ldots, 1/n_m] \in \mathbb{R}^m$, and $\text{diag}(\mathbf{c}) \in \mathbb{R}^{m \times m}$ is a diagonal matrix with the elements of vector $\mathbf{c}$ on the main diagonal. Specifically, for the stage of SGD update (i.e., $(t+1) \bmod \tau \neq 0$), $\mathbf{W}_t$ is the identity matrix because there is no communication between edge devices after SGD update; for the stage of intra-cluster model aggregation (i.e., $(t+1) \bmod \tau = 0$ and $(t+1) \bmod q\tau \neq 0$), $\mathbf{B}^\intercal \text{diag}(\mathbf{c}) \mathbf{B}$ captures the model averaging within each cluster independently after SGD update; and for the stage of inter-cluster model aggregation (i.e., $(t+1) \bmod q\tau = 0$), $\mathbf{B}^\intercal \text{diag}(\mathbf{c}) \mathbf{H}^{\pi}\mathbf{B}$ captures the model aggregation within each cluster followed by $\pi$ steps of gossip averaging across clusters.  


To facilitate the convergence analysis, we first introduce the quantities of interests. Multiplying $\mathbf{1}_{n}/n$ on both sides in \eqref{eq_update_rule}, we get 
\begin{equation}
\mathbf{X}_{t+1}\frac{\mathbf{1}_n}{n} =  \mathbf{X}_t \frac{\mathbf{1}_n}{n} -\eta\mathbf{G}_t \frac{\mathbf{1}_n}{n},
\end{equation}
where $\mathbf{W}_t$ disappears due to the fact that $\mathbf{1}_n/n$ is a right eigenvector of $\mathbf{B}^\intercal \text{diag}(\mathbf{c}) \mathbf{H}^{\pi}\mathbf{B}$ and $\mathbf{B}^\intercal \text{diag}(\mathbf{c}) \mathbf{B}$ with eigenvalue of 1. Define the average model as
\begin{equation}\label{eq:aver_model}
\mathbf{u}_t = \mathbf{X}_t \frac{\mathbf{1}_n}{n}. 
\end{equation}
After rearranging, one can obtain
\begin{equation}\label{eq_diff_ave}
\mathbf{u}_{t+1} =\mathbf{u}_t-\frac{\eta}{n}\sum_{k=1}^n\mathbf{g}_k(\mathbf{x}_t^{(k)}). 
\end{equation}
Note that the averaged local model $\mathbf{u}_t$ is updated via performing the perturbed SGD contributed by all devices. In the following, we will focus on the convergence of the averaged model $\mathbf{u}_t$, which is a common practice in distributed optimization literature \cite{nedic2018network}. 

%


\subsection{Convergence Results}\label{subsec_conv_resu}

We now provide the main theoretical results of the paper in Theorem~\ref{th:convergence} and Corollary~\ref{coro}. We only provide the proof sketch here and include the detailed proofs in the appendices. Define the following constants:
\begin{gather}\label{def_Omega12}
\Omega_{1}=\frac{\zeta^{2\pi}}{1-\zeta^{2\pi}}, \quad \Omega_{2}=\frac{1}{1-\zeta^{2\pi}}+\frac{2}{1-\zeta^{\pi}}+\frac{\zeta^{\pi}}{(1-\zeta^{\pi})^{2}},
\end{gather}
and $n\times n$ matrix $\mathbf{A} = \mathbf{1}_{n}\mathbf{1}_{n}^\intercal/n$. For the sake of presentation, we use $\mathbf{V}$ to denote $\mathbf{B}^\intercal \text{diag}(\mathbf{c}) \mathbf{B}$ in the following.

\begin{lemma}[Convergence Decomposition] Under Assumptions~\ref{ass:smoothness}, \ref{ass:gradient}, and \ref{ass:lowerbounded}, if the learning rate $\eta\leq\frac{1}{L}$, the iterates of Algorithm~\ref{algorithm-3} satisfy: \label{lemma_bound_ave_grad}
\begin{align}
&\frac{1}{T}\sum_{t=0}^{T-1}\mathbb{E}\|\nabla F(\mathbf{u}_t)\|^2  \leq   \underbrace{\frac{2(F(\mathbf{x}_1)-F_{\inf})}{\eta T}+\frac{\eta L\sigma^2}{n}}_{\text{fully sync SGD}} \nonumber \\
&+\underbrace{\frac{2 L^2}{nT}(\sum_{t=0}^{T-1}\mathbb{E}\|\mathbf{X}_t(\mathbf{V}-\mathbf{A})\|_{\textup{F}}^2 + \sum_{t=0}^{T-1}\mathbb{E}\|\mathbf{X}_t(\mathbf{I}-\mathbf{V})\|_{\textup{F}}^2)}_{\text{residual error}}. \nonumber
\end{align}
\end{lemma}
\begin{IEEEproof}
According to the Lipschitz Assumption~\ref{ass:smoothness}, we have:
\begin{align}
&\mathbb{E}[F(\mathbf{u}_{t+1})-F(\mathbf{u}_t)]\notag\\
&\leq \mathbb{E}[\langle\nabla F(\mathbf{u}_t),\mathbf{u}_{t+1}-\mathbf{u}_t\rangle] + \frac{L}{2}\mathbb{E}\|\mathbf{u}_{t+1}-\mathbf{u}_t\|^2\notag\\
&\labelrel={eq_03} -\eta\mathbb{E}[\langle\nabla F(\mathbf{u}_t),\overline{\mathbf{G}}(\mathbf{X}_t)\rangle] + \frac{\eta^2L}{2}\mathbb{E}\|\overline{\mathbf{G}}(\mathbf{X}_t)\|^2,\label{eq_conv}
\end{align}
where~\eqref{eq_03} holds due to~\eqref{eq_diff_ave}, and we use $\overline{\mathbf{G}}(\mathbf{X}_t)$ to denote $(1/n)\sum_{k=1}^n\mathbf{g}_k(\mathbf{x}_{t}^{(k)})$. For the first term in \eqref{eq_conv}, we have
\begin{align}
    &\mathbb{E}[\langle\nabla F(\mathbf{u}_t),\overline{\mathbf{G}}(\mathbf{X}_t)\rangle] \labelrel={eq_01} \langle\nabla F(\mathbf{u}_t),\sum_{k=1}^n\frac{1}{n}\nabla F_k(\mathbf{x}_t^{(k)})\rangle\notag\\
    & \labelrel={eq_02} \frac{1}{2}\|\nabla F(\mathbf{u}_t)\|^2 + \frac{1}{2}\|\sum_{k=1}^n\frac{1}{n}\nabla F_k(\mathbf{x}_t^{(k)})\|^2\notag\\
    &\quad- \frac{1}{2}\|\nabla F(\mathbf{u}_t)-\sum_{k=1}^n\frac{1}{n}\nabla F_k(\mathbf{x}_t^{(k)})\|^2,\label{bound_dif_grad}
\end{align}
where~\eqref{eq_01} follows from Assumption~\ref{ass:gradient}, and~\eqref{eq_02} uses $\|\bm{a}\|^2+\|\bm{b}\|^2-\|\bm{a}-\bm{b}\|^2=2\langle\bm{a},\bm{b}\rangle$ for $\bm{a},\bm{b}\in\mathbb{R}^d$.

For the second term in \eqref{eq_conv}, we have
\begin{align}
    \mathbb{E}\|\overline{\mathbf{G}}(\mathbf{X}_t)\|^2
     &\labelrel={eq_04}\mathbb{E}\|\sum_{k=1}^n\frac{1}{n}(\mathbf{g}_k(\mathbf{x}_t^{(k)})-\nabla F_k(\mathbf{x}_t^{(k)}))\|^2 \notag\\
    &\qquad+ \mathbb{E}\|\sum_{k=1}^n\frac{1}{n}\nabla F_k(\mathbf{x}_t^{(k)})\|^2\notag\\
    & \labelrel={eq_05}  \sum_{k=1}^n\frac{1}{n^2}\mathbb{E}\|\mathbf{g}_k(\mathbf{x}_t^{(k)})-\nabla F_k(\mathbf{x}_t^{(k)})\|^2\notag\\
    &\qquad + \mathbb{E}\|\sum_{k=1}^n\frac{1}{n}\nabla F_k(\mathbf{x}_t^{(k)})\|^2\notag\\
    & \labelrel\leq{in_02} \frac{\sigma^2}{n}+\mathbb{E}\|\sum_{k=1}^n\frac{1}{n}\nabla F_k(\mathbf{x}_t^{(k)})\|^2.\label{bound_ave_local_gradient}
\end{align}
Here, \eqref{eq_04} follows from $\mathbb{E}[\|\bm{a}\|^2]=\mathbb{E}[\|\bm{a}-\mathbb{E}(\bm{a})\|^2]+\|\mathbb{E}(\bm{a})\|^2$ with $\bm{a}\in\mathbb{R}^d$, \eqref{eq_05} is due to the unbiased gradient estimation in Assumption~\ref{ass:gradient}, and \eqref{in_02} follows from the bounded variance in Assumption~\ref{ass:gradient}. 
Applying~\eqref{bound_dif_grad} and~\eqref{bound_ave_local_gradient} to~\eqref{eq_conv}, we obtain:
\begin{multline}
     \mathbb{E}[F(\mathbf{u}_{t+1})-F(\mathbf{u}_t)] 
    \leq  \frac{\eta^2L}{2}\left[\frac{\sigma^2}{n}+\mathbb{E}\|\sum_{k=1}^n\frac{1}{n}\nabla F_k(\mathbf{x}_t^{(k)})\|^2\right]\notag\\
     +\frac{\eta}{2}\mathbb{E}\|\nabla F(\mathbf{u}_t)-\sum_{k=1}^n{\frac{1}{n}}\nabla F_k(\mathbf{x}_t^{(k)})\|^2  - \frac{\eta}{2}\mathbb{E}\|\nabla F(\mathbf{u}_t)\|^2\notag\\
    - \frac{\eta}{2}\mathbb{E}\|\sum_{i=1}^n{\frac{1}{n}}\nabla F_k(\mathbf{x}_t^{(k)})\|^2.\notag
\end{multline}
Rearranging the above inequality, we have:
\begin{align}
 \mathbb{E}\|\nabla F(\mathbf{u}_t)\|^2 
 &=  \frac{2(\mathbb{E} F(\mathbf{u}_t)-\mathbb{E} F(\mathbf{u}_{t+1}))}{\eta}+\frac{\eta L\sigma^2}{n}  \notag\\
 &\quad+\mathbb{E}\|\nabla F(\mathbf{u}_t)-\sum_{k=1}^n\frac{1}{n}\nabla F_k(\mathbf{x}_t^{(k)})\|^2\notag\\ 
 &\quad-(1-\eta L )\mathbb{E}\|\sum_{k=1}^n\frac{1}{n}\nabla F_k(\mathbf{x}_t^{(k)})\|^2\notag\\
&\labelrel\leq{in_2} \frac{2(\mathbb{E} F(\mathbf{u}_t)-\mathbb{E} F(\mathbf{u}_{t+1}))}{\eta}+\frac{\eta L\sigma^2}{n} \notag\\
&\quad+ \mathbb{E}\|\nabla F(\mathbf{u}_t)-\sum_{k=1}^n\frac{1}{n}\nabla F_k(\mathbf{x}_t^{(k)})\|^2,\label{in_glob_each}
\end{align}
where~\eqref{in_2} holds when $\eta\leq\frac{1}{L}$. For the third term in~\eqref{in_glob_each}, according to Assumption~\ref{ass:smoothness}, we have:
\begin{align}
    &\|\nabla F(\mathbf{u}_t)-\sum_{k=1}^n\frac{1}{n}\nabla F_k(\mathbf{x}_t^{(k)})\|^2 = \|\sum_{k=1}^n\frac{1}{n}\nabla F_k(\mathbf{u}_t)\notag\\
    & -\sum_{k=1}^n\frac{1}{n}\nabla F_k(\overline{\mathbf{x}}_t^{(k)})
    +\sum_{k=1}^n\frac{1}{n}\nabla F_k(\overline{\mathbf{x}}_t^{(k)})-\sum_{k=1}^n\frac{1}{n}\nabla F_k(\mathbf{x}_t^{(k)})\|^2\notag\\
    & \leq 2\sum_{k=1}^n\frac{1}{n}\|\nabla F_k(\mathbf{u}_t)-\nabla F_k(\overline{\mathbf{x}}_t^{(k)})\|^2\notag\\
    &\quad+2\sum_{k=1}^n\frac{1}{n}\|\nabla F_k(\overline{\mathbf{x}}_t^{(k)})-\nabla F_k(\mathbf{x}_t^{(k)})\|^2\notag\\
    & \leq 2\sum_{k=1}^n\frac{1}{n} L^2\|\mathbf{u}_t-\overline{\mathbf{x}}_t^{(k)}\|^2+2\sum_{k=1}^n\frac{1}{n} L^2\|\overline{\mathbf{x}}_t^{(k)}-\mathbf{x}_t^{(k)}\|^2.\notag
\end{align}
From the definition of Frobenius norm and \eqref{eq:aver_model}, we get:
\begin{align} 
    \sum_{k=1}^n\frac{1}{n} L^2\mathbb{E}\|\mathbf{u}_t-\overline{\mathbf{x}}_t^{(k)}\|
    & = \frac{1}{n}L^2\mathbb{E}\|\mathbf{u}_t\mathbf{1}_{n}^\intercal-\mathbf{X}_t\mathbf{V}\|_{\textup{F}}^2\notag\\
    & = \frac{1}{n}L^2\mathbb{E}\|\mathbf{X}_t\mathbf{1}_{n}\mathbf{1}_{n}^\intercal-\mathbf{X}_t\mathbf{V}\|_{\textup{F}}^2\notag\\
    & = \frac{1}{n}L^2\mathbb{E}\|\mathbf{X}_t(\mathbf{V}-\mathbf{A})\|_{\textup{F}}^2. \label{in_3} 
\end{align}
Similarly, we have:
\begin{align} 
    \sum_{k=1}^n\frac{1}{n} L^2\mathbb{E}\|\overline{\mathbf{x}}_t^{(k)}-\mathbf{x}_t^{(k)}\|
     = \frac{1}{n}L^2\mathbb{E}\|\mathbf{X}_t(\mathbf{I}-\mathbf{V})\|_{\textup{F}}^2. \label{in_3_1} 
\end{align}
Combining~\eqref{in_glob_each},~\eqref{in_3} and~\eqref{in_3_1} with Assumption~\ref{ass:lowerbounded}, taking the total expectation and averaging over all iterations, then we have Lemma~\ref{lemma_bound_ave_grad}.

\end{IEEEproof}

Lemma~\ref{lemma_bound_ave_grad} aims to provide the composition of the total convergence error bound. The $\emph{residual error}$ provides hints on how to derive the convergence properties of CE-FedAvg. Specifically, the first term $\|\mathbf{X}_t(\mathbf{V}-\mathbf{A})\|_{\textup{F}}^2$ represents the inter-cluster error between the global average model $\mathbf{X}_t\mathbf{A}$ and the edge server models $\mathbf{X}_t\mathbf{V}$. The second term $\|\mathbf{X}_t(\mathbf{I}-\mathbf{V})\|_{\textup{F}}^2$ represents the intra-cluster error between the device models $\mathbf{X}_t\mathbf{I}$ and the edge server models $\mathbf{X}_t\mathbf{V}$. Next, we will provide the upper bounds for these two terms.
\begin{lemma}[Bounded inter-cluster error]\label{lemma_bound_inter_final}
Under Assumptions \ref{ass:smoothness}, \ref{ass:gradient}, \ref{ass:mixing} and \ref{ass:gradient_global}, the iterates of Algorithm~\ref{algorithm-3} satisfy:
\begin{align*}
&\frac{1}{nT}\sum_{t=0}^{T-1}\mathbb{E}\|\mathbf{X}_t(\mathbf{V}-\mathbf{A})\|_{\textup{F}}^2 \leq \frac{2\eta^2(\Omega_1q\tau+\frac{m-1}{n}q\tau)\sigma^2}{1-4\eta^2L^2q^2\tau^2\Omega_2}\nonumber\\
+ & \frac{4\eta^2q^2\tau^2\Omega_2}{1-4\eta^2L^2q^2\tau^2\Omega_2}\left(\epsilon^2+\frac{L^2}{nT}\sum_{t=0}^{T-1}\mathbb{E}\|\mathbf{X}_t(\mathbf{I}-\mathbf{V})\|_{\textup{F}}^2\right). 
\end{align*}
\end{lemma}
\begin{IEEEproof}
The proof is provided in Appendix \ref{proof_lemma2} in the supplementary text.
\end{IEEEproof}
Lemma~\ref{lemma_bound_inter_final} shows that the inter-cluster error contains the intra-cluster error. Next, we will bound the intra-cluster error.
\begin{lemma}[Bounded intra-cluster error] \label{lemma_bound_intra_final} 
Under Assumptions \ref{ass:smoothness}, \ref{ass:gradient}, \ref{ass:mixing}, \ref{ass:gradient_local}, the iterates of Algorithm~\ref{algorithm-3} satisfy:
\begin{equation*}
\frac{1}{nT}\sum_{t=0}^{T-1}\mathbb{E}\|\mathbf{X}_t(\mathbf{I}-\mathbf{V})\|_{\textup{F}}^2 \leq \frac{(\frac{n-m}{n})\eta^2\tau \sigma^2}{1-2\eta^2L^2\tau^2}+\frac{2\eta^2\tau^2 \sum_{i=1}^m\frac{n_i}{n}\epsilon_i^2}{1-2\eta^2L^2\tau^2}.
\end{equation*}
\end{lemma}
\begin{IEEEproof}
The proof is provided in Appendix \ref{proof_lemma3} in the supplementary text.
\end{IEEEproof}

Lemma \ref{lemma_bound_intra_final} gives the upper bound of intra-cluster error. Combining Lemmas~\ref{lemma_bound_ave_grad}, \ref{lemma_bound_inter_final} and \ref{lemma_bound_intra_final} and choosing a proper learning rate, we can derive the following convergence bound:

\begin{theorem}[Convergence of CE-FedAvg]\label{th:convergence}
Let Assumptions~\ref{ass:smoothness}--\ref{ass:gradient_global} hold, and let $\Omega_1$, $\Omega_2$, $L$, $\sigma$, $\epsilon$, $\epsilon_i$ be as defined therein. If the learning rate satisfies
\begin{equation}\label{eq_learning_rate}
\eta \leq \min\{\frac{1}{2L\tau}, \frac{1}{2\sqrt{2\Omega_2}Lq\tau}\},
\end{equation}
then for any $T > 0$, the iterates of Algorithm~\ref{algorithm-3} for CE-FedAvg satisfy
\begin{align}\label{eq:convergence}
\frac{1}{T}\sum_{t=0}^{T-1}&\mathbb{E}\|\nabla F(\mathbf{u}_t)\|^2  \leq   \frac{2(F(\mathbf{x}_1)-F_{\inf})}{\eta T}+\frac{\eta L\sigma^2}{n} \nonumber \\
&+ 8\eta^2L^2(\Omega_1q\tau+\frac{m-1}{n}q\tau)\sigma^2+16\eta^2L^2q^2\tau^2\Omega_2 \epsilon^2 \nonumber \\
&+8\frac{n-m}{n}\eta^2L^2\tau \sigma^2+16L^2\eta^2\tau^2  \sum_{i=1}^m\frac{n_i}{n}\epsilon_i^2.
\end{align}
\end{theorem}
\begin{IEEEproof}
Substituting the results in Lemmas \ref{lemma_bound_inter_final} and \ref{lemma_bound_intra_final} into Lemma \ref{lemma_bound_ave_grad}, we have:
\begin{multline}\label{th1_upp}
\frac{1}{T}\sum_{t=0}^{T-1}\mathbb{E}\|\nabla F(\mathbf{u}_t)\|^2  \leq   \frac{2(F(\mathbf{x}_1)-F_{\inf})}{\eta T}+\frac{\eta L\sigma^2}{n}\\
+ \frac{4\eta^2L^2(\Omega_1q\tau+\frac{m-1}{n}q\tau)\sigma^2}{1-4\eta^2L^2q^2\tau^2\Omega_2}+\frac{8\eta^2L^2q^2\tau^2\Omega_2 \epsilon^2}{1-4\eta^2L^2q^2\tau^2\Omega_2}\\
+ \left(\frac{4\eta^2L^2q^2\tau^2\Omega_2}{1-4\eta^2L^2q^2\tau^2\Omega_2}+1\right)\bigg(\frac{2(\frac{n-m}{n})\eta^2L^2\tau \sigma^2}{1-2\eta^2L^2\tau^2}\\
+\frac{4\eta^2L^2\tau^2 \sum_{i=1}^m\frac{n_i}{n}\epsilon_j^2}{1-2\eta^2L^2\tau^2}\bigg).
\end{multline}
When $\eta\leq \min\{\frac{1}{2L\tau}, \frac{1}{2\sqrt{2\Omega_2}Lq\tau}\}$, we have:
\begin{gather}
\frac{1}{1-4\eta^2L^2q^2\tau^2\Omega_2}\leq2,\label{eq_th1_eta_upp1}\\
\frac{4\eta^2L^2q^2\tau^2\Omega_2}{1-4\eta^2L^2q^2\tau^2\Omega_2}\leq 1, \quad
\frac{1}{1-2\eta^2L^2\tau^2}\leq 2.\label{eq_th1_eta_upp2}
\end{gather}
Putting \eqref{eq_th1_eta_upp1} and \eqref{eq_th1_eta_upp2} back into \eqref{th1_upp}, we arrive at the conclusion. 
\end{IEEEproof}

Further, by setting the learning rate to be $\eta = \frac{1}{L}\sqrt{\frac{n}{T}}$, we can obtain the following corollary:
\begin{corollary}\label{coro}
For CE-FedAvg, under Assumptions \ref{ass:smoothness}-\ref{ass:gradient_global},  if the learning rate is $\eta = \frac{1}{L}\sqrt{\frac{n}{T}}$ when $T>4\tau^2n\max\{1,2\Omega_2 q^2\}$, then 
\begin{gather*}
    \frac{1}{T}\sum_{t=0}^{T-1}\mathbb{E}\|\nabla F(\mathbf{u}_t)\|^2  \leq O(\frac{1}{\sqrt{T}})+O(\frac{q\tau+\tau}{T})+O(\frac{q^2\tau^2+\tau^2}{T}).
\end{gather*}
\end{corollary}

Corollary~\ref{coro} provides some notable insights. First, the last two terms show the trade-off between communication cost and convergence. While smaller communication periods $q$ and $\tau$ speed up the convergence and reduce the convergence error, they also increase the overall communication cost. Second, the error increases w.r.t. the magnitude of $q^2\tau^2$. Thus, the convergence rate of $O(1/\sqrt{T})$ can be guaranteed by ensuring the total iteration number satisfies $T>q^4\tau^4$.


\subsection{Comparison of Iteration Complexity}\label{sec_complexity}

In the following, we consider the extreme cases of CE-FedAvg and show that our analysis recovers the results of prior algorithms that can be treated as special cases of CE-FedAvg.  

\begin{itemize}
    \item \noindent\textbf{Comparison to Hier-FAvg.} When the topology of edge backhaul is fully connected, the value of $\zeta$ becomes 0, and the model update rule of CE-FedAvg is essentially the same as that of Hier-FAvg. Therefore, our convergence result in Theorem~\ref{th:convergence} reduce to those of Hier-FAvg in \cite{wang2022demystifying}. Meanwhile, Theorem~\ref{th:convergence} shows that the fully connected network topology gives the fastest convergence speed in terms of iteration complexity among all the connected topologies because it has the smallest values of $\Omega_1$ and $\Omega_2$.
    

    \item \noindent\textbf{Comparison to FedAvg.} When $m=1$ and $q=1$, all devices communicate with a single edge server after $\tau$ local iterations and $\epsilon = 0$. In this case, the proposed CE-FedAvg algorithm reduces to FedAvg, and the iteration complexity of CE-FedAvg reduces to $O(\frac{1}{\eta T})+O(\frac{\eta\sigma^2}{n})+O(\eta^2\tau\sigma^2)+O(\eta^2\tau^2\epsilon_i^2)$. This coincides with the complexity of FedAvg given in \cite{yu2019linear}. 

    \item \noindent\textbf{Comparison to Decentralized Local SGD.} When $n=m$ and $\tau = 1$, each edge server only coordinates one device and communicates with neighboring servers after $q$ iterations and $\epsilon_i = 0$. The proposed CE-FedAvg algorithm reduces to decentralized local SGD, and the iteration complexity of CE-FedAvg reduces to $O(\frac{1}{\eta T})+O(\frac{\eta\sigma^2}{n})+O(\eta^2q\sigma^2)+O(\eta^2q^2\epsilon^2)$. This coincides with the complexity of decentralized local SGD given in \cite{wang2021cooperative}. 
\end{itemize}

\begin{subsection}{Discussions}\label{sec:discussion}
In the following, we compare our main results with prior work and analyze the impacts of cluster-level data distribution and cluster size on the algorithmic convergence of CE-FedAvg. 

\begin{remark}[\textbf{Comparison with SE-FEEL}]\label{remark_comp_se_feel}
We compare our convergence result with that of a concurrent work SE-FEEL \cite{sun2021semi} that analyzes CE-FedAvg only under the global divergence assumption~\ref{ass:gradient_single_glob}. 
%
%
Specifically, \cite{sun2021semi} provides a convergence rate of
\[
O(\frac{1}{\eta T})+O(\frac{\eta\sigma^2}{n})+O(\eta^2q\tau\sigma^2)+O(\eta^2q^2\tau^2\hat{\epsilon}^2).
\]
According to the above result, $q$ and $\tau$ have the same effect on the convergence bound, which cannot show any benefit of intra-cluster model aggregation. In comparison, our work shows a much tighter convergence rate of
\[
O(\frac{1}{\eta T})+O(\frac{\eta\sigma^2}{n})+O(\eta^2\tau(q+1)\sigma^2)+O(\eta^2\tau^2(q^2\epsilon^2+\sum_{i=1}^m\epsilon_i^2)).  
\]
We can observe that both intra-cluster aggregation period $\tau$ and inter-cluster aggregation period $q\tau$ affect the convergence bound. In particular, given a fixed inter-cluster aggregation period $q \tau$, more frequent intra-cluster aggregation (i.e., a smaller $\tau$) leads to faster convergence and lower convergence error. This clearly shows the benefit of intra-cluster model aggregation in CE-FedAvg. 
\end{remark}

\begin{remark}[\textbf{Effect of cluster size}]\label{remark_clust_size}
We analyze the impact of cluster size on the convergence of CE-FedAvg under a fixed number of devices $n$. For the IID setting (i.e., $\epsilon^2=\epsilon_{i}^2=0, \forall i$), the terms containing $m$ in \eqref{eq:convergence} is
\begin{gather}\label{SGD_noise_m}
\frac{4\eta^2L^2\sigma^2\tau(2q-1)}{n}m.
\end{gather}
As $4\eta^2L^2\sigma^2\tau(2q-1)/n$ is always positive, a smaller value of $m$ leads to a lower convergence error bound. %
For the non-IID setting, cluster size $m$ can also affect the inter-cluster and intra-cluster divergences (i.e., Assumptions~\ref{ass:gradient_local} and \ref{ass:gradient_global}). 
For simplicity, assume all clusters have the same number of devices, i.e., $n_{i}=n/m, \forall i\in[m]$. We have the following lemma:
\begin{lemma}\label{lemma_eps}
Under equal cluster sizes, the inter-cluster divergence can be written as:
\begin{multline}\label{eq_glob_div_dec}
    \frac{1}{m}\sum_{i=1}^m\|\nabla f_i(\mathbf{x}) - \nabla F(\mathbf{x})\|^2  = \\
    \frac{m}{n^2}\sum_{i=1}^m\|\sum_{k\in \mathcal{S}_i}\nabla F_{k}(\mathbf{x})\|^2 -\frac{1}{n^2}\|\sum_{k=1}^{n}\nabla F_{k}(\mathbf{x})\|^2.
\end{multline}
\end{lemma}
\begin{IEEEproof}
If the number of edge devices in each cluster is equal, then $n_1=\ldots=n_m=\frac{n}{m}$. Then we can write the inter-cluster divergence as follows.
\begin{align*}
    &\frac{n_i}{n}\sum_{i=1}^m\|\nabla f_i(\mathbf{x}) - \nabla F(\mathbf{x})\|^2  = \frac{1}{m}\sum_{i=1}^m\|\nabla f_i(\mathbf{x}) - \nabla F(\mathbf{x})\|^2\notag\\
    & = \frac{1}{m}\sum_{i=1}^m\|\nabla f_i(\mathbf{x})\|^2 + \|\nabla F(\mathbf{x})\|^2-2\langle \frac{1}{m}\sum_{i=1}^{m}\nabla f_i(\mathbf{x}), \nabla F(\mathbf{x})\rangle\notag\\
    & \labelrel={eq_lo_gl_cross:equality} \frac{1}{m}\sum_{i=1}^m\|\frac{m}{n}\sum_{k\in \mathcal{S}_i}\nabla F_{k}(\mathbf{x})\|^2-\|\nabla F(\mathbf{x})\|^2\\
    & = \frac{m}{n^2}\sum_{i=1}^m\|\sum_{k\in \mathcal{S}_i}\nabla F_{k}(\mathbf{x})\|^2-\frac{1}{n^2}\|\sum_{k=1}^{n}\nabla F_{k}(\mathbf{x})\|^2,\notag
\end{align*}
where~\eqref{eq_lo_gl_cross:equality} holds due to the definition $\nabla F(\mathbf{x}) = \frac{1}{m}\sum_{i=1}^{m}\nabla f_i(\mathbf{x})$.
\end{IEEEproof}

Suppose we combine any $\rho > 1$ existing clusters (assume the cluster index $i = 1, \ldots, \rho$ without loss of generality) into a new cluster. According to the Cauchy–Schwarz inequality, we have
\begin{gather}\label{in_cluster_m_rho}
\sum_{i=1}^\rho\left\|\sum_{k\in \mathcal{S}_i}\nabla F_{k}(\mathbf{x})\right\|^2 \geq \frac{1}{\rho}\left\|\sum_{k\in \bigcup_{i = 1}^{\rho} \mathcal{S}_i}\nabla F_{k}(\mathbf{x})\right\|^2.
\end{gather}
Therefore, by Lemma~\ref{lemma_eps}, given the same set of devices and random grouping, it is easy to see that in the RHS of \eqref{eq_glob_div_dec}, the first term decreases as $m$ decreases while the second term remains the same. Therefore, the inter-cluster divergence decreases as $m$ decreases, corresponding to faster convergence in CE-FedAvg.
\end{remark}

\begin{remark}[\textbf{Effect of cluster-level data distribution}] \label{remark_clus_data_dist}
We investigate the impact of cluster-level data distribution (IID and non-IID) on the convergence of CE-FedAvg. According to \eqref{eq_global_inter_intra} which shows that the global divergence can be decomposed into the inter-cluster and intra-cluster divergences, we can obtain the following:
\begin{equation}\label{eq_eps_inter_intra}
\hat{\epsilon}^2 = \epsilon^2+\sum_{i=1}^{m}\frac{n_i}{n}\epsilon_i^2.
\end{equation}
Therefore, given certain data distributions on devices (i.e., the global divergence $\hat{\epsilon}^2$ is fixed), decreasing the inter-cluster divergence $\epsilon^2$ will increase the intra-cluster divergence $\sum_{i=1}^{m}n_i\epsilon_i^2/n$. According to \eqref{eq:convergence} of Theorem~\ref{th:convergence}, since $16 \eta^2L^2q^2\tau^2\Omega_2>8\eta^2L^2\tau^2$, this will lead to a lower total convergence bound. In particular, when the cluster-level data distribution is IID (i.e., $\epsilon^2=0$), the convergence bound in Theorem~\ref{th:convergence} is the smallest, and CE-FedAvg converges at the fastest speed.
\end{remark} 
\end{subsection}

\section{Experiments}

\subsection{Experimental Setup}

We consider a CFEL system with 64 devices and 8 edge servers. Each edge server is connected with 8 devices, and edge servers are interconnected over edge backhaul with a ring topology. In the experiments, we consider the image classification task on two common FL datasets: FEMNIST \cite{caldas2019leaf} and CIFAR-10 \cite{krizhevsky2009learning}. The FEMNIST dataset is the federated splitting version of EMNIST dataset which includes 3,550 users. We randomly sample 64 users to simulate the non-IID data distribution for experimentation. Each user's local data is divided into $90\%$ and $10\%$ for training and testing, respectively. The common testing dataset is composed of the testing data from all devices. The model trained on FEMNIST is a CNN with two $3 \times 3$ convolutional layers (each with 32 channels and ReLu activation followed with $2 \times 2$ max pooling), a full connected layer with 1024 units and ReLu activation, and a final softmax output layer (6,603,710 total parameters) \cite{li2019abnormal}. The CIFAR-10 dataset contains 50,000 training images and 10,000 testing images. To simulate the non-IID data distribution, by default, the 50,000 training images are partitioned across devices following the Dirichlet distribution \cite{hsu2019measuring} with concentration parameter 0.5. We train a modified VGG-11 (9,750,922 total parameters) on CIFAR-10. The original 10,000 testing images are used as the common testing dataset. For each dataset, the common testing set is used to evaluate the generalization performances of the trained models.

For CE-FedAvg, we set the mixing matrix $\mathbf{H}$ according to Assumption~\ref{ass:mixing} and the number of gossip steps in each global aggregation round $\pi = 10$ by default. To demonstrate the effectiveness of CE-FedAvg, we compare it with three baselines: FedAvg \cite{mcmahan2017communication}, Hier-FAvg \cite{liu2020client} and Local-Edge. For fair comparison, the baseline algorithms are adapted as follows: 
\begin{itemize}
    \item \emph{FedAvg:} In every global round, each device performs $q \tau$ iterations of SGD update and uploads its updated model to the cloud for global aggregation. This corresponds to the traditional cloud-based FL framework. 
    
    \item \emph{Hier-FAvg:} In every global round, each device first alternatively performs $\tau$ iterations of SGD update and uploads its updated model to the associated edge server for local aggregation for $q - 1$ times. Next, each device performs $\tau$ iterations of SGD update and uploads its updated model to the cloud for global aggregation. This corresponds to the hierarchical FL framework. 
    
    
    

    \item \emph{Local-Edge:} In every global round, each device alternatively performs $\tau$ iterations of SGD update and uploads its updated model to the associated edge server for local aggregation for $q$ times without collaboration between edge servers. This corresponds to the edge-based FL framework.  
    
\end{itemize}

For all experiments, we use mini-batch SGD with momentum of 0.9 to train the local model with batch size of 50. The learning rate of each algorithm is tuned from $\{0.01, 0.05, 0.1\}$ for CIAFR-10 and $\{0.1, 0.06, 0.03, 0.01\}$ for FEMNIST using grid search. Following the implementation in \cite{reddi2020adaptive}, instead of doing $\tau$ local training steps per device, we perform $\tau$ epochs of training over each device's dataset. Moreover, to account for varying numbers of gradient steps per device, we weight the average of device models by each device's local sample size. We run each experiment with 5 random seeds and report the average. All algorithms are implemented using Pytorch on an Ubuntu server with 4 NVIDIA RTX 8000 GPUs. 

We estimate the total training time as the sum of computing time and communication time. We use thop\footnote{https://pypi.org/project/thop/} to measure the computation workload in terms of the number of floating point operations (FLOPs). The number of FLOPs needed for each training sample per iteration is 920.67 MFLOPs for VGG-11 on CIFAR-10 and 13.30 MFLOPs for CNN on FEMNIST, respectively. The edge devices are assumed to be iPhone X whose processing capacity is 691.2 GFLOPS. Following \cite{sun2021semi}, we assume the edge servers are connected in a ring topology via high-speed links with bandwidth of 50 Mbps unless otherwise specified. The devices and edge servers are connected via wireless links whose bandwidth is 10 Mbps per device, and the bandwidth between each device and the cloud is set to be 1 Mbps \cite{yuan2020hierarchical}.

\subsection{Experimental Results}

\begin{figure}[t]
    \subfloat[FEMNIST]{{\includegraphics[width=0.24\textwidth]{ {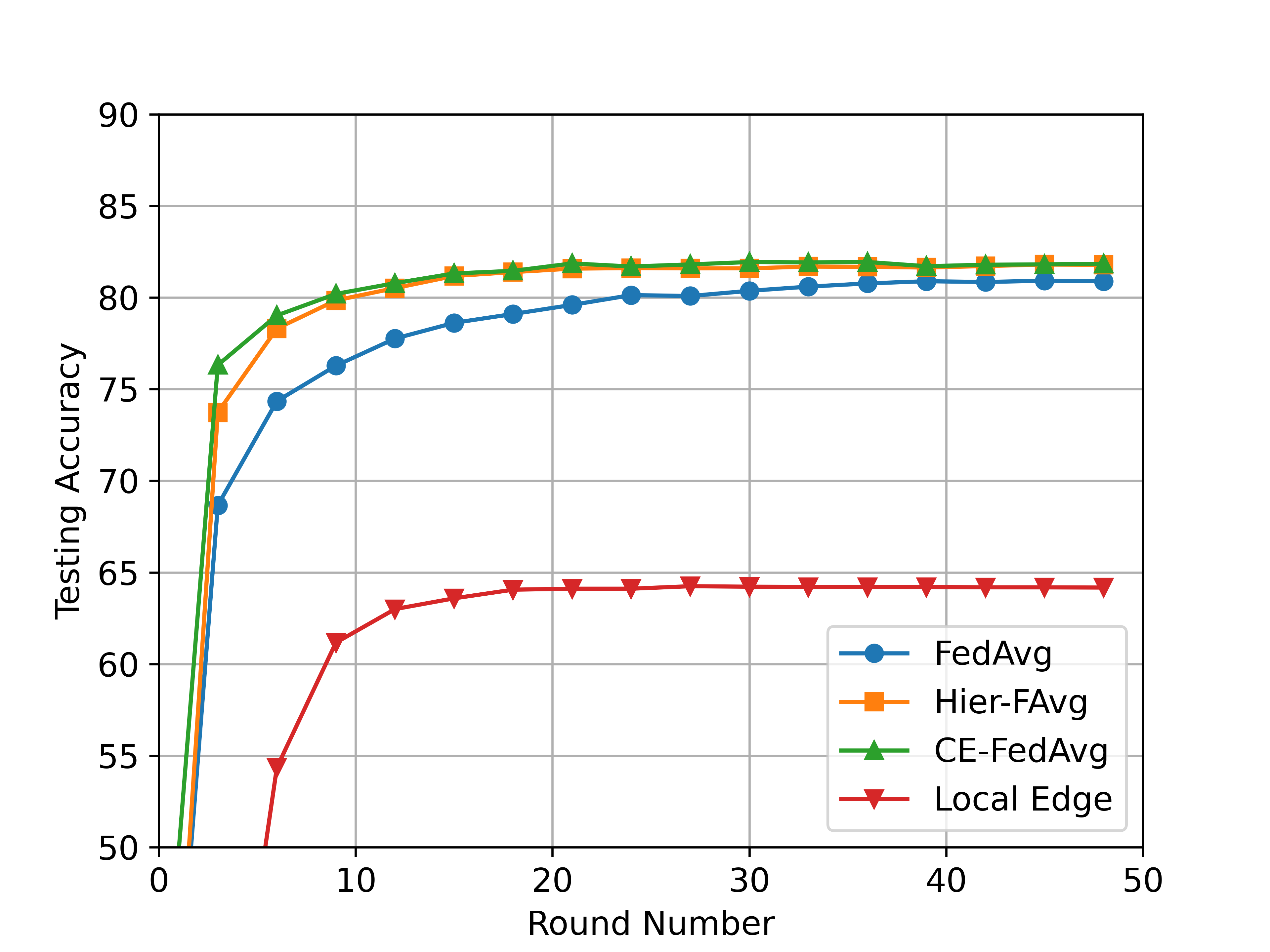}} }} 
    \subfloat[FEMNIST]{{\includegraphics[width=0.24\textwidth]{ {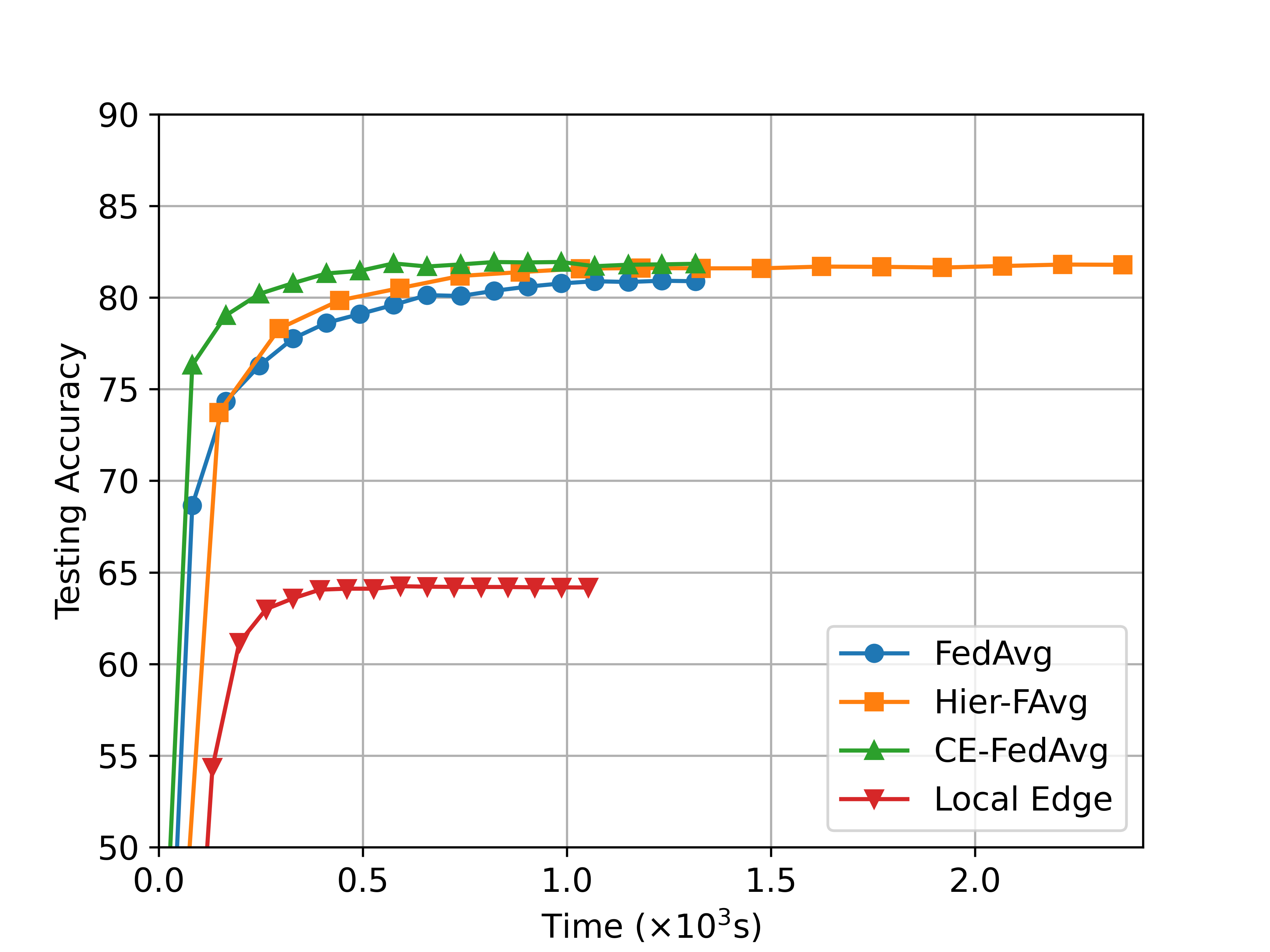}} }} \\
    \subfloat[CIFAR-10]{{\includegraphics[width=0.24\textwidth]{ {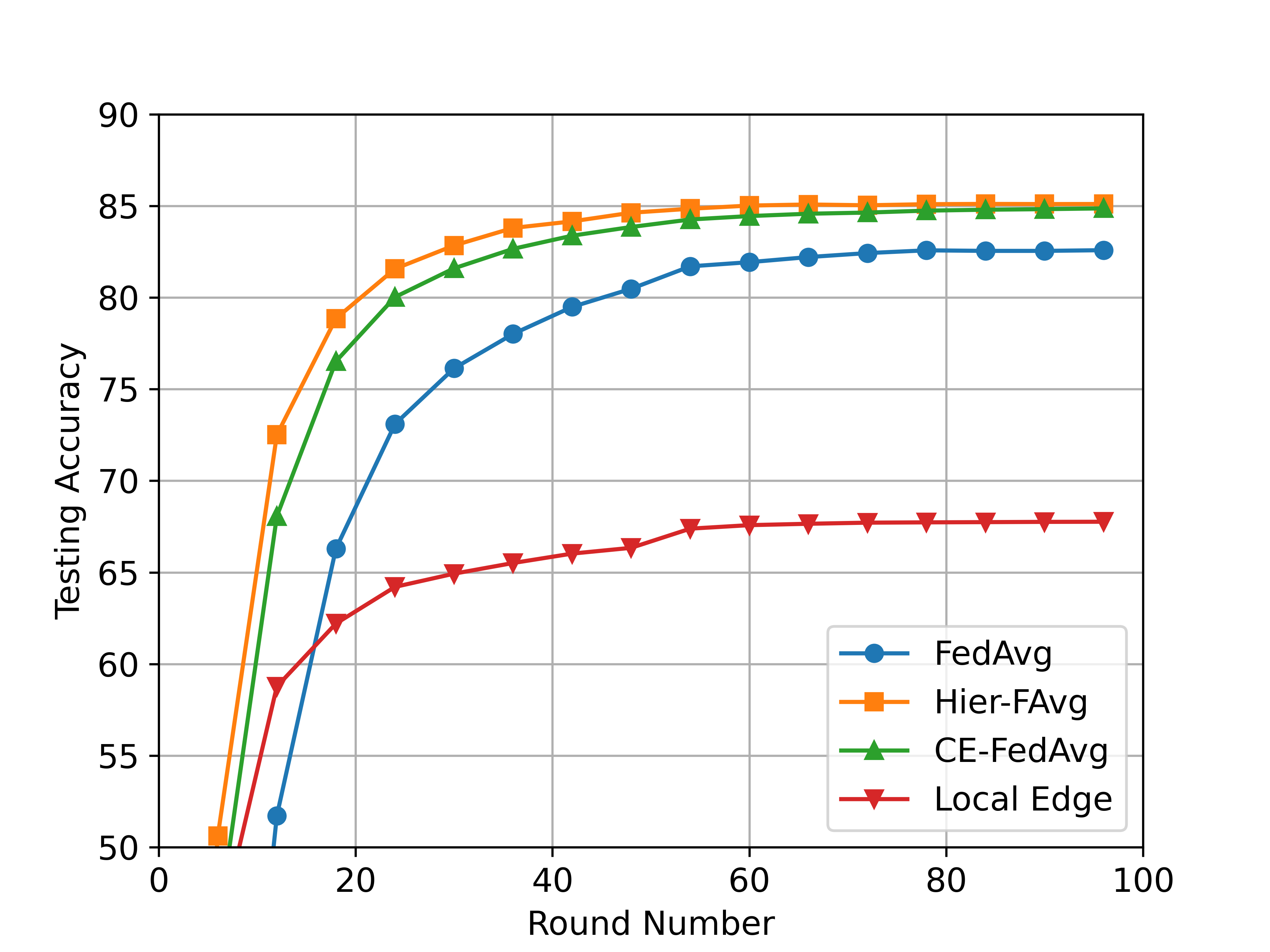}} }} 
    \subfloat[CIFAR-10]{{\includegraphics[width=0.24\textwidth]{ {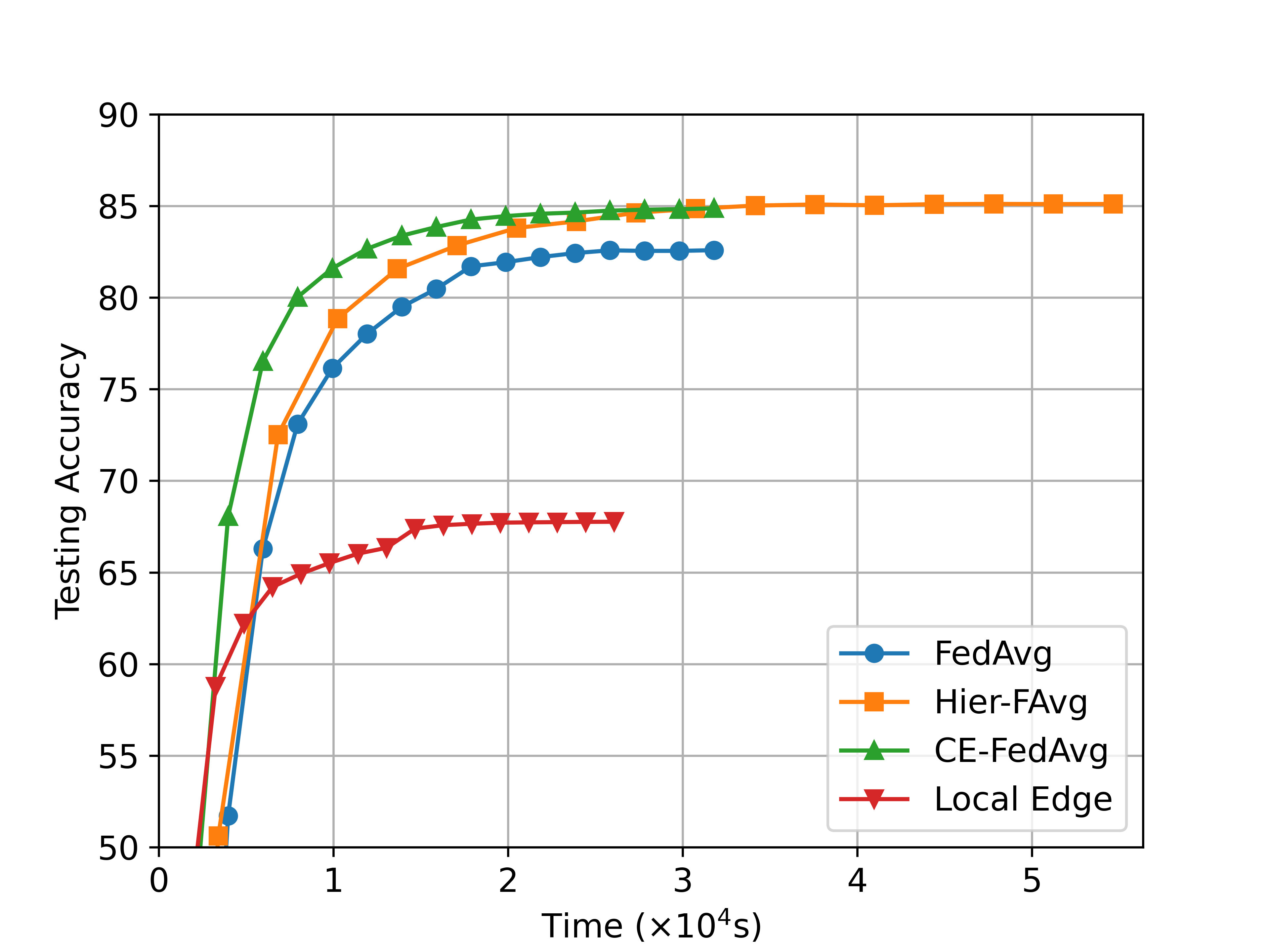}} }} 
    \caption{Convergence rate and runtime comparisons of CE-FedAvg and the baseline algorithms when $\tau= 2$ and $q = 8$ for FEMNIST and CIFAR-10 datasets. (a) and (c) show how the accuracy changes over global round; (b) and (d) show how the accuracy changes over runtime.}
    \label{fig:test_acc_baseline}%
\end{figure}

We first compare the convergence speed and runtime of CE-FedAvg and the baseline algorithms while fixing $\tau = 2$ and $q = 8$. For CE-FedAvg and Local-Edge, we measure the average test accuracy of edge models at each global round, and for FedAvg and Hier-FAvg, we measure the test accuracy of cloud model at each global round. Fig.~\ref{fig:test_acc_baseline} shows the convergence process. From the figure, we can observe that in terms of global round number, Hier-FAvg generally converges faster than CE-FedAvg by aggregating all local models centrally at the cloud. 

Both Hier-FAvg and CE-FedAvg converge faster than FedAvg by using local model aggregation before global model aggregation. Furthermore, Local-Edge converges to a much lower model accuracy because a smaller amount of data is used to train each edge model. On the other hand, in terms of runtime, CE-FedAvg can achieve a better time-to-accuracy than all baseline algorithms by leveraging a distributed network of cooperative edge servers to perform fast local and global model aggregations. Specifically, On FEMNIST, the runtime of CE-FedAvg required to achieve a target test accuracy of $80\%$ is $62.5\%$ and $58.3\%$ less than that of FedAvg and Hier-FedAvg, respectively. On CIFAR-10, the runtime of CE-FedAvg required to achieve a target test accuracy of $80\%$ is $50.0\%$ and $41.8\%$ less than that of FedAvg and Hier-FedAvg, respectively.

\begin{figure}[t]
    \subfloat[FEMNIST]{{\includegraphics[width=0.24\textwidth]{ {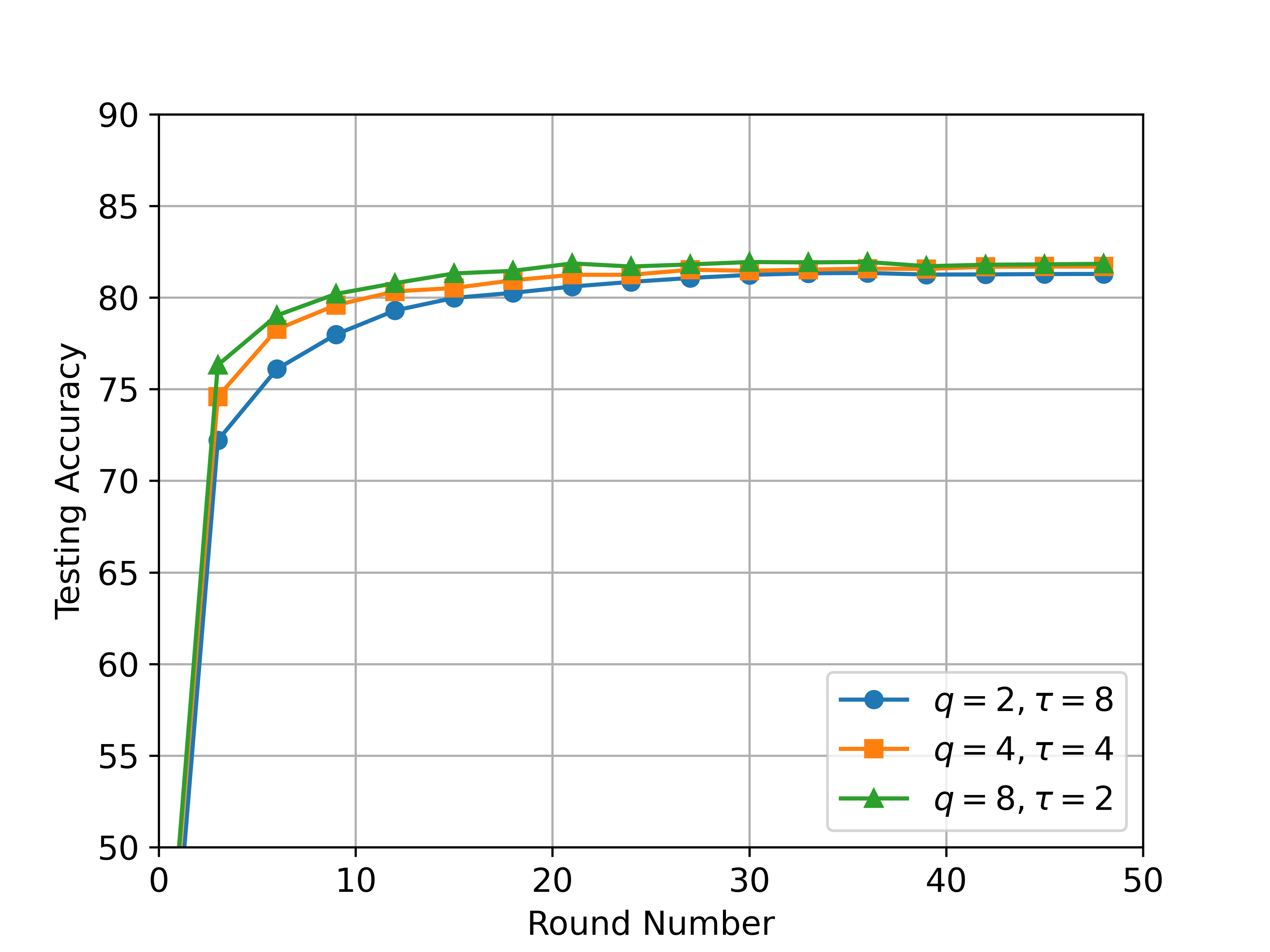}} }} 
    \subfloat[FEMNIST]{{\includegraphics[width=0.24\textwidth]{ {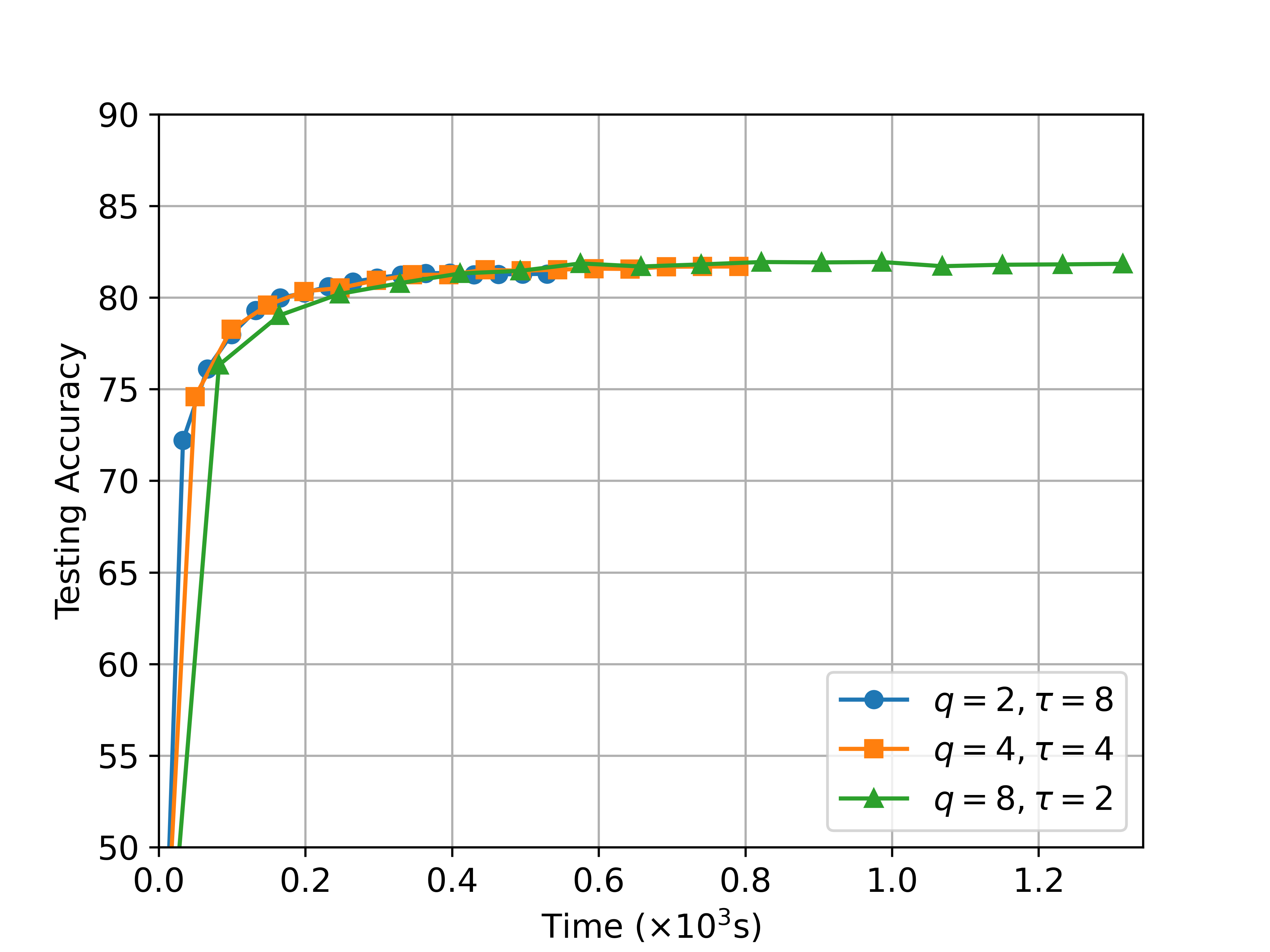}} }} \\
    \subfloat[CIFAR-10]{{\includegraphics[width=0.24\textwidth]{ {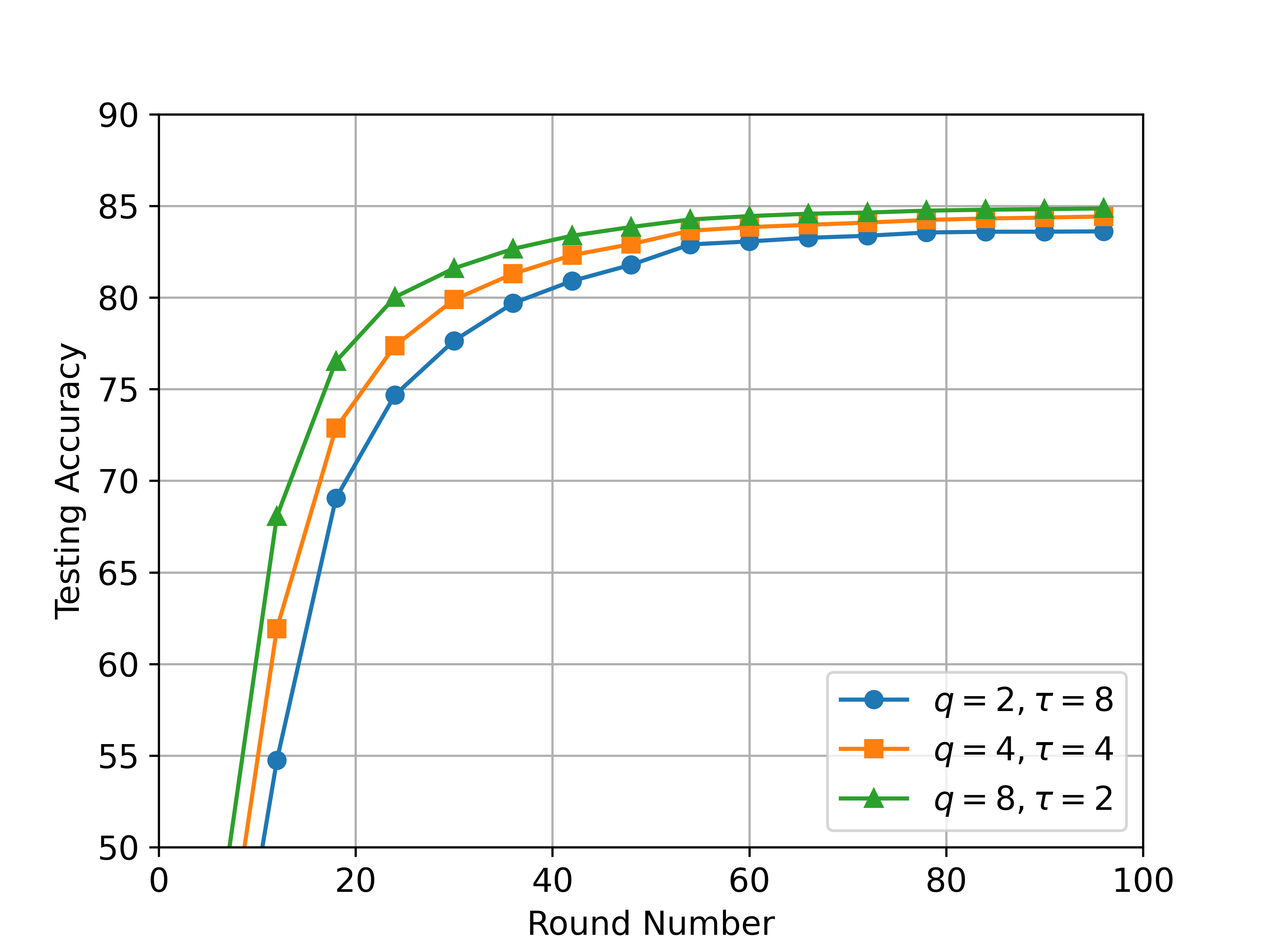}} }} 
    \subfloat[CIFAR-10]{{\includegraphics[width=0.24\textwidth]{ {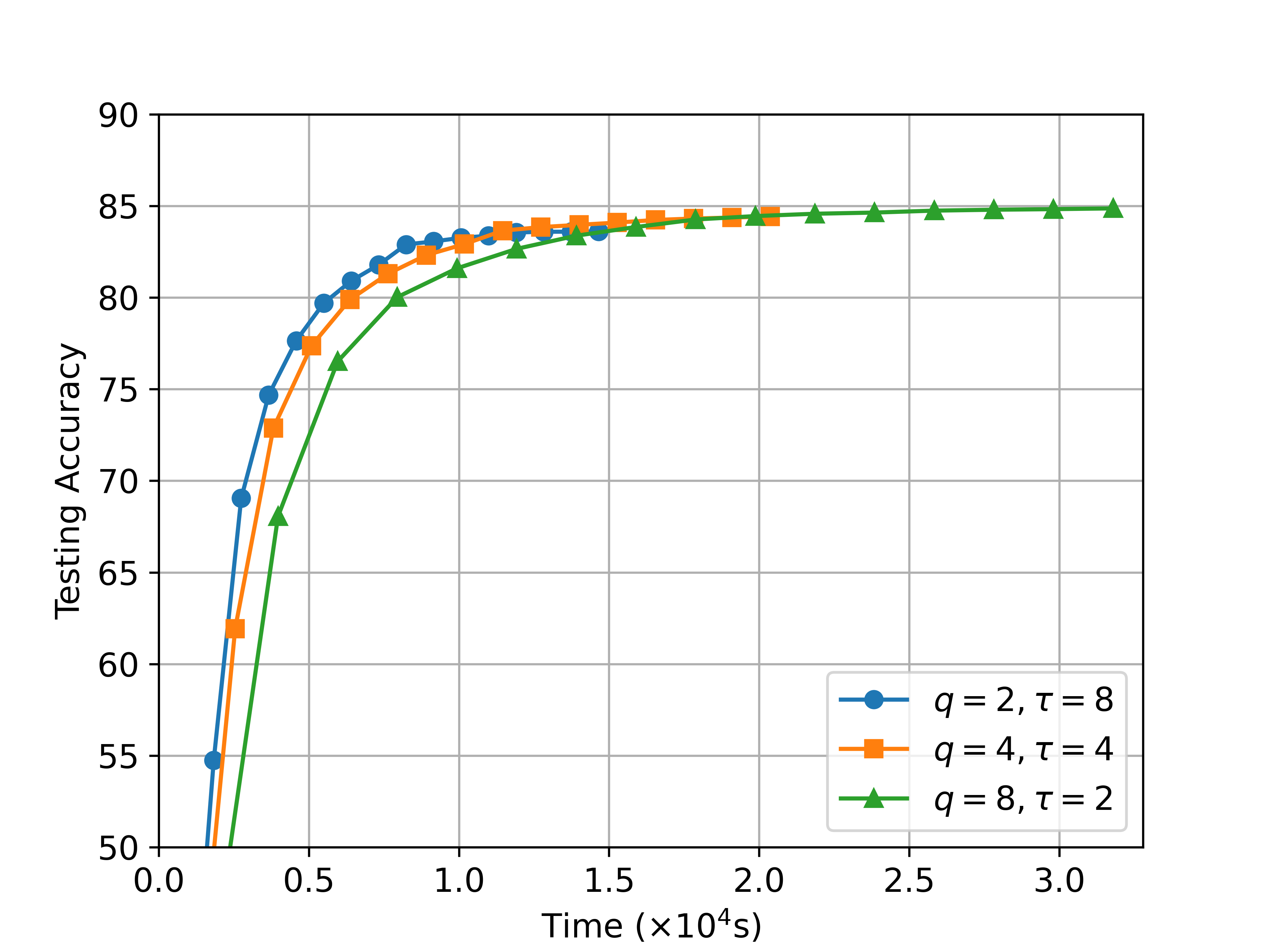}} }} 
    \caption{Convergence rate and runtime comparisons of CE-FedAvg for CIFAR-10 and FEMNIST datasets under different $\tau$ when $q \tau = 16$.}
    \label{fig:test_acc_cifar_tau_q}%
\end{figure}
\begin{figure}[t]
    \subfloat[FEMNIST]{{\includegraphics[width=0.24\textwidth]{ {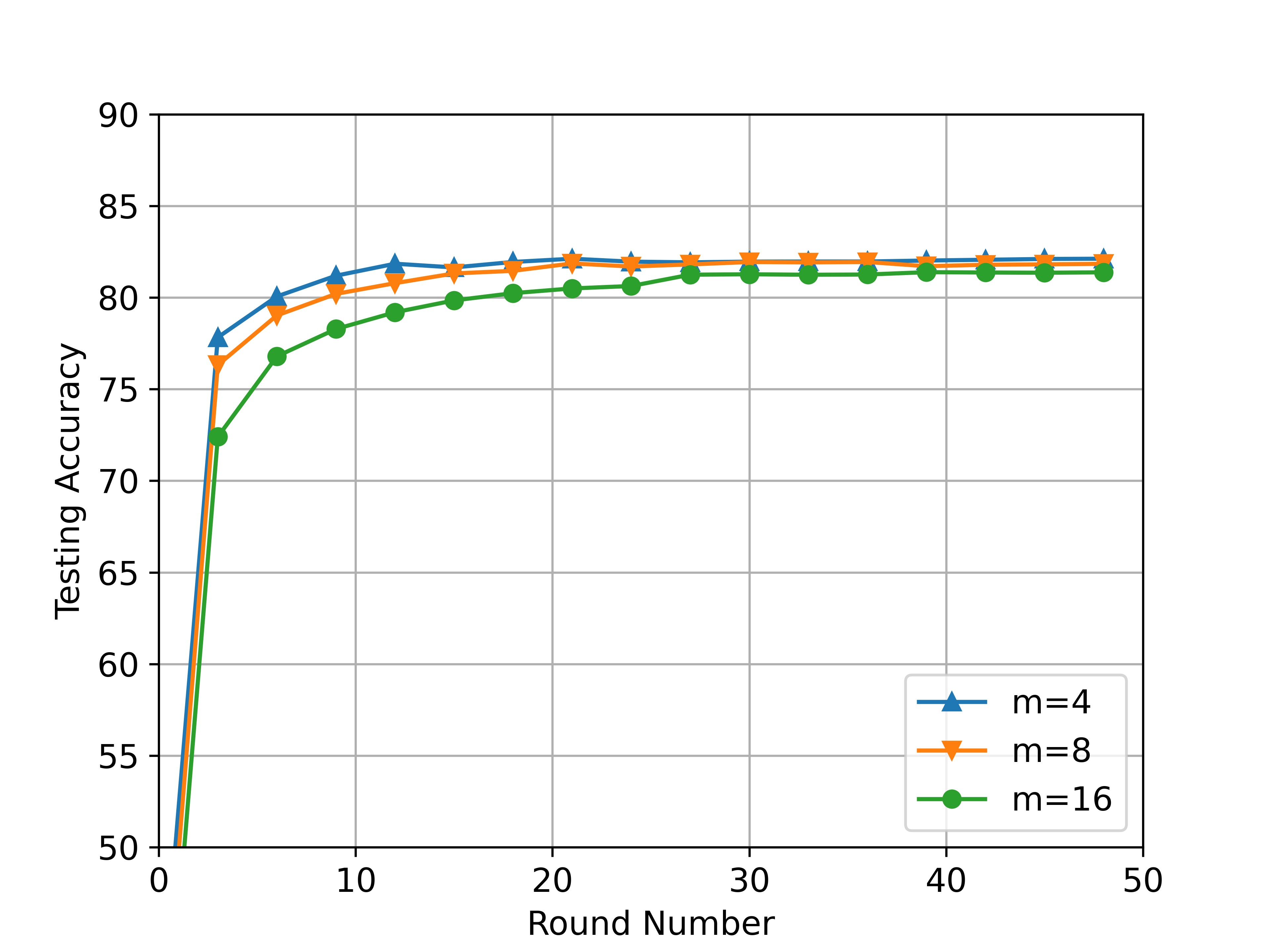}} }}
    \subfloat[CIFAR-10]{{\includegraphics[width=0.24\textwidth]{ {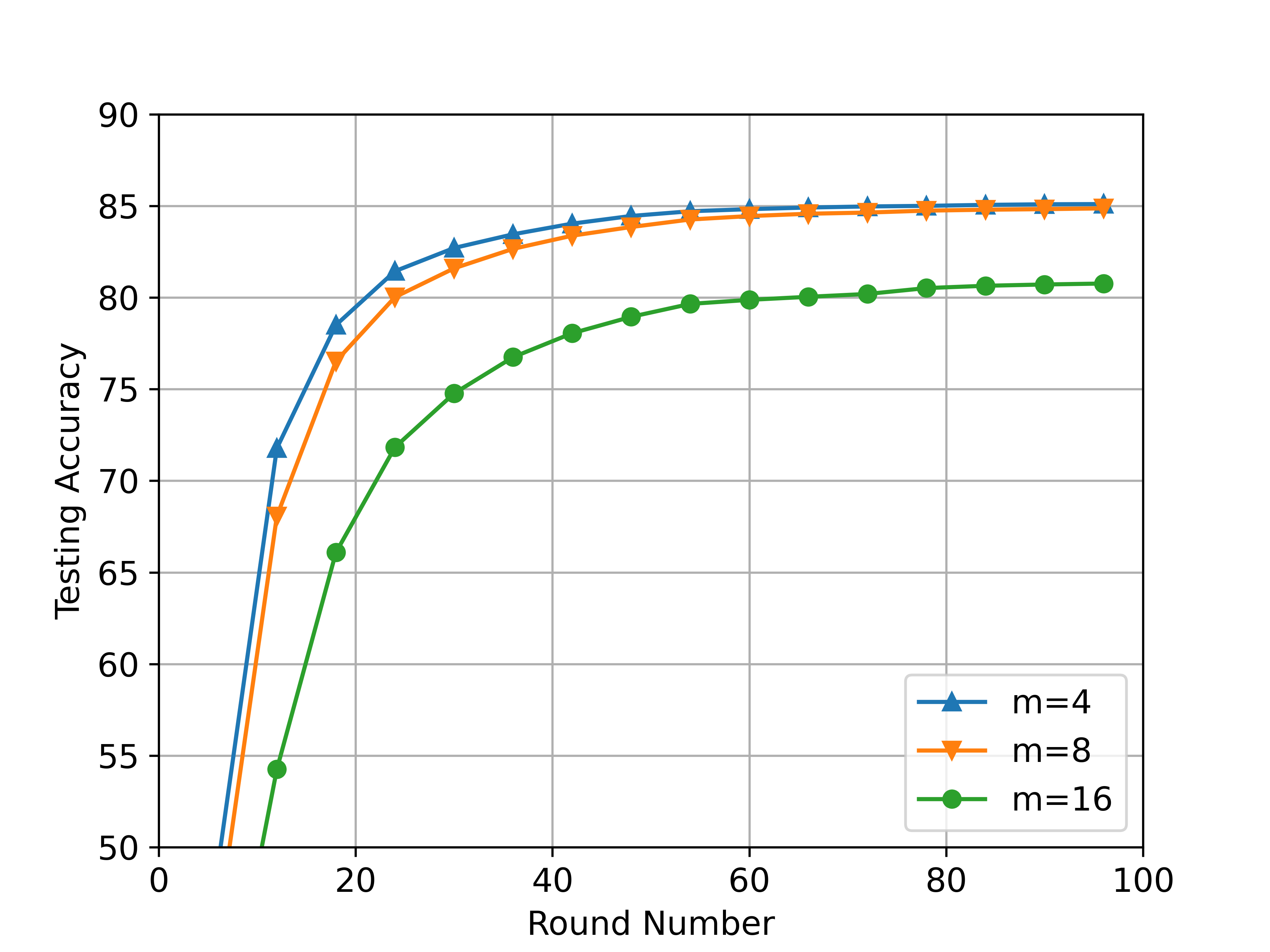}} }} 
    \caption{Testing accuracy vs. round number of CE-FedAvg under different cluster number $m$ for CIFAR-10 and FEMNIST datasets when fixing $n= 64$, $\tau = 2$, and $q = 8$.}%
    \label{fig:test_acc_groups}%
\end{figure}

Next, we vary $\tau$ from $\{2, 4, 8\}$ while fixing $q \tau = 16$ and compare the performances of CE-FedAvg on FEMNIST and CIFAR-10 in Fig.~\ref{fig:test_acc_cifar_tau_q}. From the figure, we can observe that CE-FedAvg can converge faster in terms of global round number as $\tau$ decreases. This demonstrates the benefit of frequent local model aggregation in improving the convergence speed, matching the theoretical analysis in Remark~\ref{remark_comp_se_feel}. However, in terms of runtime, a smaller $\tau$ incurs longer communication delay for local model aggregation in each global round, which could lead to inferior performance on time-to-accuracy. For instance, to achieve a target test accuracy of $80\%$ on FEMNIST, the time needed for $\tau=2$ is $24.6\%$ and $24.2\%$ more than that of $\tau=4$ and $\tau=8$, respectively. To achieve a target test accuracy of $80\%$ on CIFAR-10,  the time needed for $\tau=2$ is $4\%$ and $24\%$ more than that of $\tau=4$ and $\tau=8$, respectively. 

Then, we investigate how the cluster number $m$ affects the performance of CE-FedAvg. We select $m = \{4, 8, 16\}$ while fixing the total number of devices $n = 64$, corresponding to $\{16, 8, 4\}$ randomly assigned devices per cluster. Fig.~\ref{fig:test_acc_groups} depicts the testing accuracy vs. round number. As can be observed from the figure, decreasing $m$ leads to better convergence because more devices cooperate with each other during each edge round, and the divergence of their models is smaller. This is consistent with the analysis in Remark~\ref{remark_clust_size}.

\begin{figure}[t]
\centering
\includegraphics[width=0.3\textwidth]{ {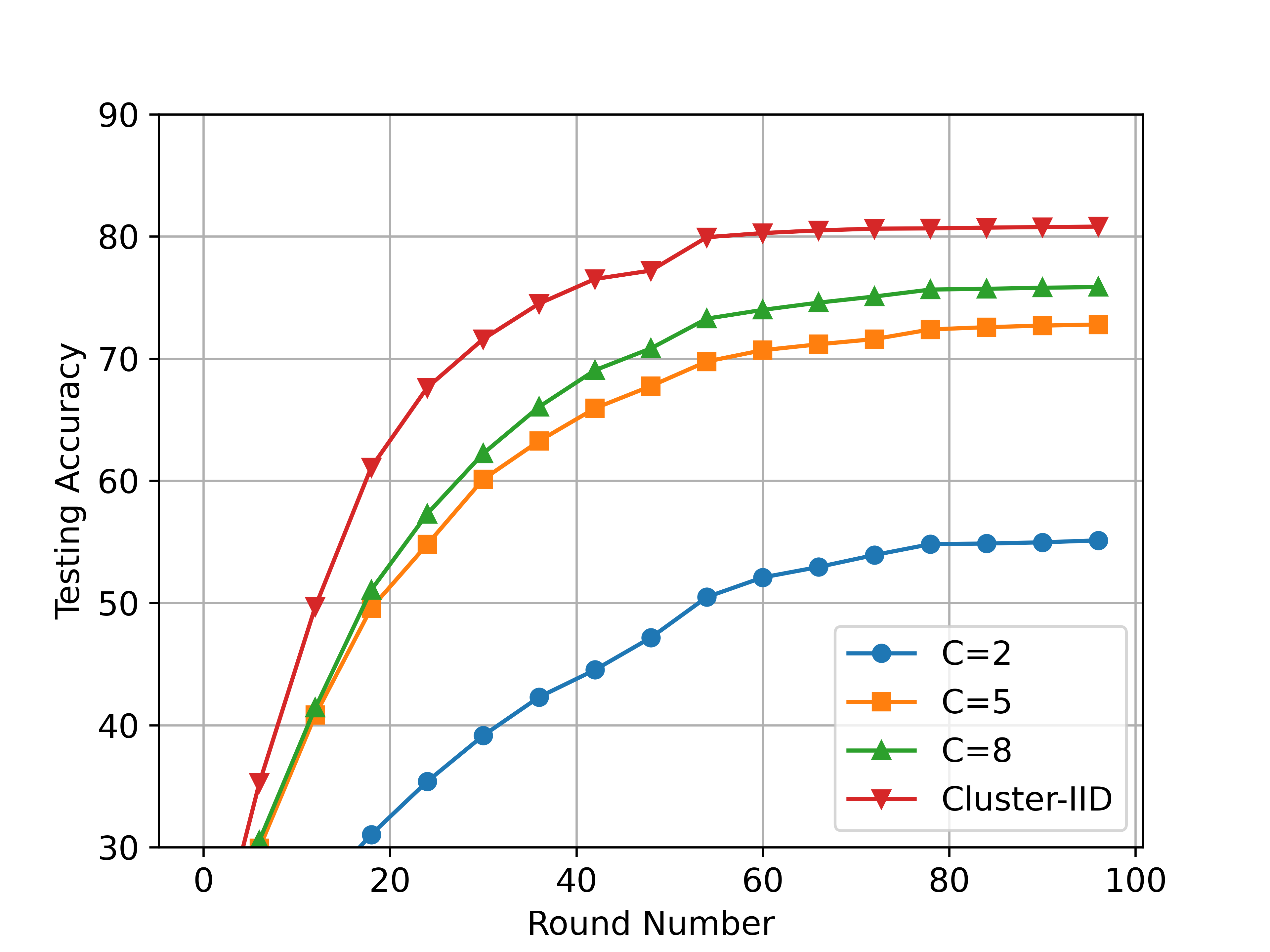}}
\caption{Testing accuracy vs. round number for CE-FedAvg on CIFAR-10 dataset under different cluster-level data distributions when fixing $n= 64$, $\tau = 2$, and $q = 8$. Here $C$ denotes the number of label classes each cluster has.}
\label{fig:test_acc_distribution}%
\end{figure}
\begin{figure}[t]
\centering
\includegraphics[width=0.3\textwidth]{ {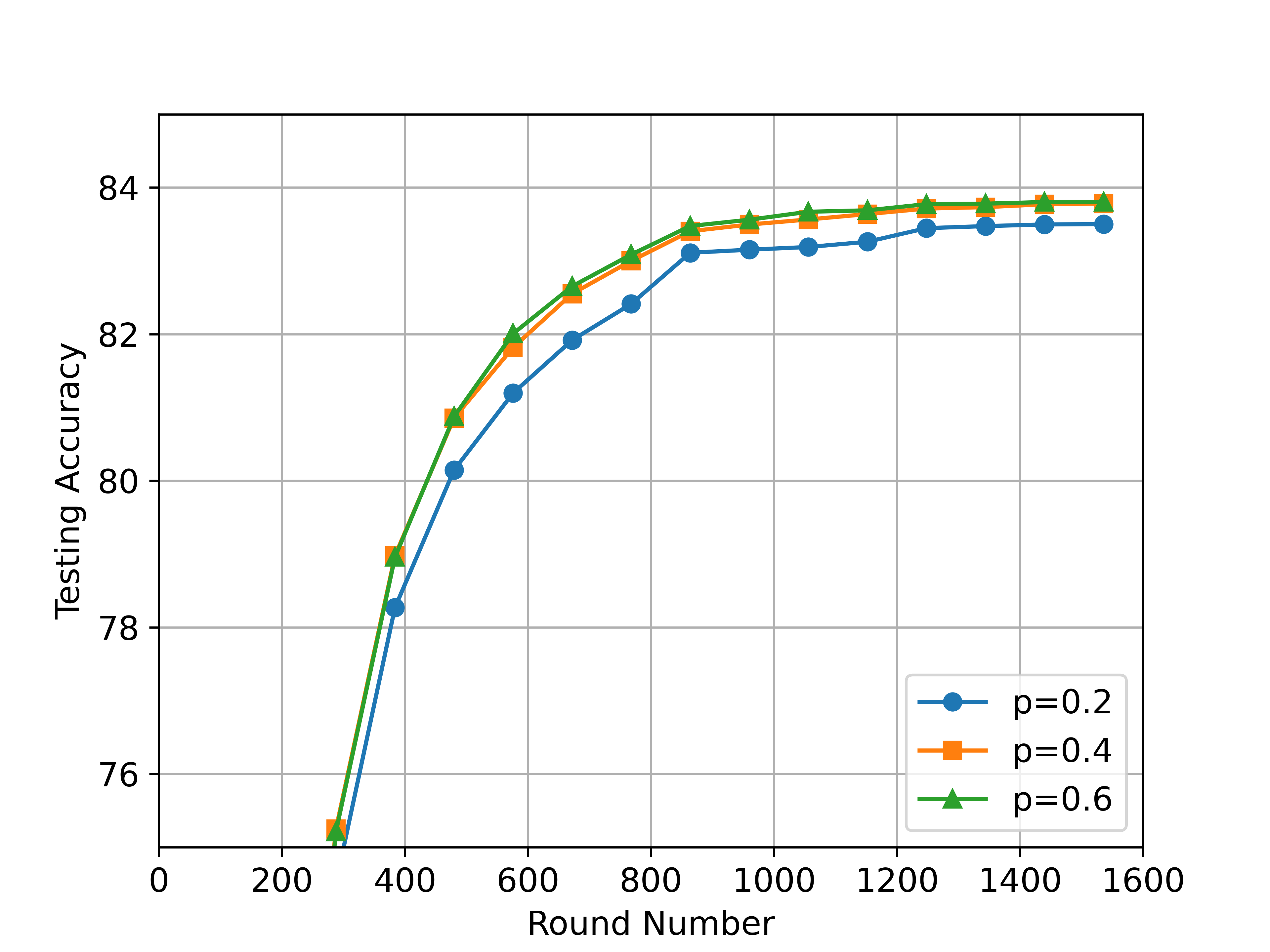}}
\caption{Testing accuracy vs. round number for CE-FedAvg on CIFAR-10 dataset under different edge backhaul topologies when fixing $n= 64$, $\tau = 1$, $q = 1$, and $\pi=1$.}
\label{fig:test_acc_topology}%
\end{figure}
After that, we study the impact of cluster-level data distribution on the performance of CE-FedAvg. In CFEL, there are two levels of non-IID data distribution: device-level and cluster-level, corresponding to the intra-cluster and inter-cluster divergence properties. Note that even though the data distribution of device exhibits heterogeneity, the data distribution of cluster can be homogeneous. %
%
%
Specifically, we consider the following two cases on cluster-level data distribution for CIFAR-10 dataset:
\begin{itemize}
    \item \emph{Cluster IID:} The 50,000 training images are first evenly partitioned in an IID fashion across $m=8$ clusters with each cluster having 6250 images. Then, within each cluster, we sort the 6250 images by label, evenly divide them into 16 shards, and then assign each of 8 devices 2 shards such that devices will only have images of two labels. The data distribution among clusters is IID in this case. 
    
    \item \emph{Cluster Non-IID:} We first sort the 50,000 training images by label, evenly divide them into $C \times 8$ shards, and then assign $C$ shards to each of the 8 clusters such that each cluster roughly has images of $C$ labels. We set $C = \{2, 5, 8\}$ in the experiment. Then, within each cluster, we sort the assigned images by label, evenly divide them into 16 shards, and then assign each of 8 clusters 2 shards such that devices will only have images of two labels. The data distribution among clusters is non-IID in this case.
\end{itemize}
We compare the performance of CE-FedAvg under the above cases in Fig.~\ref{fig:test_acc_distribution}. The result shows that CE-FedAvg converges much faster under the Cluster IID than Cluster Non-IID case. Therefore, if the grouping strategy of devices can be controlled in practice, it makes sense to group devices to follow the IID fashion across clusters to accelerate the convergence and reduce the runtime of CE-FedAvg. Furthermore, as $C$ increases, the inter-cluster divergence increases while the global divergence is fixed, and the convergence speed of CE-FedAvg will decrease correspondingly, matching our theoretical analysis in Remark~\ref{remark_clus_data_dist}.

Finally, we evaluate the convergence of CE-FedAvg under varying edge backhaul topologies in Fig.~\ref{fig:test_acc_topology}. We generate random network topologies by Erd\H{o}s-R\'enyi model with edge probability $p=\{0.2, 0.4, 0.6\}$. As observed in the figure, a more connected network topology (i.e., a larger value of $p$ and smaller value of $\xi$) generally accelerates the convergence and leads to a higher model accuracy achieved after 1500 communication rounds in CE-FedAvg. This matches our theoretical results in Theorem~\ref{th:convergence}.

\section{Conclusion}

In this paper, we propose CFEL, a novel FL framework that integrates a distributed network of cooperative edge servers for fast model aggregation and achieves scalable and low-latency model learning at mobile edge networks. We further develop a new federated optimization algorithm called CE-FedAvg under the proposed CFEL and analyze its convergence properties under the general non-convex and non-IID setting. Our extensive experiments on FL benchmark datasets  demonstrate that compared with prior FL frameworks, CFEL can largely reduce the training time to achieve a target model accuracy. For future work, we will investigate computational heterogeneity and rigorous privacy protection in CFEL.


%



\ifCLASSOPTIONcompsoc
  \section*{Acknowledgments}
\else
  \section*{Acknowledgment}
\fi
The work was supported in part by the U.S. National Science Foundation under grants CNS-2047761, CNS-2106761, and CMMI-2222670.


\ifCLASSOPTIONcaptionsoff
  \newpage
\fi



%
\bibliographystyle{IEEEtran}
\bibliography{IEEEabrv,guo_career.bib}







\newpage
\onecolumn
\title{Scalable and Low-Latency Federated Learning with Cooperative Mobile Edge Networking (Supplementary)}



\appendices
\section{Proof Preliminaries}
For ease of notation, we define the averaged stochastic gradient, the averaged mini-batch gradient and the averaged gradient of average model as:
\begin{equation*}
    \overline{\mathbf{G}}(\mathbf{X}_{t})=\mathbf{G}_t\frac{\mathbf{1}_{n}}{n}=\frac{1}{n}\sum_{k=1}^n\mathbf{g}_k(\mathbf{x}_{t}^{(k)}),\quad
    \overline{\nabla F}(\mathbf{X}_{t}) = \mathbb{E}[\overline{\mathbf{G}_t}] = \frac{1}{n}\sum_{k=1}^n\nabla F_k(\mathbf{x}_{t}^{(k)}), \quad \nabla F(\mathbf{u}_{t}) =  \frac{1}{n}\sum_{k=1}^n\nabla F_k(\mathbf{u}_{t}).\\
\end{equation*}
We also define the gradient matrices for local models $\mathbf{x}_t^{{k}}$, cluster models $\overline{\mathbf{x}}_{t}^{(k)}$ and average model $\mathbf{u}_t$ as:
\begin{align*}
    \mathcal{H}(\mathbf{X}_{t}) & = \mathbb{E}[\mathbf{G}(\mathbf{X}_{t})]  = [\nabla F_1(\mathbf{x}_{t}^{(1)}), \nabla F_2(\mathbf{x}_{t}^{(2)}), \ldots ,\nabla F_n(\mathbf{x}_{t}^{(n)})] \in \mathbb{R}^{d \times n},\\
    \mathcal{J}(\mathbf{X}_{t}) & = [\nabla F_1(\overline{\mathbf{x}}_{t}^{(1)}), \nabla F_2(\overline{\mathbf{x}}_{t}^{(2)}), \ldots ,\nabla F_n(\overline{\mathbf{x}}_{t}^{(n)})] \in \mathbb{R}^{d \times n},\\
     \mathcal{I}(\mathbf{X}_{t}) & = [\nabla F_1(\mathbf{u}_{t}), \nabla F_2(\mathbf{u}_{t}), \ldots ,\nabla F_n(\mathbf{u}_{t})] \in \mathbb{R}^{d \times n}.
\end{align*}
Here, $\overline{\mathbf{x}}_{t}^{(k)}$ is the average model within cluster $i$ where device $k\in i$. For ease of representation, we use $\mathcal{H}_t, \mathcal{J}_t, \mathcal{I}_t$ to denote $\mathcal{H}(\mathbf{X}_{t})$, $\mathcal{J}(\mathbf{X}_{t})$ and $\mathcal{I}(\mathbf{X}_{t})$, respectively.

\section{Useful Lemmas}
We use $\mathbf{Z}$ to denote $\mathbf{B}^\intercal \text{diag}(\mathbf{c}) \mathbf{H}^{\pi}\mathbf{B}$. Recall that $\mathbf{V}=\mathbf{B}^\intercal \text{diag}(\mathbf{c}) \mathbf{H}^{\pi}\mathbf{B}$ and $\mathbf{A}=\mathbf{1}_{n}\mathbf{1}_{n}^\intercal/n.$ 

\setcounter{lemma}{4}
\setcounter{lemma}{4}
\begin{lemma}\label{lemma_eig_zv}
Let the matrices $\mathbf{Z}$ and $\mathbf{V}$ be defined therein. Then we have:
\begin{itemize}
    \item $\mathbf{1}_{n}$ is a right eigenvector of $\mathbf{Z}$ and $\mathbf{V}$ with eigenvalue 1.
    \item $\mathbf{1}_n^\intercal$ is a left eigenvector of $\mathbf{Z}$ and $\mathbf{V}$ with eigenvalue 1.
    \item The eigenvalues of $\mathbf{Z}$ are the same as the eigenvalues of $\mathbf{H}^{\pi}$.
\end{itemize}
\end{lemma}

\begin{lemma}
\label{lemma_comm_A_W}
Let matrices $\mathbf{A}$ and $\mathbf{W}_t$ be defined therein. Then we have:
\begin{align*}
    \mathbf{W}_t\mathbf{A}=\mathbf{A}\mathbf{W}_t=\mathbf{A}.
\end{align*}
\end{lemma}

\begin{lemma}
\label{lemma_eig_Mat}
Let the matrices $\mathbf{Z}$, $\mathbf{V}$, integers $l$ and $\pi$ be defined therein. Then we have:
\begin{equation*}
    \|\mathbf{Z}^{l}-\mathbf{A}\|_{\textup{op}}=\zeta^{l\pi},\quad \|\mathbf{V}-\mathbf{A}\|_{\textup{op}}=1,
\end{equation*}
where $\zeta =  \max \{{|\lambda_2(\mathbf{H})|,|\lambda_n(\mathbf{H})|}\}$, and $\lambda_{i}(\cdot)$ denote the $i$-th largest eigenvalue of a matrix.
\end{lemma}

\begin{lemma}
\label{lemma_tr_Fa}
Suppose $\mathbf{C}\in\mathbb{R}^{d\times n}$, $\mathbf{D}\in\mathbb{R}^{n\times n}$ are two matrices, then we have:
\begin{align*}
    \|\mathbf{C}\mathbf{D}\|_{\textup{F}} \leq \|\mathbf{C}\|_{\textup{F}}\|\mathbf{D}\|_{\textup{F}}.
\end{align*}
\end{lemma}

\begin{lemma}
\label{lemma_comm_Z_V}
Let the matrices $\mathbf{Z}$ and $\mathbf{V}$ be defined therein. Then we have:
\begin{align*}
    \mathbf{Z}\mathbf{V}=\mathbf{V}\mathbf{Z}=\mathbf{Z}.
\end{align*}
\end{lemma}

\begin{lemma}
\label{lemma_Fa_op}
Suppose $\mathbf{C}\in\mathbb{R}^{d\times n}$, $\mathbf{D}\in\mathbb{R}^{n\times n}$ are two matrices, then we have:
\begin{align*}
   \|\mathbf{CD}\|_{\textup{F}}  \leq \|\mathbf{C}\|_{\textup{F}}\|\mathbf{D}\|_{\textup{op}}.
\end{align*}
\end{lemma}

Lemmas~\ref{lemma_eig_zv}-\ref{lemma_eig_Mat},~\ref{lemma_tr_Fa}-\ref{lemma_Fa_op} are provided by \cite{sun2021semi} and \cite{castiglia2020multi}, respectively. 

\begin{lemma}
\label{lemma_i_v_op}
Let the matrix $\mathbf{V}$ be defined therein. Then we have:
\begin{align*}
    \|\mathbf{I}-\mathbf{V}\|_{\textup{op}}=1,\quad \|\mathbf{I}-\mathbf{A}\|_{\textup{op}}=1.
\end{align*}
\end{lemma}

\begin{IEEEproof}
According to the definition of the matrix operator norm, we have:
\begin{align*}
    \|\mathbf{I}-\mathbf{V}\|_{\textup{op}} = & \sqrt{\lambda_{\max}(\mathbf{I}-\mathbf{V})^\intercal(\mathbf{I}-\mathbf{V})}\notag\\
     \labelrel={eq_i_v_sqr} & \sqrt{\lambda_{\max}(\mathbf{I}-\mathbf{V})},
\end{align*}
where~\eqref{eq_i_v_sqr} follows from $\mathbf{V}^2=\mathbf{V}$. By using Lemma 5 in \cite{castiglia2020multi}, we have $ \|\mathbf{I}-\mathbf{V}\|_{\textup{op}} =1$. Similarly, we can obtain $\|\mathbf{I}-\mathbf{A}\|_{\textup{op}}=1$. 
\end{IEEEproof}

\section{INTERMEDIATE RESULTS}

In this section, we use notation $\mathbb{E}$ to denote the expectation $\mathbb{E}_{\theta_k}$ over mini-batch $\theta_k$.
\begin{lemma}[Inter-cluster residual error decomposition]
\label{lemma:inter_t12} 
Under Assumptions~\ref{ass:smoothness} and~\ref{ass:gradient}, we have:
\begin{align*}
    \mathbb{E}\frac{1}{n}\|\mathbf{X}_t(\mathbf{V}-\mathbf{A})\|_{\textup{F}}^2\leq& \underbrace{\frac{2\eta^2}{n}\mathbb{E}\Biggl|\Biggl| \sum_{\alpha=0}^{l -1}(\mathbf{Y}_{\alpha}-\mathbf{P}_{\alpha})(\mathbf{Z}^{l -\alpha}-\mathbf{A})+(\mathbf{Y}_{l}-\mathbf{P}_{l })(\mathbf{V}-\mathbf{A})\Biggl|\Biggl|_{\textup{F}}^2}_{T_1}\\ &+ \underbrace{\frac{2\eta^2}{n}\mathbb{E}\Biggl|\Biggl| \sum_{\alpha=0}^{l -1}\mathbf{P}_{\alpha}(\mathbf{Z}^{l -\alpha}-\mathbf{A})+\mathbf{P}_{l }(\mathbf{V}-\mathbf{A}))\Biggl|\Biggl|_{\textup{F}}^2}_{T_2},
\end{align*}
where $\mathbf{Y}_{\alpha}, \mathbf{Y}_{l}, \mathbf{P}_{\alpha}, \mathbf{P}_{l}$ are defined in \eqref{def_ypqr_1} and \eqref{def_ypqr_ell}.
\end{lemma}
\begin{IEEEproof}
According to the update rule
$\mathbf{X}_{t+1}=(\mathbf{X}_t-\eta\mathbf{G}_t)\mathbf{W}_t$,
we have:
\begin{align*}
    \mathbf{X}_t(\mathbf{V}-\mathbf{A}) & =(\mathbf{X}_{t-1}-\eta \mathbf{G}_{t-1})\mathbf{W}_{t-1}(\mathbf{V}-\mathbf{A})\notag\\
    & \labelrel={eq_inter_stoc} \mathbf{X}_{t-1}(\mathbf{V}-\mathbf{A})\mathbf{W}_{t-1}-\eta\mathbf{G}_{t-1}\mathbf{W}_{t-1}(\mathbf{V}-\mathbf{A})\\
    & = (\mathbf{X}_{t-2}-\eta \mathbf{G}_{t-2})(\mathbf{V}-\mathbf{A})\mathbf{W}_{t-2}\mathbf{W}_{t-1}-\eta\mathbf{G}_{t-1}\mathbf{W}_{t-1}(\mathbf{V}-\mathbf{A})\notag\\
    & = \mathbf{X}_{t-2}(\mathbf{V}-\mathbf{A})\mathbf{W}_{t-2}\mathbf{W}_{t-1}-\eta \mathbf{G}_{t-2}\mathbf{W}_{t-2}\mathbf{W}_{t-1}(\mathbf{V}-\mathbf{A})-\eta\mathbf{G}_{t-1}\mathbf{W}_{t-1}(\mathbf{V}-\mathbf{A})\notag,
\end{align*}
where~\eqref{eq_inter_stoc} follows Lemma~\ref{lemma_comm_A_W} and the following property of doubly stochastic matrix: $\mathbf{V}\mathbf{W}_{t-1} = \mathbf{V}\mathbf{W}_{t-1} = \mathbf{W}_{t-1}$. Then, by induction, we have:
\begin{equation*}
    \mathbf{X}_t(\mathbf{V}-\mathbf{A})=\mathbf{X}_0(\mathbf{V}-\mathbf{A})\prod_{u=0}^{t-1}\mathbf{W}_{u}-\eta \sum_{c=1}^{t-1}\mathbf{G}_{c}\prod_{u=c}^{t-1}\mathbf{W}_{u}\left(\mathbf{V}-\mathbf{A}\right).
\end{equation*}
Here, $u, c$ are the indices for global iteration. Since all clients were initialized with the same model, $\mathbf{X}_0(\mathbf{V}-\mathbf{A})=0$. Then, the squared norm of the inter-cluster residual error can be written as:
\begin{equation}
    \mathbb{E}\frac{1}{n}\|\mathbf{X}_t(\mathbf{V}-\mathbf{A})\|_{\textup{F}}^2 = \frac{\eta^2}{n}\mathbb{E}\|\sum_{c=1}^{t-1}\mathbf{G}_c \mathbf{\Phi}_{c,t-1}(\mathbf{V}-\mathbf{A})\|_{\textup{F}}^2, \label{eq_g_c_update}
\end{equation}
where $\mathbf{\Phi}_{c,t-1} := \prod_{u=c}^{t-1}\mathbf{W}_{u}$. %

Recall that $t=l q\tau+r\tau+s$, where $l \in [0, p - 1]$ is the global round index, $r \in [0, q - 1] $ is the edge round index, and $s \in [0, \tau - 1]$ is the local iteration index. Since $\mathbf{V}^l =\mathbf{V}$ and $\mathbf{VZ}=\mathbf{ZV}=\mathbf{Z}$ by Lemma~\ref{lemma_comm_Z_V}, we have:
\begin{align}
    \mathbf{\Phi}_{c,t-1} = 
        \begin{cases}
            \mathbf{I}, & \ l q\tau+r \tau< c < l q\tau+r \tau+s\\
            \mathbf{V}, & \ l q\tau< c\leq l q\tau+r \tau\\
            \mathbf{Z}, & \ (l -1)q\tau< c \leq l q\tau\\
            \mathbf{Z}^2, & \ (l -2)q\tau< c \leq (l -1)q\tau\\
            \vdots & \\
            \mathbf{Z}^l.  & \ 1\leq c \leq q\tau
        \end{cases}\label{def_Phi_1}
\end{align}

For ease of presentation, for integer $\alpha \in [0,\ldots,l-1]$, we define the accumulated stochastic gradient matrices within one global update period as: 
\begin{equation}\label{def_ypqr_1}
    \mathbf{Y}_{\alpha}  =\sum_{c=\alpha q\tau+1}^{(\alpha+1)q\tau}\mathbf{G}_c, \quad \mathbf{P}_{\alpha}=\sum_{c=\alpha q\tau+1}^{(\alpha+1)q\tau}\mathcal{J}_c,\quad  \mathbf{Q}_{\alpha} = \sum_{c=\alpha q\tau+1}^{(\alpha+1)q\tau}\mathcal{I}_c,\quad  
    \mathbf{R}_{\alpha} = \sum_{c = \alpha q\tau+1}^{(\alpha+1)q\tau}\mathcal{H}_c. 
\end{equation}
Similarly, let 
\begin{equation}\label{def_ypqr_ell}
    \mathbf{Y}_{l}=\sum_{c=l q\tau+1}^{l q\tau+r \tau+s-1}\mathbf{G}_c, \quad \mathbf{P}_{l}=\sum_{c =l q\tau+1}^{l q\tau+r \tau+s-1}\mathcal{J}_c, \quad \mathbf{Q}_{l }=\sum_{c=l q\tau+1}^{l q\tau+r \tau+s-1}\mathcal{I}_c,\quad \mathbf{R}_{l}=\sum_{c=l q\tau+1}^{l q\tau+r \tau+s-1}\mathcal{H}_c.
\end{equation}
Thus, we obtain:
\begin{align*}
    \sum_{c=1}^{q\tau}\mathbf{G}_c\Phi_{c,t-1}(\mathbf{V}-\mathbf{A}) & =\mathbf{Y}_0\mathbf{Z}^l(\mathbf{V} -\mathbf{A}),\\
    \sum_{c=q\tau+1}^{2q\tau}\mathbf{G}_c\Phi_{c,t-1}(\mathbf{V}-\mathbf{A}) & =\mathbf{Y}_1\mathbf{Z}^{l -1}(\mathbf{V}-\mathbf{A}),\\
    & \ldots \notag\\
    \sum_{c=(l -1)q\tau+1}^{l q\tau}\mathbf{G}_c\Phi_{c,t-1}(\mathbf{V}-\mathbf{A}) & =\mathbf{Y}_{l -1}\mathbf{Z}(\mathbf{V}-\mathbf{A}),\\
    \sum_{c=l q\tau+1}^{l q\tau+r \tau+s-1}\mathbf{G}_c\Phi_{c,t-1}(\mathbf{V}-\mathbf{A}) & =\mathbf{Y}_{l}(\mathbf{V}-\mathbf{A}).
\end{align*}

By summing all of them together and according to Lemmas~\ref{lemma_comm_A_W} and~\ref{lemma_comm_Z_V}, we obtain:
\begin{align}\label{eq:sum_of_ups}
    \sum_{c=1}^{t-1}\mathbf{G}_{c}\mathbf{\Phi}_{c,t-1}(\mathbf{V}-\mathbf{A}) = \sum_{\alpha=0}^{l -1}\mathbf{Y}_{\alpha}(\mathbf{Z}^{l -\alpha}-\mathbf{A}) + \mathbf{Y}_{l}(\mathbf{V}-\mathbf{A}).
\end{align}

Plugging~\eqref{eq:sum_of_ups} into~\eqref{eq_g_c_update}, we obtain: 
\begin{align*}
    \mathbb{E}\frac{1}{n}&\|\mathbf{X}_t(\mathbf{I}-\mathbf{A})\|_{\textup{F}}^2 =  \frac{\eta^2}{n}\mathbb{E}\|\sum_{\alpha=0}^{l -1}\mathbf{Y}_{\alpha}(\mathbf{Z}^{l -\alpha}-\mathbf{A})+\mathbf{Y}_{l}(\mathbf{V}-\mathbf{A})\|_{\textup{F}}^2\\
    = &\frac{\eta^2}{n}\mathbb{E}\Biggl|\Biggl| \sum_{\alpha=0}^{l -1}(\mathbf{Y}_{\alpha}-\mathbf{P}_{\alpha})(\mathbf{Z}^{l -\alpha}-\mathbf{A})+(\mathbf{Y}_{l }-\mathbf{P}_{l})(\mathbf{V}-\mathbf{A}) + \sum_{\alpha=0}^{l -1}\mathbf{P}_{\alpha}(\mathbf{Z}^{l -\alpha}-\mathbf{A})+\mathbf{P}_{l }(\mathbf{V}-\mathbf{A})\Biggr]\Biggr|\!\Biggr|_{\textup{F}}^2\\
    \leq & \underbrace{\frac{2\eta^2}{n}\mathbb{E}\Biggl|\Biggl| \sum_{\alpha=0}^{l -1}(\mathbf{Y}_{\alpha}-\mathbf{P}_{\alpha})(\mathbf{Z}^{l -\alpha}-\mathbf{A})+(\mathbf{Y}_{l}-\mathbf{P}_{l })(\mathbf{V}-\mathbf{A})\Biggl|\Biggl|_{\textup{F}}^2}_{T_1} + \underbrace{\frac{2\eta^2}{n}\mathbb{E}\Biggl|\Biggl| \sum_{\alpha=0}^{l -1}\mathbf{P}_{\alpha}(\mathbf{Z}^{l -\alpha}-\mathbf{A})+\mathbf{P}_{l }(\mathbf{V}-\mathbf{A}))\Biggl|\Biggl|_{\textup{F}}^2}_{T_2}.
\end{align*}
\end{IEEEproof}

\begin{lemma}[Bounded $\emph{T}_1$]
\label{lemma_bound_t1}
Under Assumptions~\ref{ass:smoothness},~\ref{ass:gradient} and~\ref{ass:mixing}, we have:
\begin{equation}\label{in_bound_t1_final}
\sum_{t=0}^{T-1}T_{1} \leq 2\eta^2\sigma^2T\left( \Omega_{1}q\tau+\frac{m-1}{n}q\tau\right)  + 4\eta^2L^2q^2\tau^2 \Omega_2 \sum_{t=0}^{T-1}\frac{1}{n}\|\mathbf{X}_t(\mathbf{V}-\mathbf{I})\|_{\textup{F}}^2.
\end{equation}
\end{lemma}
\begin{IEEEproof}
According to the definition of $T_1$ in Lemma~\ref{lemma:inter_t12}, we have:
\begin{align}
    \emph{T}_1  = & \frac{2\eta^2}{n}\mathbb{E}\Biggl|\Biggl| \sum_{\alpha=0}^{l -1}(\mathbf{Y}_{\alpha}-\mathbf{P}_{\alpha})(\mathbf{Z}^{l -\alpha}-\mathbf{A})+(\mathbf{Y}_{l}-\mathbf{P}_{l })(\mathbf{V}-\mathbf{A})\Biggl|\Biggl|_{\textup{F}}^2\notag\\
    = & \frac{2\eta^2}{n}\mathbb{E}\Biggl|\Biggl| \sum_{\alpha=0}^{l -1}(\mathbf{Y}_{\alpha}-\mathbf{R}_{\alpha})(\mathbf{Z}^{l -\alpha}-\mathbf{A})+\sum_{\alpha=0}^{l -1}(\mathbf{R}_{\alpha}-\mathbf{P}_{\alpha})(\mathbf{Z}^{l -\alpha}-\mathbf{A}) +(\mathbf{Y}_{l}-\mathbf{R}_{l})(\mathbf{V}-\mathbf{A})+(\mathbf{R}_{l}-\mathbf{P}_{l})(\mathbf{V}-\mathbf{A})\Biggl|\Biggl|_{\textup{F}}^2\nonumber\\
    \leq & \underbrace{\frac{4\eta^2}{n}\mathbb{E}\Biggl|\Biggl| \sum_{\alpha=0}^{l -1}(\mathbf{Y}_{\alpha}-\mathbf{R}_{\alpha})(\mathbf{Z}^{l -\alpha}-\mathbf{A})+(\mathbf{Y}_{l}-\mathbf{R}_{l})(\mathbf{V}-\mathbf{A})\Biggl|\Biggl|_{\textup{F}}^2}_{T_3}\nonumber\\
    & +\underbrace{\frac{4\eta^2}{n}\mathbb{E}\Biggl|\Biggl|\sum_{\alpha=0}^{l -1}(\mathbf{R}_{\alpha}-\mathbf{P}_{\alpha})(\mathbf{Z}^{l -\alpha}-\mathbf{A})+(\mathbf{R}_{l}-\mathbf{P}_{l})(\mathbf{V}-\mathbf{A})\Biggl|\Biggl|_{\textup{F}}^2}_{T_4}.\label{eq_T1_34}
\end{align}

For deriving the upper bound of $\emph{T}_1$, we need to bound $\emph{T}_3$ and $\emph{T}_4$. Firstly, we bound $\emph{T}_3$ as
\begin{align}
     \emph{T}_3= & \frac{4\eta^2}{n}\mathbb{E}\Biggl|\Biggl| \sum_{\alpha=0}^{l -1}(\mathbf{Y}_{\alpha}-\mathbf{R}_{\alpha})(\mathbf{Z}^{l -\alpha}-\mathbf{A})+(\mathbf{Y}_{l}-\mathbf{R}_{l})(\mathbf{V}-\mathbf{A})\Biggl|\Biggl|_{\textup{F}}^2\notag\\
     = & \frac{4\eta^2}{n} \sum_{\alpha=0}^{l -1}\mathbb{E}\|(\mathbf{Y}_{\alpha}-\mathbf{R}_{\alpha})(\mathbf{Z}^{l -\alpha}-\mathbf{A})\|_{\textup{F}}^2+\frac{4\eta^2}{n}\mathbb{E}\|(\mathbf{Y}_l-\mathbf{R}_l)(\mathbf{V}-\mathbf{A})\|_{\textup{F}}^2\nonumber\notag\\
    & + \underbrace{\frac{4\eta^2}{n}\sum_{\alpha=0}^{l -1}\sum_{\alpha^{\prime}=0,\alpha^{\prime}\neq \alpha}^{l -1}\mathbb{E}\underbrace{\left<(\mathbf{Y}_\alpha-\mathbf{R}_\alpha)(\mathbf{Z}^{l -\alpha}-\mathbf{A}), (\mathbf{Y}_{\alpha^{\prime}}-\mathbf{R}_{\alpha^{\prime}})(\mathbf{Z}^{l -{\alpha^{\prime}}}-\mathbf{A}) \right>}_{TR}}_{TR_0}\nonumber\\
    & + \underbrace{\frac{8\eta^2}{n}\sum_{\alpha=0}^{l -1}\mathbb{E}\left<(\mathbf{Y}_{l}-\mathbf{R}_{l })(\mathbf{V}-\mathbf{A}), (\mathbf{Y}_{\alpha}-\mathbf{R}_{\alpha})(\mathbf{Z}^{l -{\alpha}}-\mathbf{A})  \right>}_{TR_1}.\label{eq_T3_CROSS}
\end{align}
According to Assumption~\ref{ass:gradient}, the gradient estimation of each device is unbiased (i.e., $\mathbb{E}\mathbf{Y}_{\alpha}=\mathbf{R}_{\alpha},  \mathbb{E}\mathbf{Y}_{l}=\mathbf{R}_{l}$). Then all the cross terms in~\eqref{eq_T3_CROSS} become zeros. Therefore, we have:
\begin{align}
    \emph{T}_3 = & \frac{4\eta^2}{n} \sum_{\alpha=0}^{l -1}\mathbb{E}\|(\mathbf{Y}_{\alpha}-\mathbf{R}_{\alpha})(\mathbf{Z}^{l -\alpha}-\mathbf{A})\|_{\textup{F}}^2+\frac{4\eta^2}{n}\mathbb{E}\|(\mathbf{Y}_{l}-\mathbf{R}_{l})(\mathbf{V}-\mathbf{A})\|_{\textup{F}}^2\notag\\
    \labelrel\leq{T3_FA_OP} & \frac{4\eta^2}{n} \sum_{\alpha=0}^{l -1}\mathbb{E}\|\mathbf{Y}_{\alpha}-\mathbf{R}_{\alpha}\|_{\textup{F}}^2\|\mathbf{Z}^{l -\alpha}-\mathbf{A}\|_{\textup{op}}^2+\frac{4\eta^2}{n}\mathbb{E}\|(\mathbf{Y}_{l}-\mathbf{R}_{l })(\mathbf{V}-\mathbf{A})\|_{\textup{F}}^2\notag\\
    \labelrel\leq{t3_op_eig} & \frac{4\eta^2}{n} \sum_{\alpha=0}^{l -1}\mathbb{E}\|\mathbf{Y}_{\alpha}-\mathbf{R}_{\alpha}\|_{\textup{F}}^2\zeta^{2\pi(l-\alpha)} + \frac{4\eta^2}{n}\|(\mathbf{Y}_{l}-\mathbf{R}_{l})(\mathbf{V}-\mathbf{A})\|_{\textup{F}}^2,\label{T3_OP_EIG}
\end{align}
where~\eqref{T3_FA_OP} follows from Lemma~\ref{lemma_Fa_op}, and~\eqref{t3_op_eig} follows from Lemma~\ref{lemma_eig_Mat}. 

For any $\alpha \in [0, \ldots, l - 1]$, from the definitions of $\mathbf{Y}_{\alpha}$ and $\mathbf{R}_{\alpha}$ in~\eqref{def_ypqr_1}, we get:
\begin{align}
    \mathbb{E}\frac{1}{n}\|\mathbf{Y}_{\alpha}-\mathbf{R}_{\alpha}\|_{\textup{F}}^2 & = \frac{1}{n}\mathbb{E}\Biggl|\Biggl|\sum_{c=\alpha q\tau+1}^{(\alpha+1)q\tau}(\mathbf{G}_c-\mathcal{H}(\mathbf{X}_c))\Biggl|\Biggl|_{\textup{F}}^2\notag\\
    & \labelrel={eq_def_f_norm} \sum_{k=1}^n\frac{1}{n}\mathbb{E}\|\sum_{c=\alpha q\tau+1}^{(\alpha +1)q\tau}(\mathbf{g}_k(\mathbf{x}_c^{(k)})-\nabla F_k(\mathbf{x}_c^{(k)}))\|^2\notag\\
    & \labelrel={eq_GRAD_YR} \sum_{k=1}^n\frac{1}{n}\sum_{c =\alpha q\tau+1}^{(\alpha +1)q\tau}\mathbb{E}\|\mathbf{g}_k(\mathbf{x}_c^{(k)})-\nabla F_k(\mathbf{x}_c^{(k)})\|^2\notag\\
    & \labelrel\leq{eq_BOUND_YR} \sum_{k=1}^n\frac{1}{n}\sum_{c =\alpha q\tau+1}^{(\alpha +1)q\tau}\sigma^2\notag\\
    & = q\tau\sigma^2,\label{bound_YF}
\end{align}
where~\eqref{eq_def_f_norm} follows the definition of Frobenius norm, and  \eqref{eq_GRAD_YR} and \eqref{eq_BOUND_YR} follow from Assumption~\ref{ass:gradient}.
For the global round $l$, we have:
\begin{align}
    & \mathbb{E}\frac{1}{n}\|(\mathbf{Y}_l -\mathbf{R}_{l})(\mathbf{V}-\mathbf{A})\|_{\textup{F}}^2 \\
    & = \mathbb{E}\frac{1}{n}\|(\mathbf{Y}_{l}-\mathbf{R}_{l})\mathbf{V}-(\mathbf{Y}_{l}-\mathbf{R}_{l})\mathbf{A})\|_{\textup{F}}^2\notag\\
    & = \sum_{i=1}^{m}\frac{n_i}{n}\mathbb{E}\|\sum_{c=l q\tau+1}^{l q\tau+r\tau+s-1}[(\frac{1}{n_i}\sum_{k\in\mathcal{S}_i}\mathbf{g}_k(\mathbf{x}_c^{(k)})-\frac{1}{n_i}\sum_{k\in\mathcal{S}_i}\nabla F_k(\mathbf{x}_{c}^{(k)})) - (\frac{1}{n}\sum_{j=1}^m\sum_{k\in\mathcal{S}_{j}}\mathbf{g}_k(\mathbf{x}_c^{(k)})-\frac{1}{n}\sum_{j=1}^m\sum_{k\in\mathcal{S}_{j}}\nabla F_k(\mathbf{x}_{c}^{(k)})) ]\|^2\notag\\
    & \labelrel={eq_t3_var} \sum_{i=1}^{m}\frac{n_i}{n}\mathbb{E}\|\sum_{c=l q\tau+1}^{l q\tau+r\tau+s-1}\frac{1}{n_i}\sum_{k\in\mathcal{S}_i}(\mathbf{g}_k(\mathbf{x}_c^{(k)})-\nabla F_k(\mathbf{x}_{c}^{(k)}))\|^2 -\mathbb{E} \|\sum_{c=l q\tau+1}^{l q\tau+r\tau+s-1}[\frac{1}{n}\sum_{j=1}^m\sum_{k\in\mathcal{S}_{j}}(\mathbf{g}_k(\mathbf{x}_c^{(k)})-\nabla F_k(\mathbf{x}_c^{(k)}))] \|^2\notag\\
    & \labelrel={eq_t3_indep} \sum_{c=l q\tau+1}^{l q\tau+r\tau+s-1}[\sum_{i=1}^{m}\frac{n_i}{n}\frac{1}{n_i^2}\sum_{k\in\mathcal{S}_i}\mathbb{E}\|\mathbf{g}_k(\mathbf{x}_c^{(k)})-\nabla F_k(\mathbf{x}_{c}^{(k)})\|^2-\frac{1}{n^2}\sum_{j=1}^m\sum_{k\in\mathcal{S}_{j}}\mathbb{E}\|\mathbf{g}_k(\mathbf{x}_c^{(k)})-\nabla F_k(\mathbf{x}_{c}^{(k)}) \|^2]\notag\\
    & = \sum_{c=l q\tau+1}^{l q\tau+r\tau+s-1}\frac{1}{n}\sum_{i=1}^{m}(\frac{1}{n_i}-\frac{1}{n})\sum_{k\in\mathcal{S}_i}\mathbb{E}\|\mathbf{g}_k(\mathbf{x}_c^{(k)})-\nabla F_k(\mathbf{x}^{(k)})\|^2\notag\\
    & \leq \frac{m-1}{n}(r\tau+s-1)\sigma^2,\label{bound_YFlpha1}
\end{align}
where~\eqref{eq_t3_var} follows from $\sum_{i=1}^{m} a_{i}\|\bm{b}_{i}-\overline{\bm{b}}\|^2=\sum_{i=1}^{m} a_{i}\|\bm{b}_{i}\|^2-\|\overline{\bm{b}}\|^2$ for any $\bm{b}_i\in\mathbb{R}^d$, $0\leq a_i\leq 1$, and $\sum_{i=1}^{m} a_{i}=1$, and \eqref{eq_t3_indep} holds due to Assumption~\ref{ass:gradient}. Plugging~\eqref{bound_YF} and~\eqref{bound_YFlpha1} back into~\eqref{T3_OP_EIG}, we get:
\begin{equation*}
    \emph{T}_3\leq 4q\tau\eta^2\sigma^2\sum_{\alpha=0}^{l -1} \zeta^{2\pi(l -\alpha)} +4\frac{m-1}{n}(r\tau+s-1)\eta^2\sigma^2.
\end{equation*}

Recalling Lemma~\ref{lemma_bound_ave_grad}, our goal is to calculate $\sum_{t=0}^{T-1}\emph{T}_3$, which can be decomposed into $\sum_{l=0}^{p-1}\sum_{r=0}^{q-1}\sum_{s=0}^{\tau-1}\emph{T}_3.$ Firstly, we sum over all iterates in $l $-th global round period, for $r=0, \ldots, q-1$ and $s=0, \ldots, \tau-1$, we have:
\begin{align*}
    \sum_{r=0}^{q-1}\sum_{s=0}^{\tau-1}\emph{T}_3 & \leq 4q^2\tau^2\eta^2\sigma^2\sum_{\alpha=0}^{l -1} \zeta^{2\pi(l -\alpha)} + 2\frac{m-1}{n}\tau^2q(q-1)\eta^2\sigma^2 + 2\frac{m-1}{n}q\tau(\tau-1)\eta^2\sigma^2\notag\\
    & \labelrel\leq{bound_sum_T3_zeta} 2q\tau\eta^2\sigma^2\left( \frac{\zeta^{2\pi}}{1-\zeta^{2\pi}}q\tau+
    \frac{m-1}{n}q\tau\right),
\end{align*}
where~\eqref{bound_sum_T3_zeta} uses the following fact:
\begin{align*}
        \sum_{\alpha=0}^{l-1} \zeta^{2\pi(l -\alpha)} \leq \sum_{\alpha=-\infty}^{l-1}\zeta^{2\pi(l -\alpha)} \leq \frac{\zeta^{2\pi}}{1-\zeta^{2\pi}}.
\end{align*}
Secondly, summing over all global round periods from $l =0,\ldots,p-1$, we have:
\begin{align}
    \sum_{l =0}^{p-1}\sum_{r=0}^{q-1}\sum_{s=0}^{\tau-1}\emph{T}_3 &\leq 2\eta^2\sigma^2 T\left( \frac{\zeta^{2\pi}}{1-\zeta^{2\pi}}q\tau+\frac{m-1}{n}q\tau\right).\label{bound_t3_overtime}
\end{align}
Referring to $\Omega_1=\frac{\zeta^{2\pi}}{1-\zeta^{2\pi}}$, then we get:
\begin{align*}
    \sum_{t=0}^{T-1}\emph{T}_3 &\leq 2\eta^2\sigma^2T\left( \Omega_{1}q\tau+\frac{m-1}{n}q\tau\right).
\end{align*}

Next, we bound $\emph{T}_4$ as follows:
\begin{align}
    \emph{T}_4 = & \frac{4\eta^2}{n}\mathbb{E}\Biggl|\Biggl|\sum_{\alpha=0}^{l -1}(\mathbf{R}_{\alpha}-\mathbf{P}_{\alpha})(\mathbf{Z}^{l -\alpha}-\mathbf{A})+(\mathbf{R}_{l }-\mathbf{P}_{l })(\mathbf{V}-\mathbf{A})\Biggl|\Biggl|_{\textup{F}}^2\notag\\
    = & \frac{4\eta^2}{n} \sum_{\alpha=0}^{l -1}\mathbb{E}\|(\mathbf{R}_{\alpha}-\mathbf{P}_{\alpha})(\mathbf{Z}^{l -\alpha}-\mathbf{A})\|_{\textup{F}}^2+\frac{4\eta^2}{n}\|(\mathbf{R}_{l}-\mathbf{P}_{l})(\mathbf{V}-\mathbf{A})\|_{\textup{F}}^2\nonumber\\
    & + \underbrace{\frac{4\eta^2}{n}\sum_{\alpha=0}^{l -1}\sum_{\alpha^{\prime}=0,\alpha^{\prime}\neq \alpha}^{l -1}\mathbb{E}\underbrace{\left<(\mathbf{R}_{\alpha}-\mathbf{P}_{\alpha})(\mathbf{Z}^{l -\alpha}-\mathbf{A}), (\mathbf{R}_{\alpha^{\prime}}-\mathbf{P}_{\alpha^{\prime}})(\mathbf{Z}^{l-{\alpha^{\prime}}}-\mathbf{A}) \right>}_{TR^{\prime}}}_{TR_0^{\prime}}\nonumber\\
    & + \underbrace{\frac{8\eta^2}{n}\sum_{\alpha=0}^{l -1}\mathbb{E}\left<(\mathbf{R}_{l }-\mathbf{P}_{l})(\mathbf{V}-\mathbf{A}), (\mathbf{R}_\alpha-\mathbf{P}_\alpha)(\mathbf{Z}^{l -\alpha}-\mathbf{A})  \right>}_{TR_1^{\prime}}.\label{in_T4}
\end{align}
$\emph{TR}^{\prime}$ can be bounded as:
\begin{align}
    \emph{TR}^{\prime} \labelrel\leq{in_T4_TR'} & \mathbb{E}\|(\mathbf{R}_\alpha-\mathbf{P}_\alpha)(\mathbf{Z}^{l -\alpha}-\mathbf{A})\|_{\textup{F}} \mathbb{E}\|(\mathbf{R}_{\alpha^{\prime}}-\mathbf{P}_{\alpha^{\prime}})(\mathbf{Z}^{l -\alpha^{\prime}}-\mathbf{A})\|_{\textup{F}}\notag\\
    \labelrel\leq{in_T4_FA_OP_1} & \mathbb{E}\|\mathbf{R}_\alpha-\mathbf{P}_\alpha\|_{\textup{F}}\|\mathbf{Z}^{l -\alpha}-\mathbf{A}\|_{\textup{op}} \mathbb{E}\|\mathbf{R}_{\alpha^{\prime}}-\mathbf{P}_{\alpha^{\prime}}\|_{\textup{F}}\|\mathbf{Z}^{l -{\alpha^{\prime}}}-\mathbf{A}\|_{\textup{op}}\notag\\
    \labelrel\leq{in_T4_OP_EIG_1} & \zeta^{(2l-\alpha-{\alpha^{\prime}})\pi}\mathbb{E}\|\mathbf{R}_\alpha-\mathbf{P}_\alpha\|_{\textup{F}}\mathbb{E}\|\mathbf{R}_{\alpha^{\prime}}-\mathbf{P}_{\alpha^{\prime}}\|_{\textup{F}}\notag\\
    \leq & \frac{1}{2}\zeta^{(2l-\alpha-{\alpha^{\prime}})\pi}[\mathbb{E}\|\mathbf{R}_\alpha-\mathbf{P}_\alpha\|_{\textup{F}}+\mathbb{E}\|\mathbf{R}_{\alpha^{\prime}}-\mathbf{P}_{\alpha^{\prime}}\|_{\textup{F}}],\notag
\end{align}
where~\eqref{in_T4_TR'} follows from Cauchy-Schwarz inequality,~\eqref{in_T4_FA_OP_1} follows from Lemma~\ref{lemma_Fa_op} and~\eqref{in_T4_OP_EIG_1} follows from Lemma~\ref{lemma_eig_Mat}.
Using the same techniques, we get:
\begin{equation*}
    \emph{TR}_1^{\prime} \leq \frac{4\eta^2}{n}\sum_{\alpha=0}^{l -1}\zeta^{(l-{\alpha})\pi}[\mathbb{E}\|\mathbf{R}_{l}-\mathbf{P}_{l}\|_{\textup{F}}+\mathbb{E}\|\mathbf{R}_{\alpha}-\mathbf{P}_{\alpha}\|_{\textup{F}}].
\end{equation*}
By summing the above inequalities of $\emph{TR}_0^{\prime}$ and $\emph{TR}_1^{\prime}$, we have:
\begin{align}
    \emph{TR}_0^{\prime}+\emph{TR}_1^{\prime} \leq & \frac{2\eta^2}{n}\sum_{\alpha=0}^{l -1}\sum_{{\alpha^{\prime}}=0,\alpha^{\prime}\neq \alpha}^{l-1}\zeta^{(2l-\alpha-{\alpha^{\prime}})\pi}[\mathbb{E}\|\mathbf{R}_\alpha-\mathbf{P}_\alpha\|_{\textup{F}}+\mathbb{E}\|\mathbf{R}_{\alpha^{\prime}}-\mathbf{P}_{\alpha^{\prime}}\|_{\textup{F}}] \notag\\
    &+ \frac{4\eta^2}{n}\sum_{\alpha=0}^{l -1}\zeta^{(l-\alpha)\pi}[\mathbb{E}\|\mathbf{R}_{l}-\mathbf{P}_{l }\|_{\textup{F}}+\mathbb{E}\|\mathbf{R}_\alpha-\mathbf{P}_\alpha\|_{\textup{F}}].\notag
\end{align}
Using the symmetric property of indices of $\alpha^{\prime}$ and $\alpha$, we obtain:
\begin{align}
    \emph{TR}_0^{\prime}+\emph{TR}_1^{\prime} \leq & \frac{4\eta^2}{n}\sum_{\alpha=0}^{l -1}\sum_{\alpha^{\prime}=0,\alpha^{\prime}\neq \alpha}^{l -1}\zeta^{(2l-\alpha-{\alpha^{\prime}})\pi}\mathbb{E}\|\mathbf{R}_\alpha-\mathbf{P}_\alpha\|_{\textup{F}} + \frac{4\eta^2}{n}\sum_{\alpha=0}^{l -1}\zeta^{(l-\alpha)\pi}[\mathbb{E}\|\mathbf{R}_{l}-\mathbf{P}_{l }\|_{\textup{F}}+\mathbb{E}\|\mathbf{R}_\alpha-\mathbf{P}_\alpha\|_{\textup{F}}].\label{in_TR'_SUM}
\end{align}

Plugging~\eqref{in_TR'_SUM} into~\eqref{in_T4}, $\emph{T}_4$ can be bounded by
\begin{align}
    \emph{T}_4 \leq & \frac{4\eta^2}{n} \sum_{\alpha=0}^{l -1}\mathbb{E}\|(\mathbf{R}_{\alpha}-\mathbf{P}_{\alpha})(\mathbf{Z}^{l -\alpha}-\mathbf{A})\|_{\textup{F}}^2+\frac{4\eta^2}{n}\mathbb{E}\|(\mathbf{R}_{l }-\mathbf{P}_{l })(\mathbf{V}-\mathbf{A})\|_{\textup{F}}^2\notag\\
    & + \frac{4\eta^2}{n}\sum_{\alpha=0}^{l -1}\sum_{\alpha^{\prime}=0,\alpha^{\prime}\neq \alpha}^{l -1}\zeta^{(2l-\alpha-{\alpha^{\prime}})\pi}\mathbb{E}\|\mathbf{R}_\alpha-\mathbf{P}_\alpha\|_{\textup{F}}^2 + \frac{4\eta^2}{n}\sum_{\alpha=0}^{l -1}\zeta^{(l-\alpha)\pi}[\mathbb{E}\|\mathbf{R}_{l}-\mathbf{P}_{l}\|_{\textup{F}}^2+\mathbb{E}\|\mathbf{R}_\alpha-\mathbf{P}_\alpha\|_{\textup{F}}^2]\notag\\
    \labelrel\leq{in_T4_FA_OP_2} &\frac{4\eta^2}{n} \sum_{\alpha=0}^{l -1}\mathbb{E}\|\mathbf{R}_{\alpha}-\mathbf{P}_{\alpha}\|_{\textup{F}}^2\|\mathbf{Z}^{l -\alpha}-\mathbf{A}\|_{\textup{op}}^2+\frac{4\eta^2}{n}\mathbb{E}\|\mathbf{R}_{l }-\mathbf{P}_{l}\|_{\textup{F}}^2\|\mathbf{V}-\mathbf{A}\|_{\textup{op}}^2\notag\\
    & + \frac{4\eta^2}{n}\sum_{\alpha=0}^{l -1}\zeta^{(l-\alpha)\pi}\mathbb{E}\|\mathbf{R}_\alpha-\mathbf{P}_\alpha\|_{\textup{F}}^2\sum_{\alpha^{\prime}=0,\alpha^{\prime}\neq \alpha}^{l -1}\zeta^{(l-\alpha^{\prime})\pi} + \frac{4\eta^2}{n}\sum_{\alpha=0}^{l -1}\zeta^{(l-\alpha)\pi}\mathbb{E}\|\mathbf{R}_{l}-\mathbf{P}_{l}\|_{\textup{F}}^2+\frac{4\eta^2}{n}\sum_{\alpha=0}^{l -1}\zeta^{(l-\alpha)\pi}\mathbb{E}\|\mathbf{R}_\alpha-\mathbf{P}_\alpha\|_{\textup{F}}^2\notag\\
    \labelrel\leq{in_T4_OP_EIG_2} & \frac{4\eta^2}{n} \sum_{\alpha=0}^{l -1}\mathbb{E}\|\mathbf{R}_{\alpha}-\mathbf{P}_{\alpha}\|_{\textup{F}}^2\zeta^{2(l-\alpha)\pi}+\frac{4\eta^2}{n}\mathbb{E}\|\mathbf{R}_{l}-\mathbf{P}_{l}\|_{\textup{F}}^2\notag\\
    & + \frac{4\eta^2}{n}\sum_{\alpha=0}^{l -1}\zeta^{(l-\alpha)\pi}\mathbb{E}\|\mathbf{R}_\alpha-\mathbf{P}_\alpha\|_{\textup{F}}^2\frac{\zeta^{\pi}}{1-\zeta^{\pi}} + \frac{4\eta^2}{n}\mathbb{E}\|\mathbf{R}_{l}-\mathbf{P}_{l}\|_{\textup{F}}^2\frac{1}{1-\zeta^{\pi}}+\frac{4\eta^2}{n}\sum_{\alpha=0}^{l -1}\zeta^{(l-\alpha)\pi}\mathbb{E}\|\mathbf{R}_\alpha-\mathbf{P}_\alpha\|_{\textup{F}}^2\notag\\
    = & \frac{4\eta^2}{n} \sum_{\alpha=0}^{l -1}(\zeta^{2(l-\alpha)\pi}+\frac{\zeta^{(l-\alpha+1)\pi}}{1-\zeta^{\pi}}+\zeta^{(l-\alpha)\pi})\mathbb{E}\|\mathbf{R}_{\alpha}-\mathbf{P}_{\alpha}\|_{\textup{F}}^2+\frac{4\eta^2}{n}(\frac{2-\zeta^{\pi}}{1-\zeta^{\pi}})\mathbb{E}\|\mathbf{R}_{l}-\mathbf{P}_{l}\|_{\textup{F}}^2,\label{bound_T4_RP}
\end{align}
where~\eqref{in_T4_FA_OP_2} follows from Lemma~\ref{lemma_Fa_op},~\eqref{in_T4_OP_EIG_2} follows from Lemma~\ref{lemma_eig_Mat} and the rules of series. Then for any $\alpha \in [0,\ldots, l - 1]$, we have:
\begin{align}
    \mathbb{E}\frac{1}{n}\|\mathbf{R}_{\alpha}-\mathbf{P}_{\alpha}\|_{\textup{F}}^2 & = \mathbb{E}\frac{1}{n}\Biggl|\Biggl|\sum_{c=\alpha q\tau+1}^{(\alpha+1)q\tau}(\mathcal{H}_c-\mathcal{J}_c)\Biggl|\Biggl|_{\textup{F}}^2\notag\\
    & = \sum_{k=1}^n\frac{1}{n}\mathbb{E}\Biggl|\Biggl|\sum_{c=\alpha q\tau+1}^{(\alpha+1)q\tau}(\nabla F_{k}(\mathbf{x}_c^{(k)}) - \nabla F_{k}(\overline{\mathbf{x}}_{c}^{(k)}))\Biggl|\Biggl|^2\notag\\
    & \labelrel\leq{in_BOUND_FE_C_D} \sum_{k=1}^n\frac{1}{n} q\tau\sum_{c=\alpha q\tau+1}^{(\alpha+1)q\tau}L^2\|\overline{\mathbf{x}}_c^{(k)}-\mathbf{x}_c^{(k)}\|^2\notag\\
    & = L^2q\tau\sum_{c=\alpha q\tau+1}^{(\alpha+1)q\tau}\frac{1}{n}\|\mathbf{X}_c(\mathbf{V}-\mathbf{I})\|_{\textup{F}}^2,\label{bound_RP}
\end{align}
where~\eqref{in_BOUND_FE_C_D} follows from Assumption~\ref{ass:smoothness}. Using the similar techniques, for the global round $l$, we obtain:
\begin{align}
    \mathbb{E}\frac{1}{n}\|\mathbf{R}_{l}-\mathbf{P}_{l}\|_{\textup{F}}^2 \leq L^2(r\tau+s-1)\sum_{c=l q\tau+1}^{l q\tau+r \tau+s-1}\frac{1}{n}\|\mathbf{X}_c(\mathbf{V}-\mathbf{I})\|_{\textup{F}}^2,\label{bound_RP_ALPHA1}
\end{align}
Plugging~\eqref{bound_RP} and~\eqref{bound_RP_ALPHA1} back into~\eqref{bound_T4_RP}, we get:
\begin{align}
    \emph{T}_4 \leq & 4\eta^2 L^2 q\tau \sum_{\alpha =0}^{l -1}(\zeta^{(2(l-\alpha ))\pi}+\frac{\zeta^{(l-\alpha +1)\pi}}{1-\zeta^{\pi}}+\zeta^{(l-\alpha )\pi})\sum_{c=\alpha q\tau+1}^{(\alpha +1)q\tau} \frac{1}{n}\|\mathbf{X}_c(\mathbf{V}-\mathbf{I})\|_{\textup{F}}^2\notag\\
    & +4\eta^2L^2(r\tau+s-1)(\frac{2-\zeta^{\pi}}{1-\zeta^{\pi}})\sum_{c=l q\tau+1}^{l q\tau+r \tau+s-1}\frac{1}{n}\|\mathbf{X}_c(\mathbf{V}-\mathbf{I})\|_{\textup{F}}^2.\notag
\end{align}
Our goal is to calculate $\sum_{t=0}^{T-1}\emph{T}_4$, which can be decomposed into $\sum_{l=0}^{p-1}\sum_{r=0}^{q-1}\sum_{s=0}^{\tau-1}\emph{T}_4.$ Summing over all iterations in the $l$-th global round for $r=0, \ldots ,q-1$ and $s=0, \ldots ,\tau-1$, we have:
\begin{align}
    \sum_{r=0}^{q-1}\sum_{s=0}^{\tau-1}\emph{T}_4  \leq  & 4\eta^2L^2q^2\tau^2 \sum_{\alpha =0}^{l -1}(\zeta^{2(l-\alpha )\pi}+\frac{\zeta^{(l-\alpha +1)\pi}}{1-\zeta^{\pi}}+\zeta^{(l-\alpha )\pi})\sum_{c=\alpha q\tau+1}^{(\alpha +1)q\tau} \frac{1}{n}\|\mathbf{X}_c(\mathbf{V}-\mathbf{I})\|_{\textup{F}}^2\notag\\
    & +2\eta^2L^2(\frac{2-\zeta^{\pi}}{1-\zeta^{\pi}})\sum_{c=l q\tau+1}^{(l+1)q\tau-1}\frac{1}{n}\|\mathbf{X}_c(\mathbf{V}-\mathbf{I})\|_{\textup{F}}^2(q^2\tau^2-q\tau).\label{bound_T4_ZETA_1}
\end{align}
Here, for $\alpha \in [0,\dots,l]$, we denote:
\begin{equation}\label{eq_Gamma}
    \Gamma_\alpha  =  \zeta^{2(l -\alpha )\pi}+2\zeta^{(l -\alpha )\pi}+\frac{\zeta^{(l -\alpha +1)\pi}}{1-\zeta^{\pi}}.
\end{equation}
We can find that $\Gamma_l  = \frac{3-\zeta^{\pi}}{1-\zeta^{\pi}}>\frac{2-\zeta^{\pi}}{1-\zeta^{\pi}}$. Thus, we can further bound~\eqref{bound_T4_ZETA_1} as
\begin{align}
    \sum_{r=0}^{q-1}\sum_{s=0}^{\tau-1}\emph{T}_4 \leq 4\eta^2L^2q^2\tau^2\sum_{\alpha =0}^{l }\Gamma_\alpha \sum_{c=l q\tau+1}^{(l+1)q\tau}\frac{1}{n}\|\mathbf{X}_c(\mathbf{V}-\mathbf{I})\|_{\textup{F}}^2.\notag
\end{align}
Then, summing over all global rounds for $l =0,\ldots,p-1$, we have:
\begin{align}
    \sum_{l =0}^{p-1}\sum_{r=0}^{q-1}\sum_{s=0}^{\tau-1}\emph{T}_4 \leq & 4\eta^2L^2 q^2\tau^2\sum_{l=0}^{p-1}\sum_{\alpha =0}^{l }\Gamma_\alpha \sum_{c=\alpha  q\tau+1}^{(\alpha +1)q\tau}\frac{1}{n}\|\mathbf{X}_c(\mathbf{V}-\mathbf{I})\|_{\textup{F}}^2\notag\\
    = & 4\eta^2L^2 q^2\tau^2\sum_{\alpha =0}^{p-1}\sum_{l=\alpha }^{p-1 }\Gamma_{l}\sum_{c=\alpha  q\tau+1}^{(\alpha +1)q\tau}\frac{1}{n}\|\mathbf{X}_c(\mathbf{V}-\mathbf{I})\|_{\textup{F}}^2.    \label{eq_delta_f}
\end{align}
Expanding the summation in\eqref{eq_delta_f}, we get:
\begin{align}
    \sum_{l =\alpha }^{p-1}\left(  \zeta^{2(l -\alpha )\pi}+2\zeta^{(l -\alpha )\pi}+\frac{\zeta^{(l -\alpha +1)\pi}}{1-\zeta^{\pi}}\right) &\leq \sum_{l =\alpha }^{\infty}\left(  \zeta^{2(l -\alpha )\pi}+2\zeta^{(l -\alpha )\pi}+\frac{\zeta^{(l -\alpha +1)\pi}}{1-\zeta^\pi}\right)\notag\\
    & \leq \frac{1}{1-\zeta^{2\pi}}+\frac{2}{1-\zeta^{\pi}}+\frac{\zeta^{\pi}}{(1-\zeta^{\pi})^{2}}.\notag
\end{align}
Let $\Omega_{2}=\frac{1}{1-\zeta^{2\pi}}+\frac{2}{1-\zeta^{\pi}}+\frac{\zeta^{\pi}}{(1-\zeta^{\pi})^2}$. Then we have:
\begin{align}
   \sum_{t=0}^{T-1}\emph{T}_4\leq 4\eta^2L^2q^2\tau^2 \Omega_2 \sum_{t=0}^{T-1}\frac{1}{n}\|\mathbf{X}_t(\mathbf{V}-\mathbf{I})\|_{\textup{F}}^2.\label{bound_t4_overtime}
\end{align}
Finally, by plugging~\eqref{bound_t3_overtime} and~\eqref{bound_t4_overtime} into~\eqref{eq_T1_34}, we arrive at:
\begin{equation}
\sum_{t=0}^{T-1}T_{1} \leq 2\eta^2\sigma^2T\left( \Omega_{1}q\tau+\frac{m-1}{n}q\tau\right)  + 4\eta^2L^2q^2\tau^2 \Omega_2 \sum_{t=0}^{T-1}\frac{1}{n}\|\mathbf{X}_t(\mathbf{V}-\mathbf{I})\|_{\textup{F}}^2. \notag
\end{equation}
\end{IEEEproof}

\begin{lemma}[Bounded $\emph{T}_2$]
\label{lemma_BOUND_T2}
Under Assumptions~\ref{ass:smoothness},~\ref{ass:gradient},~\ref{ass:mixing} and~\ref{ass:gradient_local}, we have:
\begin{equation*}
    \sum_{t=0}^{T-1} T_{2} \leq 4\eta^2L^2q^2\tau^2 \Omega_2\sum_{t=0}^{T-1} \frac{1}{n}\|\mathbf{X}_t(\mathbf{V}-\mathbf{A})\|_{\textup{F}}^2+4\eta^2q^2\tau^2 \Omega_2 T\epsilon^2. 
\end{equation*}
\end{lemma}
\begin{IEEEproof}
We first bound the term $T_2$ as follows:
\begin{align}
   \emph{T}_2 = & \frac{2\eta^2}{n}\mathbb{E}\Biggl|\Biggl| \sum_{\alpha = 0 }^{l -1}\mathbf{P}_{\alpha }(\mathbf{Z}^{l -\alpha }-\mathbf{A})+\mathbf{P}_{l }(\mathbf{V}-\mathbf{A}))\Biggl|\Biggl|_{\textup{F}}^2\notag\\
   = & \frac{2\eta^2}{n}\mathbb{E}\Biggl|\Biggl| \sum_{\alpha = 0 }^{l -1}(\mathbf{P}_{\alpha }-\mathbf{Q}_{\alpha })(\mathbf{Z}^{l -\alpha }-\mathbf{A})+\sum_{\alpha = 0 }^{l -1}\mathbf{Q}_{\alpha }(\mathbf{Z}^{l -\alpha }-\mathbf{A})+(\mathbf{P}_{l}-\mathbf{Q}_{l})(\mathbf{V}-\mathbf{A})+\mathbf{Q}_{l}(\mathbf{V}-\mathbf{A})\Biggl|\Biggl|_{\textup{F}}^2\notag\\
    \leq & \underbrace{\frac{4\eta^2}{n}\mathbb{E}\Biggl|\Biggl| \sum_{\alpha = 0 }^{l -1}(\mathbf{P}_{\alpha }-\mathbf{Q}_{\alpha })(\mathbf{Z}^{l -\alpha }-\mathbf{A})+(\mathbf{P}_{l}-\mathbf{Q}_{l})(\mathbf{V}-\mathbf{A})\Biggl|\Biggl|_{\textup{F}}^2}_{T_5}\notag\\
    & +\underbrace{\frac{4\eta^2}{n}\mathbb{E}\Biggl|\Biggl|\sum_{\alpha = 0 }^{l -1}\mathbf{Q}_{\alpha }(\mathbf{Z}^{l -\alpha }-\mathbf{A})+\mathbf{Q}_{l}(\mathbf{V}-\mathbf{A})\Biggl|\Biggl|_{\textup{F}}^2}_{T_6}.\label{eq_T2_56}
\end{align}
Then the first term $\emph{T}_5$ can be bounded as follows:
\begin{align}
    \emph{T}_5 = & \frac{4\eta^2}{n}\mathbb{E}\Biggl|\Biggl|\sum_{\alpha =0}^{l -1}(\mathbf{P}_{\alpha }-\mathbf{Q}_{\alpha })(\mathbf{Z}^{l -\alpha }-\mathbf{A})+(\mathbf{P}_{l}-\mathbf{Q}_{l})(\mathbf{V}-\mathbf{A})\Biggl|\Biggl|_{\textup{F}}^2\notag\\
    = & \frac{4\eta^2}{n} \sum_{\alpha =0}^{l -1}\mathbb{E}\|(\mathbf{P}_{\alpha }-\mathbf{Q}_{\alpha })(\mathbf{Z}^{l -\alpha }-\mathbf{A})\|_{\textup{F}}^2+4\eta^2\|(\mathbf{P}_{l}-\mathbf{Q}_{l})(\mathbf{V}-\mathbf{A})\|_{\textup{F}}^2\nonumber\\
    & + \underbrace{\frac{4\eta^2}{n}\sum_{\alpha =0}^{l -1}\sum_{\alpha ^{\prime}=0,\alpha ^{\prime}\neq \alpha }^{l -1}\mathbb{E}\underbrace{\left<(\mathbf{P}_\alpha -\mathbf{Q}_\alpha )(\mathbf{Z}^{l -\alpha }-\mathbf{A}), (\mathbf{P}_{\alpha ^{\prime}}-\mathbf{Q}_{\alpha ^{\prime}})(\mathbf{Z}^{l -{\alpha ^{\prime}}}-\mathbf{A}) \right>}_{TR^{\prime\prime}}}_{TR_0^{\prime\prime}}\nonumber\\
    & + \underbrace{\frac{8\eta^2}{n}\sum_{\alpha =0}^{l -1}\mathbb{E}\left<(\mathbf{P}_{l}-\mathbf{Q}_{l})(\mathbf{V}-\mathbf{A}), (\mathbf{P}_\alpha -\mathbf{Q}_\alpha )(\mathbf{Z}^{l -\alpha }-\mathbf{A})  \right>}_{TR_1^{\prime\prime}}.\label{in_T5}
\end{align}
Using the same techniques as in bounding $\emph{T}_{4}$, we get:
\begin{align}\label{bound_T5_PQ}
    \emph{T}_5 \leq 4\eta^2 \sum_{\alpha =0}^{l -1}(\zeta^{2(l-\alpha )\pi}+\frac{\zeta^{(l-\alpha +1)\pi}}{1-\zeta^{\pi}}+\zeta^{(l-\alpha )\pi})\mathbb{E}\frac{1}{n}\|(\mathbf{P}_{\alpha }-\mathbf{Q}_{\alpha })\|_{\textup{F}}^2+4\eta^2(\frac{2-\zeta^{\pi}}{1-\zeta^{\pi}})\mathbb{E}\frac{1}{n}\|(\mathbf{P}_{l}-\mathbf{Q}_{l})\|_{\textup{F}}^2.
\end{align}
Then for any $\alpha \in [0, l - 1]$, we have
\begin{align}
    \mathbb{E}\frac{1}{n}\|\mathbf{P}_{\alpha }-\mathbf{Q}_{\alpha }\|_{\textup{F}}^2 
    & = \mathbb{E}\frac{1}{n}\Biggl|\Biggl|\sum_{c=\alpha q\tau+1}^{(\alpha +1)q\tau}(\mathcal{J}_c-\mathcal{I}_c)\Biggl|\Biggl|_{\textup{F}}^2\notag\\
    & = \sum_{k=1}^n\frac{1}{n}\mathbb{E}\Biggl|\Biggl|\sum_{c=\alpha q\tau+1}^{(\alpha +1)q\tau}(\nabla F_{k}(\overline{\mathbf{x}}_c^{(k)}) - \nabla F_k(\mathbf{u}_c))\Biggl|\Biggl|^2\notag\\
    & \labelrel\leq{in_BOUND_PQ_C_G} \sum_{k=1}^n\frac{1}{n}q\tau\sum_{c=\alpha q\tau+1}^{(\alpha +1)q\tau}L^2\|\overline{\mathbf{x}}_c^{(k)}-\mathbf{u}_c\|^2\notag\\
    & = L^{2}q\tau\sum_{c=\alpha q\tau+1}^{(\alpha +1)q\tau}\frac{1}{n}\|\mathbf{X}_c(\mathbf{V}-\mathbf{A})\|_{\textup{F}}^{2}, \label{bound_PQ}
\end{align}
where~\eqref{in_BOUND_PQ_C_G} follows from Assumption~\ref{ass:smoothness}. Using similar techniques for the global round $l$, we obtain:
\begin{align}
    \mathbb{E}\|\mathbf{P}_{l}-\mathbf{Q}_{l}\|_{\textup{F}}^2 \leq L^2(r\tau+s-1)\sum_{c=l q\tau+1}^{l q\tau+r \tau+s-1}\frac{1}{n}\|\mathbf{X}_c(\mathbf{V}-\mathbf{A})\|_{\textup{F}}^2.\label{bound_PQ_ALPHA1}
\end{align}
Plugging~\eqref{bound_PQ} and~\eqref{bound_PQ_ALPHA1} back into~\eqref{bound_T5_PQ}, we get:
\begin{align}
    \emph{T}_5 \leq & 4\eta^2L^2q\tau \sum_{\alpha =0}^{l -1}(\zeta^{2(l-\alpha )\pi}+\frac{\zeta^{(l-\alpha +1)\pi}}{1-\zeta^{\pi}}+\zeta^{(l-\alpha )\pi})\sum_{c=\alpha q\tau+1}^{(\alpha +1)q\tau}\frac{1}{n}\|\mathbf{X}_c(\mathbf{V}-\mathbf{A})\|_{\textup{F}}^2\notag\\
    &+4\eta^2L^2(\frac{2-\zeta^{\pi}}{1-\zeta^{\pi}})(r\tau+s-1)\sum_{c=l q\tau+1}^{l q\tau+r\tau+s-1}\frac{1}{n}\|\mathbf{X}_c(\mathbf{V}-\mathbf{A})\|_{\textup{F}}^2.\notag
\end{align}
Our goal is to calculate $\sum_{t=0}^{T-1}\emph{T}_5$, which can be decomposed into $\sum_{l=0}^{p-1}\sum_{r=0}^{q-1}\sum_{s=0}^{\tau-1}\emph{T}_5.$ We sum over all iterates in $l $-th global round period, for $r=0, \ldots ,q-1$ and $s=0, \ldots ,\tau-1$:
\begin{align*}
    \sum_{r=0}^{q-1}\sum_{s=0}^{\tau-1}\emph{T}_5  \leq  & 4\eta^2L^2q\tau \sum_{\alpha=0}^{l -1}(\zeta^{2(l-\alpha)\pi}+\frac{\zeta^{(l-\alpha+1)\pi}}{1-\zeta^{\pi}}+\zeta^{(l-\alpha)\pi})\sum_{\alpha q\tau+1}^{(\alpha+1)q\tau}\frac{1}{n}\|\mathbf{X}_c(\mathbf{V}-\mathbf{A})\|_{\textup{F}}^2\notag\\
    & +2\eta^2L^2(\frac{2-\zeta^{\pi}}{1-\zeta^{\pi}})\sum_{c=l q\tau+1}^{(l+1)q\tau-1}\frac{1}{n}\|\mathbf{X}_c(\mathbf{V}-\mathbf{A})\|_{\textup{F}}^2(q^2\tau^2-q\tau).
\end{align*}
By the same method in $\emph{T}_4$, we have: 
\begin{align}
   \sum_{t=0}^{T-1}\emph{T}_5 \leq 4\eta^2L^2q^2\tau^2 \Omega_2\sum_{t=0}^{T-1} \frac{1}{n}\|\mathbf{X}_t(\mathbf{V}-\mathbf{A})\|_{\textup{F}}^2.\label{bound_t5_overtime}
\end{align}
The second term $\emph{T}_6$ can be bounded as:
\begin{align}
    \emph{T}_6 = & \frac{4\eta^2}{n}\mathbb{E}\Biggl|\Biggl|\sum_{\alpha=0}^{l -1}\mathbf{Q}_{\alpha }(\mathbf{Z}^{l -\alpha}-\mathbf{A})+\mathbf{Q}_{l}(\mathbf{V}-\mathbf{A})\Biggl|\Biggl|_{\textup{F}}^2\notag\\
    = & \frac{4\eta^2}{n} \sum_{\alpha=0}^{l -1}\mathbb{E}\|\mathbf{Q}_{\alpha }(\mathbf{Z}^{l -\alpha}-\mathbf{A})\|_{\textup{F}}^2+4\eta^2\|\mathbf{Q}_{l}(\mathbf{V}-\mathbf{A})\|_{\textup{F}}^2\nonumber\\
    & + \underbrace{\frac{4\eta^2}{n}\sum_{\alpha =0}^{l -1}\sum_{\alpha ^{\prime}=0,\alpha ^{\prime}\neq \alpha }^{l -1}\mathbb{E}\underbrace{\left<\mathbf{Q}_\alpha (\mathbf{Z}^{l -\alpha }-\mathbf{A}), \mathbf{Q}_{\alpha ^{\prime}}(\mathbf{Z}^{l -{\alpha ^{\prime}}}-\mathbf{A}) \right>}_{TR^{\prime\prime\prime}}}_{TR_0^{\prime\prime\prime}}\nonumber\\
    & + \underbrace{\frac{8\eta^2}{n}\sum_{\alpha =0}^{l -1}\mathbb{E}\left<\mathbf{Q}_{l}(\mathbf{V}-\mathbf{A}), \mathbf{Q}_\alpha (\mathbf{Z}^{l -\alpha }-\mathbf{A})  \right>}_{TR_1^{\prime\prime\prime}}.\label{in_T6}
\end{align}
$\emph{TR}^{\prime\prime\prime}$ can be bounded as:
\begin{align}
    \emph{TR}^{\prime\prime\prime} \labelrel\leq{in_T6_TR'''} & \|\mathbf{Q}_\alpha (\mathbf{Z}^{l -\alpha }-\mathbf{A})\|_{\textup{F}} \|\mathbf{Q}_{\alpha ^{\prime}}(\mathbf{Z}^{l -{\alpha ^{\prime}}}-\mathbf{A})\|_{\textup{F}}\notag\\
    \labelrel={EQ_T6_TR'''} & \|\mathbf{Q}_\alpha (\mathbf{V}-\mathbf{A})\mathbf{Z}^{l -\alpha }\|_{\textup{F}} \|\mathbf{Q}_{\alpha ^{\prime}}(\mathbf{V}-\mathbf{A})\mathbf{Z}^{l -{\alpha ^{\prime}}}\|_{\textup{F}}\notag\\
    \labelrel\leq{in_T6_FA_OP_1} & \|\mathbf{Q}_\alpha (\mathbf{V}-\mathbf{A})\|_{\textup{F}}\|\mathbf{Z}^{l -\alpha }\|_{\textup{op}} \|\mathbf{Q}_{\alpha ^{\prime}}(\mathbf{V}-\mathbf{A})\|_{\textup{F}}\|\mathbf{Z}^{l -{\alpha ^{\prime}}}\|_{\textup{op}}\notag\\
    \labelrel\leq{in_T6_OP_EIG_1} & \zeta^{(2l-\alpha -{\alpha ^{\prime}})\pi}\|\mathbf{Q}_\alpha (\mathbf{V}-\mathbf{A})\|_{\textup{F}}\|\mathbf{Q}_{\alpha ^{\prime}}(\mathbf{V}-\mathbf{A})\|_{\textup{F}}\notag\\
    \leq & \frac{1}{2}\zeta^{(2l-\alpha -\alpha ^{\prime})\pi}[\|\mathbf{Q}_\alpha (\mathbf{V}-\mathbf{A})\|_{\textup{F}}+\|\mathbf{Q}_{\alpha ^{\prime}}(\mathbf{V}-\mathbf{A})\|_{\textup{F}}],\notag
\end{align}
where~\eqref{in_T6_TR'''} follows from Cauchy-Schwarz inequality,~\eqref{EQ_T6_TR'''} follows from Lemma~\ref{lemma_comm_A_W} and~\ref{lemma_comm_Z_V},~\eqref{in_T6_FA_OP_1} follows from Lemma~\ref{lemma_Fa_op} and~\eqref{in_T6_OP_EIG_1} follows from Lemma~\ref{lemma_eig_Mat}.
Using the same techniques, we get:
\begin{align}
    \emph{TR}_1^{\prime\prime\prime} \leq \frac{4\eta^2}{n}\sum_{\alpha =0}^{l -1}\zeta^{(l-\alpha )\pi}[\mathbb{E}\|\mathbf{Q}_{l}(\mathbf{V}-\mathbf{A})\|_{\textup{F}}+\mathbb{E}\|\mathbf{Q}_\alpha (\mathbf{V}-\mathbf{A})\|_{\textup{F}}].\notag
\end{align}
Here, summing $\emph{TR}_0^{\prime\prime\prime}$ and $\emph{TR}_1^{\prime\prime\prime}$, we have:
\begin{align}
    \emph{TR}_0^{\prime\prime\prime}+\emph{TR}_1^{\prime\prime\prime} \leq & \frac{2\eta^2}{n}\sum_{\alpha =0}^{l -1}\sum_{\alpha ^{\prime}=0,\alpha ^{\prime}\neq \alpha }^{l -1}\zeta^{(2l-\alpha -\alpha ^{\prime})\pi}[\mathbb{E}\|\mathbf{Q}_\alpha (\mathbf{V}-\mathbf{A})\|_{\textup{F}}+\mathbb{E}\|\mathbf{Q}_{\alpha ^{\prime}}(\mathbf{V}-\mathbf{A})\|_{\textup{F}}]\notag \\
    & + \frac{4\eta^2}{n}\sum_{\alpha =0}^{l -1}\zeta^{(l-\alpha )\pi}[\mathbb{E}\|\mathbf{Q}_{l}(\mathbf{V}-\mathbf{A})\|_{\textup{F}}+\mathbb{E}\|\mathbf{Q}_\alpha (\mathbf{V}-\mathbf{A})\|_{\textup{F}}].\notag
\end{align}
From the symmetric property of indices of $\alpha $ and $\alpha ^{\prime}$, we obtain:
\begin{equation}\label{in_TR'''_SUM}
    \emph{TR}_0^{\prime\prime\prime}+\emph{TR}_1^{\prime\prime\prime} \leq  \frac{4\eta^2}{n}\sum_{\alpha =0}^{l -1}\sum_{\alpha ^{\prime}=0,\alpha ^{\prime}\neq \alpha }^{l -1}\zeta^{(2l-\alpha -\alpha ^{\prime})\pi}\mathbb{E}\|\mathbf{Q}_\alpha (\mathbf{V}-\mathbf{A})\|_{\textup{F}} + \frac{4\eta^2}{n}\sum_{\alpha =0}^{l -1}\zeta^{(l-\alpha )\pi}[\mathbb{E}\|\mathbf{Q}_{l}(\mathbf{V}-\mathbf{A})\|_{\textup{F}}+\mathbb{E}\|\mathbf{Q}_\alpha (\mathbf{V}-\mathbf{A})\|_{\textup{F}}].
\end{equation}
Plugging~\eqref{in_TR'''_SUM} into~\eqref{in_T6}, $\emph{T}_6$ can be bounded by:
\begin{equation*}
    \emph{T}_6 \leq  \frac{4\eta^2}{n} \sum_{\alpha=0}^{l -1}(\zeta^{2(l-\alpha )\pi}+\frac{\zeta^{(l-\alpha +1)\pi}}{1-\zeta^{\pi}}+\zeta^{(l-\alpha )\pi})\mathbb{E}\|\mathbf{Q}_\alpha (\mathbf{V}-\mathbf{A})\|_{\textup{F}}^2 + \frac{4\eta^2}{n}(\frac{2-\zeta^{\pi}}{1-\zeta^{\pi}})\mathbb{E}\|\mathbf{Q}_{l}(\mathbf{V}-\mathbf{A})\|_{\textup{F}}^2.
\end{equation*}
Then for any $0 \leq \alpha  < l $:
\begin{align*}
    \mathbb{E}\frac{1}{n}\|\mathbf{Q}_\alpha (\mathbf{V}-\mathbf{A})\|_{\textup{F}}^2 & = \sum_{i=1}^m\sum_{k\in \mathcal{S}_i}\frac{1}{n}\mathbb{E}\Biggl|\Biggl|\sum_{c=\alpha  q\tau+1}^{(\alpha +1)q\tau}(\nabla f_{i}(\mathbf{u}_c) - \nabla F(\mathbf{u}_c))\Biggl|\Biggl|^2\notag\\
    & = \sum_{i=1}^{m}\frac{n_i}{n} \mathbb{E}\Biggl|\Biggl|\sum_{c=\alpha q\tau+1}^{(\alpha +1)q\tau}(\nabla f_{i}(\mathbf{u}_c) - \nabla F(\mathbf{u}_c))\Biggl|\Biggl|^2\notag\\
    &\labelrel\leq{in_BOUND_Q_G_G} q\tau\sum_{c=\alpha q\tau+1}^{(\alpha +1)q\tau}\epsilon^2,
\end{align*}
where~\eqref{in_BOUND_Q_G_G} follows from Assumption~\ref{ass:gradient_global} and Cauchy-Schwarz inequality. Using the similar techniques, for the global round $l$, we obtain:
\begin{align}
    \mathbb{E}\frac{1}{n}\|\mathbf{Q}_{l}(\mathbf{V}-\mathbf{A})\|_{\textup{F}}^2 \leq (r\tau+s-1)\sum_{c=l q\tau+1}^{l q\tau+r\tau+s-1}\epsilon^2.\label{bound_Q_ALPHA1}
\end{align}
Plugging~\eqref{bound_PQ} and~\eqref{bound_Q_ALPHA1} back into~\eqref{bound_T5_PQ}, we get:
\begin{align}
    \emph{T}_6 \leq 4\eta^2q\tau \sum_{\alpha=0}^{l -1}(\zeta^{2(l-\alpha )\pi}+\frac{\zeta^{(l-\alpha +1)\pi}}{1-\zeta^{\pi}}+\zeta^{(l-\alpha )\pi})\sum_{c=\alpha q\tau+1}^{(\alpha +1)q\tau}\epsilon^2+4\eta^2(\frac{2-\zeta^\pi}{1-\zeta^\pi})(r\tau+s-1)\sum_{l q\tau+1}^{l q\tau+r\tau+s-1}\epsilon^2.\notag
\end{align}
By the similar procedures in $\emph{T}_4$, we obtain:
\begin{align}
    \sum_{l =0}^{p-1}\sum_{r=0}^{q-1}\sum_{s=0}^{\tau-1}\emph{T}_6 & \leq 4\eta^2q^2\tau^2\sum_{l =0}^{p-1}\sum_{\alpha =0}^{l }\Gamma_\alpha \sum_{c=\alpha  q\tau+1}^{(\alpha +1)q\tau}\epsilon^2\notag\\
    & = 4\eta^2q^2\tau^2\sum_{\alpha =0}^{p-1}\left(\sum_{l =\alpha }^{p-1}\Gamma_l\right)\sum_{c=\alpha q\tau+1}^{(\alpha +1)q\tau}\epsilon^2.\notag
\end{align}
Then we have:
\begin{align}
   \sum_{t=0}^{T-1}\emph{T}_6 \leq 4\eta^2q^2\tau^2 \Omega_2 T\epsilon^2.\label{bound_t6_overtime}
\end{align}
Plugging~\eqref{bound_t5_overtime} and~\eqref{bound_t6_overtime} into~\eqref{eq_T2_56}, finally, we have:
\begin{equation}
\sum_{t=0}^{T-1} T_{2} \leq 4\eta^2L^2q^2\tau^2 \Omega_2\sum_{t=0}^{T-1} \frac{1}{n}\|\mathbf{X}_t(\mathbf{V}-\mathbf{A})\|_{\textup{F}}^2+4\eta^2q^2\tau^2 \Omega_2 T\epsilon^2.\notag
\end{equation} 
\end{IEEEproof}

\begin{lemma}[Intra-cluster Residual error decomposition]
\label{lemma_intra_t78} 
According to Assumptions~\ref{ass:smoothness} and \ref{ass:gradient}, we have:
\begin{align*}
\mathbb{E}\frac{1}{n}\|\mathbf{X}_t(\mathbf{I}-\mathbf{V})\|_{\textup{F}}^2\leq & \underbrace{\frac{2\eta^2}{n}\mathbb{E}\Biggl|\Biggl|(\mathbf{Y}_{l,r,s}-\mathbf{R}_{l,r,s})(\mathbf{I}-\mathbf{V})\Biggl|\Biggl|_{\textup{F}}^2}_{T_7}+ \underbrace{\frac{2\eta^2}{n}\mathbb{E}\Biggl|\Biggl| \mathbf{R}_{l,r,s}(\mathbf{I}-\mathbf{V}))\Biggl|\Biggl|_{\textup{F}}^2}_{T_8},
\end{align*}
where $\mathbf{Y}_{l,r,s}, \mathbf{P}_{l,r,s}$ are defined in~\eqref{def_ypqr_lrs}.
\end{lemma}

\begin{IEEEproof}
According to the update rule, we have:
\begin{align}
    \mathbf{X}_t(\mathbf{I}-\mathbf{V}) & =(\mathbf{X}_{t-1}-\eta \mathbf{G}_{t-1})\mathbf{W}_{t-1}(\mathbf{I}-\mathbf{V})\notag\\
    & \labelrel={eq_intra_stoc} \mathbf{X}_{t-1}(\mathbf{I}-\mathbf{V})\mathbf{W}_{t-1}-\eta\mathbf{G}_{t-1}\mathbf{W}_{t-1}(\mathbf{I}-\mathbf{V})\notag\\
    & = (\mathbf{X}_{t-2}-\eta \mathbf{G}_{t-2})(\mathbf{I}-\mathbf{V})\mathbf{W}_{t-2}\mathbf{W}_{t-1}-\eta\mathbf{G}_{t-1}\mathbf{W}_{t-1}(\mathbf{I}-\mathbf{V})\notag\\
    & = \mathbf{X}_{t-2}(\mathbf{I}-\mathbf{V})\mathbf{W}_{t-2}\mathbf{W}_{t-1}-\eta \mathbf{G}_{t-2}\mathbf{W}_{t-2}\mathbf{W}_{t-1}(\mathbf{I}-\mathbf{V})-\eta\mathbf{G}_{t-1}\mathbf{W}_{t-1}(\mathbf{I}-\mathbf{V}).\notag
\end{align}
where~\eqref{eq_intra_stoc} follows the special property of doubly stochastic matrix: $\mathbf{V}\mathbf{W}_{t-1} = \mathbf{V}\mathbf{W}_{t-1} = \mathbf{W}_{t-1}$. Then, expanding the expression, we have:
\begin{align}
    \mathbf{X}_t(\mathbf{I}-\mathbf{V}) = \mathbf{X}_0(\mathbf{I}-\mathbf{V})\prod_{u=0}^{t-1}\mathbf{W}_{u}-\eta \sum_{c=1}^{t-1}\mathbf{G}_{c}\prod_{u=c}^{t-1}\mathbf{W}_{u}\left(\mathbf{I}-\mathbf{V}\right).\notag
\end{align}
Here, $u, c$ are the indexes for global iterations. Since all clients were initialized with the same model, $\mathbf{X}_0\prod_{u=1}^{t-1}\mathbf{W}_{u}(\mathbf{V}-\mathbf{A})=0$. Then, the squared norm of intra-cluster residual error can be written as:
\begin{align}\label{eq:L_C}
    \mathbb{E}\frac{1}{n}\|\mathbf{X}_t(\mathbf{I}-\mathbf{V})\|_{\textup{F}}^2 = \frac{\eta^2}{n}\mathbb{E}\|\sum_{c=1}^{t-1}\mathbf{G}_c\mathbf{\Phi}_{c,t-1}(\mathbf{I}-\mathbf{V})\|_{\textup{F}}^2.
\end{align} 
Recall that $t=l q\tau+r\tau+s$, where $l \in [0, p - 1]$ is the global round index, $r \in [0, q - 1] $ is the edge round index, and $s \in [0, \tau - 1]$ is the local iteration index. Using the same notations as~\eqref{def_Phi_1} and~\eqref{def_ypqr_1}, we also let
\begin{equation}
 \mathbf{Y}_{l,r}=\sum_{c=l q\tau+1}^{l q\tau+r \tau}\mathbf{G}_c, \quad \mathbf{P}_{l,r} = \sum_{c=l q\tau+1}^{l q\tau+r \tau}\mathcal{J}_c,\quad \mathbf{Q}_{l,r} = \sum_{c=l q\tau+1}^{l q\tau+r \tau}\mathcal{I}_c,\quad \mathbf{R}_{l,r} = \sum_{c=l q\tau+1}^{l q\tau+r \tau}\mathcal{H}_c   
\end{equation}
\begin{equation}\label{def_ypqr_lrs}
 \mathbf{Y}_{l,r,s}=\sum_{c=l q\tau+r\tau+1}^{l q\tau+r \tau+s-1}\mathbf{G}_c, \quad \mathbf{P}_{l,r,s} = \sum_{c=l q\tau+r\tau+1}^{l q\tau+r \tau+s-1}\mathcal{J}_c,\quad \mathbf{Q}_{l,r,s} = \sum_{c=l q\tau+r\tau+1}^{l q\tau+r \tau+s-1}\mathcal{I}_c,\quad \mathbf{R}_{l,r,s} = \sum_{c=l q\tau+r\tau+1}^{l q\tau+r \tau+s-1}\mathcal{H}_c
\end{equation}

Thus, we obtain:
\begin{align}
    \sum_{c=1}^{q\tau}\mathbf{G}_c\Phi_{c,t-1}(\mathbf{I}-\mathbf{V}) & =\mathbf{Y}_0\mathbf{Z}^l(\mathbf{I} -\mathbf{V}),\notag\\
    \sum_{c=q\tau+1}^{2q\tau}\mathbf{G}_c\Phi_{c,t-1}(\mathbf{I}-\mathbf{V}) & =\mathbf{Y}_1\mathbf{Z}^{l -1}(\mathbf{I}-\mathbf{V}),\notag\\
    & \ldots, \notag\\
    \sum_{c=(l -1)q\tau+1}^{l q\tau}\mathbf{G}_c\Phi_{c,t-1}(\mathbf{I}-\mathbf{V}) & =\mathbf{Y}_{l -1}\mathbf{Z}(\mathbf{I}-\mathbf{V}),\notag\\
    \sum_{c=l q\tau+1}^{l q\tau+r \tau+s-1}\mathbf{G}_c\Phi_{c,t-1}(\mathbf{I}-\mathbf{V}) & = \mathbf{Y}_{l ,r}\mathbf{V}(\mathbf{I}-\mathbf{V}) + \mathbf{Y}_{l ,r, s}\mathbf{V}(\mathbf{I}-\mathbf{V}).\notag
\end{align}
By summing them all together and Lemmas~\ref{lemma_comm_A_W},~\ref{lemma_comm_Z_V}, we get:
\begin{align}\label{EQ:SUM_L_C}
    \sum_{c=1}^{t-1}\mathbf{G}_{c}\mathbf{\Phi}_{c,t-1}(\mathbf{I}-\mathbf{V})=\sum_{\alpha =0}^{l -1}\mathbf{Y}_{\alpha }(\mathbf{Z}^{l -\alpha }-\mathbf{Z}^{l -\alpha })+\mathbf{Y}_{l,r}(\mathbf{V}-\mathbf{V})+\mathbf{Y}_{l,r,s}(\mathbf{I}-\mathbf{V}).
\end{align}

Plugging~\eqref{EQ:SUM_L_C} into~\eqref{eq:L_C}, we obtain:
\begin{align}
    \mathbb{E}\frac{1}{n}\|\mathbf{X}_t(\mathbf{I}-\mathbf{A})\|_{\textup{F}}^2 = & \frac{\eta^2}{n}\mathbb{E}\|\mathbf{Y}_{l,r,s}(\mathbf{I}-\mathbf{V})\|_{\textup{F}}^2\notag\\
    = &\frac{\eta^2}{n}\mathbb{E}\Biggl|\Biggl|(\mathbf{Y}_{l,r,s}-\mathbf{R}_{l,r,s})(\mathbf{I}-\mathbf{V}) +\mathbf{R}_{l,r,s}(\mathbf{I}-\mathbf{V})\Biggr]\Biggr|\!\Biggr|_{\textup{F}}^2\notag\\
    \leq & \underbrace{\frac{2\eta^2}{n}\mathbb{E}\Biggl|\Biggl|(\mathbf{Y}_{l,r,s}-\mathbf{R}_{l,r,s})(\mathbf{I}-\mathbf{V})\Biggl|\Biggl|_{\textup{F}}^2}_{T_7}+ \underbrace{\frac{2\eta^2}{n}\mathbb{E}\Biggl|\Biggl| \mathbf{R}_{l,r,s}(\mathbf{I}-\mathbf{V}))\Biggl|\Biggl|_{\textup{F}}^2}_{T_8}.\notag
\end{align}
\end{IEEEproof}

\begin{lemma}[Bounded $\emph{T}_7$]
\label{lemma_BOUND_T7}
Under Assumption~\ref{ass:gradient}, we have:
\begin{equation*}
    \sum_{t=0}^{T-1} T_7 \leq \frac{n-m}{n}\eta^2\tau T\sigma^2.
\end{equation*}
\end{lemma}
\begin{IEEEproof}
For the term $\emph{T}_7$, we have:
\begin{align}
    \emph{T}_7 = & \frac{2\eta^2}{n}\mathbb{E}\|(\mathbf{Y}_{l,r,s}-\mathbf{R}_{l,r,s})(\mathbf{I}-\mathbf{V})\|_{\textup{F}}^2\notag\\
    = & \frac{2\eta^2}{n}\mathbb{E}\|(\mathbf{Y}_{l,r,s}-\mathbf{R}_{l,r,s})\mathbf{I}-(\mathbf{Y}_{l,r,s}-\mathbf{R}_{l,r,s})\mathbf{V}\|_{\textup{F}}^2\notag\\
    = & \frac{2\eta^2}{n}\sum_{i=1}^m\sum_{k\in\mathcal{S}_i}\mathbb{E}\|\sum_{c=l q\tau+r\tau+1}^{l q\tau+r\tau+s-1}[(\mathbf{g}_k(\mathbf{x}_c^{(k)})-\nabla F_k(\mathbf{x}_c^{(k)}))-(\frac{1}{n_i}\sum_{k\in\mathcal{S}_i}\mathbf{g}_k(\mathbf{x}_c^{(k)})-\frac{1}{n_i}\sum_{k\in\mathcal{S}_i}\nabla F_k(\mathbf{x}_c^{(k)})]\|^2\notag\\
    \labelrel={eq_t7_var} & \frac{2\eta^2}{n}\sum_{i=1}^m\sum_{k\in\mathcal{S}_i}\mathbb{E}\|\sum_{c=l q\tau+r\tau+1}^{l q\tau+r\tau+s-1}\mathbf{g}_k(\mathbf{x}_{c}^{(k)}) - \nabla F_k(\mathbf{x}_c^{(k)})\|^2\nonumber\\
    &-2\eta^2\sum_{i=1}^{m}\frac{n_i}{n}\mathbb{E}\|\sum_{c=l q\tau+r\tau+1}^{l q\tau+r\tau+s-1}(\frac{1}{n_i}\sum_{k\in\mathcal{S}_i}\mathbf{g}_k(\mathbf{x}_{c}^{(k)}) - \frac{1}{n_i}\sum_{k\in\mathcal{S}_i}\nabla F_k(\mathbf{x}_c^{(k)})\|^2\notag\\
    \labelrel={eq_t7_indep} & \frac{2\eta^2}{n}\sum_{i=1}^m\sum_{k\in\mathcal{S}_i}\sum_{s=l q\tau+r\tau+1}^{l q\tau+r\tau+s-1}\mathbb{E}\|\mathbf{g}_k(\mathbf{x}_{s}^{(k)}) - \nabla F_k(\mathbf{x}_s^{(k)}\|^2-2\eta^2\sum_{i=1}^{m}\frac{n_i}{n}\frac{1}{n_i^2}\sum_{k\in\mathcal{S}_i}\sum_{c=l q\tau+r\tau+1}^{l q\tau+r\tau+s-1}\mathbb{E}\|\mathbf{g}_k(\mathbf{x}_{c}^{(k)})-\nabla F_k(\mathbf{x}_s^{(k)})\|^2\notag\\
    = & \frac{2\eta^2}{n}\sum_{i=1}^{m}(1-\frac{1}{n_i})\sum_{k\in\mathcal{S}_i}\sum_{c=l q\tau+r\tau+1}^{l q\tau+r\tau+s-1}\mathbb{E}\|\mathbf{g}_k(\mathbf{x}_{c}^{(k)})-\nabla F_k(\mathbf{x}_c^{(k)})\|^2\notag\\
    \labelrel\leq{in_t7_yr} & 2\frac{n-m}{n}\eta^2(s-1)\sigma^2,\notag
\end{align}
where~\eqref{eq_t7_var} follows from $\mathbb{E}[\|\bm{a}-\mathbb{E}(\bm{a})\|^2]=\mathbb{E}[\|\bm{a}\|^2]-\|\mathbb{E}(\bm{a})\|^2$ with $\bm{a}\in\mathbb{R}^d$, and~\eqref{eq_t7_indep} holds due to Assumption~\ref{ass:gradient}.~\eqref{in_t7_yr} also follows from Assumption~\ref{ass:gradient}. Our goal is to calculate $\sum_{t=0}^{T-1}\emph{T}_7$, which can be decomposed into $\sum_{l=0}^{p-1}\sum_{r=0}^{q-1}\sum_{s=0}^{\tau-1}\emph{T}_7.$ We sum over all iterates in $l $-th global round period, for $r=0, \ldots ,q-1$ and $s=0, \ldots ,\tau-1$:
\begin{equation*}
    \sum_{r=0}^{q-1}\sum_{s=0}^{\tau-1}\emph{T}_7  \leq \frac{n-m}{n}\eta^2q\tau(\tau-1)\sigma^2.
\end{equation*}
Then, summing over all hub network update periods from $l =0,\ldots,p-1$, we have:
\begin{equation*}
    \sum_{l =0}^{p-1}\sum_{r=0}^{q-1}\sum_{s=0}^{\tau-1}\emph{T}_7 \leq \frac{n-m}{n}\eta^2(\tau-1) T\sigma^2.
\end{equation*}
Then we have:
\begin{equation*}
   \sum_{t=0}^{T-1}\emph{T}_7 \leq \frac{n-m}{n}\eta^2\tau T\sigma^2.
\end{equation*}
\end{IEEEproof}

\begin{lemma}[Bounded $\emph{T}_8$]
\label{lemma_BOUND_T8}
Under Assumptions~\ref{ass:smoothness},~\ref{ass:gradient},~\ref{ass:gradient_local}, we have:
\begin{align}
    \sum_{t=0}^{T-1} T_8 \leq 2\eta^2L^2\tau^2 \sum_{t=0}^{T-1}\frac{1}{n}\|\mathbf{X}_s(\mathbf{I}-\mathbf{V})\|_{\textup{F}}^2+2\eta^2\tau^2 T\sum_{i=1}^m\frac{n_i}{n}\epsilon_j^2.\notag
\end{align}
\end{lemma}

\begin{IEEEproof}
For the term $\emph{T}_8$, we have:
\begin{align}
    \emph{T}_8 = & \frac{2\eta^2}{n}\mathbb{E}\|\mathbf{R}_{l,r,s}(\mathbf{I}-\mathbf{V}))\|_{\textup{F}}^2\notag\\
    \leq & \frac{4\eta^2}{n}\mathbb{E}\| (\mathbf{R}_{l,r,s}-\mathbf{P}_{l,r,s})(\mathbf{I}-\mathbf{V})\|_{\textup{F}}^2 + \frac{4\eta^2}{n}\|\mathbf{P}_{l,r,s}(\mathbf{I}-\mathbf{V})\|_{\textup{F}}^2\notag\\
    \labelrel\leq{in_t8_fa_op_1} & \frac{4\eta^2}{n}\mathbb{E}\| (\mathbf{R}_{l,r,s}-\mathbf{P}_{l,r,s})\|_{\textup{F}}^2\|(\mathbf{I}-\mathbf{V})\|_{\textup{op}}^2 + \frac{4\eta^2}{n}\|\mathbf{P}_{l,r,s}(\mathbf{I}-\mathbf{V})\|_{\textup{F}}^2\notag\\
    \labelrel={eq_t8_op} & \frac{4\eta^2}{n}\mathbb{E}\| (\mathbf{R}_{l,r,s}-\mathbf{P}_{l,r,s})\|_{\textup{F}}^2 + \frac{4\eta^2}{n}\|\mathbf{P}_{l,r,s}(\mathbf{I}-\mathbf{V})\|_{\textup{F}}^2,\notag
\end{align}
where~\eqref{in_t8_fa_op_1} is due to Lemma~\ref{lemma_Fa_op},~\eqref{eq_t8_op} follows from Lemma~\ref{lemma_i_v_op}.
\begin{align}
    \mathbb{E}\frac{1}{n}\| (\mathbf{R}_{l,r,s}-\mathbf{P}_{l,r,s})\|_{\textup{F}}^2 = &\sum_{k=1}^n\frac{1}{n}\|\sum_{c=l q\tau+r\tau}^{l q\tau+r\tau+s-1}(\nabla F_k(\mathbf{x}_c^{(k)})-\nabla F_k(\overline{\mathbf{x}}_c^{(k)}))\|^2\notag\\
    \leq &\frac{s-1}{n}\sum_{k=1}^n\sum_{c=l q\tau+r\tau}^{l q\tau+r\tau+s-1}\|(\nabla F_k(\mathbf{x}_c^{(k)})-\nabla F_k(\overline{\mathbf{x}}_c^{(k)}))\|^2\notag\\
    \labelrel\leq{in_t8_liptz} & L^2(s-1)\sum_{k=1}^n\frac{1}{n}\sum_{c=l q\tau+r\tau}^{l q\tau+r\tau+s-1}\|\mathbf{x}_c^{(k)}-\overline{\mathbf{x}}_c^{(k)}\|^2\notag\\
    = & L^2(s-1)\sum_{c=l q\tau+r\tau+1}^{l q\tau+r\tau+s-1}\frac{1}{n}\|\mathbf{X}_c(\mathbf{I}-\mathbf{V})\|_{\textup{F}}^2.\notag
\end{align}
where~\eqref{in_t8_liptz} follows from Assumption~\ref{ass:smoothness}.
\begin{align}
    \mathbb{E}\frac{1}{n}\|\mathbf{P}_{l,r,s}(\mathbf{I}-\mathbf{V})\|_{\textup{F}}^2 = & \sum_{i=1}^m\frac{n_i}{n}\sum_{k\in\mathcal{S}_i}\frac{1}{n_i}\|\sum_{c=l q\tau+r\tau+1}^{l q\tau+r\tau+s-1}( \nabla F_k(\overline{\mathbf{x}}_c^{(k)})-\nabla f_i(\overline{\mathbf{x}}_c^{(k)}))\|^2\notag\\
    \leq & (s-1)\sum_{i=1}^m\frac{n_i}{n}\sum_{k\in\mathcal{S}_i}\frac{1}{n_i}\sum_{c=l q\tau+r\tau+1}^{l q\tau+r\tau+s-1}\|( \nabla F_k(\overline{\mathbf{x}}_c^{(k)})-\nabla f_i(\overline{\mathbf{x}}_c^{(k)}))\|^2\notag\\
    \labelrel\leq{in_t8_assm_intra} & (s-1)\sum_{c=l q\tau+r\tau+1}^{l q\tau+r\tau+s-1}\sum_{i=1}^m\frac{n_i}{n}\epsilon_i^2,\notag
\end{align}
where~\eqref{in_t8_assm_intra} holds due to Assumption~\ref{ass:gradient_local}. Then, we have:
\begin{align}
    \emph{T}_8\leq & 4\eta^2L^2(s-1)\sum_{c=l q\tau+r\tau+1}^{l q\tau+r\tau+s-1}\frac{1}{n}\|\mathbf{X}_c(\mathbf{I}-\mathbf{V})\|_{\textup{F}}^2+4\eta^2(s-1)\sum_{c=l q\tau+r\tau+1}^{l q\tau+r\tau+s-1}\sum_{i=1}^m\frac{n_i}{n}\epsilon_i^2.  \notag
\end{align}
Our goal is to calculate $\sum_{t=0}^{T-1}\emph{T}_8$, which can be decomposed into $\sum_{l=0}^{p-1}\sum_{r=0}^{q-1}\sum_{s=0}^{\tau-1}\emph{T}_8.$ We sum over all iterates in $l $-th global round period, for $r=0, \ldots ,q-1$ and $s=0, \ldots ,\tau-1$:
\begin{equation*}
    \sum_{r=0}^{q-1}\sum_{s=0}^{\tau-1}\emph{T}_8  \leq  2\eta^2L^2\tau(\tau-1)\sum_{c=l q\tau+1}^{(l+1) q\tau-1}\frac{1}{n}\|\mathbf{X}_c(\mathbf{I}-\mathbf{V})\|_{\textup{F}}^2+2\eta^2\tau(\tau-1)\sum_{c=l q\tau+1}^{(l+1) q\tau-1}\sum_{i=1}^m\frac{n_i}{n}\epsilon_i^2.
\end{equation*}
Then, summing over all hub network update periods from $l =0,\ldots,p-1$, we have:
\begin{equation*}
    \sum_{l =0}^{p-1}\sum_{r=0}^{q-1}\sum_{s=0}^{\tau-1}\emph{T}_8  \leq 2\eta^2L^2\tau^2 \sum_{t=0}^{T-1}\frac{1}{n}\|\mathbf{X}_t(\mathbf{I}-\mathbf{V})\|_{\textup{F}}^2+2\eta^2\tau^2 T\sum_{i=1}^m\frac{n_i}{n}\epsilon_i^2.
\end{equation*}
Then we have:
\begin{equation*}
   \sum_{t=0}^{T-1}\emph{T}_8 \leq 2\eta^2L^2\tau^2 \sum_{t=0}^{T-1}\frac{1}{n}\|\mathbf{X}_t(\mathbf{I}-\mathbf{V})\|_{\textup{F}}^2+2\eta^2\tau^2 T\sum_{i=1}^m\frac{n_i}{n}\epsilon_i^2.
\end{equation*}
\end{IEEEproof}

\section{Proof of lemma 2}\label{proof_lemma2}
The proof idea of Lemma 2 is to split $\sum_{t=0}^{T-1}\mathbb{E}\frac{1}{n}\|\mathbf{X}_t(\mathbf{V}-\mathbf{A})\|_{\textup{F}}^2$ into two terms by Lemma~\ref{lemma:inter_t12}, then Lemmas~\ref{lemma_bound_t1} and~\ref{lemma_BOUND_T2} derive the upper bounds of two terms, respectively.
\begin{IEEEproof}
Plugging Lemmas~\ref{lemma_bound_t1} and~\ref{lemma_BOUND_T2} into Lemma~\ref{lemma:inter_t12}, we have:
\begin{align}
    \sum_{t=0}^{T-1}\mathbb{E}\frac{1}{n}\|\mathbf{X}_t(\mathbf{V}-\mathbf{A})\|_{\textup{F}}^2 \leq &2\eta^2T(\Omega_1 q\tau+\frac{m-1}{n}q\tau)\sigma^2+4\eta^2q^2\tau^2\Omega_2 T\epsilon^2\notag\\ &+4\eta^2L^2q^2\tau^2\Omega_2\sum_{t=0}^{T-1}\mathbb{E}\frac{1}{n}\|\mathbf{X}_t(\mathbf{V}-\mathbf{I})\|_{\textup{F}}^2+4\eta^2L^2q^2\tau^2\Omega_2\sum_{t=0}^{T-1}\mathbb{E}\frac{1}{n}\|\mathbf{X}_t(\mathbf{V}-\mathbf{A})\|_{\textup{F}}^2\notag
\end{align}
Moving the last term to the left, we obtain:
\begin{align}
    (1-4\eta^2L^2q^2\tau^2\Omega_2)\frac{1}{T}\sum_{t=0}^{T-1}\mathbb{E}\frac{1}{n}\|\mathbf{X}_t(\mathbf{V}-\mathbf{A})\|_{\textup{F}}^2 \leq & 2\eta^2(\Omega_1q\tau+\frac{m-1}{n}q\tau)\sigma^2+4\eta^2q^2\tau^2\Omega_2 \epsilon^2 \notag\\
    & +4\eta^2L^2q^2\tau^2\Omega_2\frac{1}{T}\sum_{t=0}^{T-1}\mathbb{E}\frac{1}{n}\|\mathbf{X}_t(\mathbf{V}-\mathbf{I})\|_{\textup{F}}^2.\notag
\end{align}
Then
\begin{align}
    \frac{1}{T}\sum_{t=0}^{T-1}\mathbb{E}\frac{1}{n}\|\mathbf{X}_t(\mathbf{V}-\mathbf{A})\|_{\textup{F}}^2 \leq & \frac{2\eta^2(\Omega_1q\tau+\frac{m-1}{n}q\tau)\sigma^2}{1-4\eta^2L^2q^2\tau^2\Omega_2}\notag\\
    &+\frac{4\eta^2q^2\tau^2\Omega_2 }{1-4\eta^2L^2q^2\tau^2\Omega_2}(\epsilon^2+\frac{L^2}{T}\sum_{t=0}^{T-1}\mathbb{E}\frac{1}{n}\|\mathbf{X}_t(\mathbf{V}-\mathbf{I})\|_{\textup{F}}^2).\notag
\end{align}
\end{IEEEproof}

\section{Proof of lemma 3}\label{proof_lemma3}
The proof idea of Lemma 3 is to split $\sum_{t=0}^{T-1}\mathbb{E}\frac{1}{n}\|\mathbf{X}_t(\mathbf{I}-\mathbf{V})\|_{\textup{F}}^2$ into two terms by Lemma~\ref{lemma_intra_t78}, then Lemmas~\ref{lemma_BOUND_T7},~\ref{lemma_BOUND_T8} derive the upper bounds of two terms, respectively.
\begin{IEEEproof}
Plugging Lemmas~\ref{lemma_BOUND_T7},~\ref{lemma_BOUND_T8} into Lemma~\ref{lemma_intra_t78}, we have:
\begin{align}
    \frac{1}{T}\sum_{t=0}^{T-1}\mathbb{E}\frac{1}{n}\|\mathbf{X}_t(\mathbf{I}-\mathbf{V})\|_{\textup{F}}^2 \leq \frac{n-m}{n}\eta^2\tau \sigma^2++2\eta^2\tau ^2\sum_{i=1}^m\frac{n_i}{n}\epsilon_i^2+2\eta^2L^2\tau^2\frac{1}{T}\sum_{t=0}^{T-1}\mathbb{E}\frac{1}{n}\|\mathbf{X}_t(\mathbf{I}-\mathbf{V})\|_{\textup{F}}^2.\notag
\end{align}
Moving the last term to left, we obtain:
\begin{align}
    (1-2\eta^2L^2\tau^2)\frac{1}{T}\sum_{t=0}^{T-1}\mathbb{E}\frac{1}{n}\|\mathbf{X}_t(\mathbf{I}-\mathbf{V})\|_{\textup{F}}^2 \leq \frac{n-m}{n}\eta^2\tau \sigma^2++2\eta^2\tau^2 \sum_{i=1}^m\frac{n_i}{n}\epsilon_i^2.\notag
\end{align}
Then
\begin{align}
    \frac{1}{T}\sum_{t=0}^{T-1}\mathbb{E}\frac{1}{n}\|\mathbf{X}_t(\mathbf{I}-\mathbf{V})\|_{\textup{F}}^2 \leq \frac{(\frac{n-m}{n})\eta^2\tau \sigma^2}{1-2\eta^2L^2\tau^2}+\frac{2\eta^2\tau^2 \sum_{i=1}^m\frac{n_i}{n}\epsilon_i^2}{1-2\eta^2L^2\tau^2}.\notag
\end{align}
\end{IEEEproof}

\end{document}